\documentclass[preprintnumbers,article,amsmath,amssymb,floatfix,12pt,prd,onecolumn, superscriptaddress,nofootinbib,aps]{revtex4-2}
\usepackage{bm}
\usepackage{amsfonts}
\usepackage{latexsym}
\usepackage[latin1,utf8]{inputenc}
\usepackage{graphicx}
\usepackage{amsmath}
\usepackage{palatino}
\usepackage{mathpazo}
\usepackage{makecell}
\usepackage{textcomp}
\usepackage{float}
\usepackage{booktabs}
\usepackage{dcolumn}
\usepackage{ragged2e}
\usepackage{hyperref}
\hypersetup{colorlinks,citecolor=blue}
\hypersetup{colorlinks=true,linkcolor=blue,filecolor=magenta,urlcolor=blue}
\usepackage{amsmath}
\usepackage{natbib}
\usepackage{xcolor}
\usepackage{orcidlink}
\usepackage{epsfig}
\usepackage{caption}
\usepackage[toc]{appendix}
\usepackage{commath}
\usepackage{cancel}
\usepackage{csquotes}
\usepackage{placeins}
\usepackage{graphicx}
\usepackage{float}
\usepackage[caption=false]{subfig}

\begin{document}

%%%%%%%%%%%%%%%%%%%%%%%%%%%%%%%%%%%%%%%%%%%%%%%%%%%%%%%%%%%%%%%
 \newcommand{\hsp}{\hspace{0.1mm}}
 \newcommand{\bq}{\begin{equation}}
 \newcommand{\eq}{\end{equation}}
 \newcommand{\bqn}{\begin{eqnarray}}
 \newcommand{\eqn}{\end{eqnarray}}
 \newcommand{\nb}{\nonumber}
 \newcommand{\lb}{\label}
 \newcommand{\pp}{\partial}
 \newcommand{\om}{\ensuremath{\Omega_{M}}}
 \newcommand{\ct}{\ensuremath{c_{T}^{2}}}
 \newcommand{\mo}{\ensuremath{\mathcal{M}_{0}}}
 \newcommand{\ti}[1]{\ensuremath{\tilde{#1}}}

\title{Stable initial conditions and analytically approximate solutions of cosmological perturbations in a modified loop quantum cosmology}

\date{\today}

\author{Rui Pan}
\email{rui\_pan1@baylor.edu}
%\email{rui$\_$pan1@baylor.edu}

\author{Jamal Saeed}
\email{jamal\_saeed1@baylor.edu}
%\email{Jamal$\_$Saeed1@baylor.edu}

\author{Anzhong Wang}
 \email{anzhong\_wang@baylor.edu; the corresponding author}
% \email{anzhong$\_$wang@baylor.edu; the corresponding author}

\affiliation{GCAP-CASPER, Department of Physics and Astronomy, Baylor University, Waco, TX 76798-7316, USA}
 
\begin{abstract}
In this paper, we study cosmological perturbations  in a modified theory of loop quantum cosmologies, the so-called mLQC-I model. Our purposes are two-fold: First, using  a method developed by Birrell and Davies, we identify an initial state  in the remote contracting phase, which turns out 
to be stable, minimize particle creations and diagonalize the Hamiltonian, despite the fact that at this time some modes may be still outside of the Hubble horizon and not in their adiabatic states. Second, using the uniform asymptotic approximation  method, we obtain the first-order approximate solutions of the mode function in terms of either the Airy functions, or the first or second kind of cylindrical functions, depending on the values of the wavenumber.  In each case, the mode function contains  two integration constants, which are  uniquely determined by the initial state.  

\end{abstract}

\maketitle
\section{Introduction}
\lb{Sec:introduction}

The inflationary  cosmology, aside from solving several fundamental and conceptual problems, such as flatness, horizon, and exotic-relics, of the standard big bang cosmology, also provides a causal mechanism for generating structures in the universe and the spectrum of cosmic microwave background (CMB) anisotropies \cite{Guth:1980zm}.
These are matched to observations with unprecedented precision, especially after the latest release of  the more precise CMB measurements from the Planck satellite \cite{Planck:2018jri}, and recently from DR6 data release of the Atacama Cosmology Telescope \cite{ACT:2025fju}.

However, it is well-known that this paradigm is sensitive to the ultraviolet (UV) physics, and its successes are tightly contingent on the understanding of this UV physics \cite{Brandenberger:2012aj,Silverstein:2016ggb,Baumann:2014nda}. In particular, if the inflationary phase lasts somewhat longer than the minimal period required to solve the above mentioned problems, the length scales we observe today can originate from modes that are smaller than the Planck length during inflation. Then, the treatment of the underlying quantum field theory on a classical spacetime background becomes questionable, as now the quantum geometric effects are expected to be large, and the space and time cannot be treated  classically  any longer. This is often referred to as  {\em the trans-Planckian problem} of cosmological fluctuations \cite{Brandenberger:2012aj}.

The second problem of the inflationary paradigm is related to the existence of the big bang singularity \cite{Borde:1993xh,Borde:2001nh}, with which it is not clear how to impose initial conditions. Instead, one often ignores the pre-inflationary dynamics and sets the initial conditions at a sufficiently early time so that all the observational modes were inside the Hubble horizon. In the slow-roll inflation scenario, the spacetime becomes almost de Sitter,  and the Bunch-Davies (BD) vacuum becomes a natural choice \cite{Bunch:1978yq}. 
However, it is still an open question on how such a vacuum state can be realized dynamically in the framework of quantum cosmology (QC), considering the fact that a pre-inflationary phase always exists between the Planck and inflation scales, which are about $10^{12}$ orders of magnitude difference in terms of energy densities \cite{Ashtekar:2011ni}.  During this phase, particle creations are inevitable.

So far, various theories of quantum gravity (QG) have been proposed, and among them string/M-Theory \cite{Becker:2006dvp} and Loop Quantum Gravity (LQG) have been extensively investigated \cite{Thiemann:2007pyv,Rovelli_Vidotto_2014}.  
In particular, string/M-Theory  unifies all the interactions and could therefore provide a framework from which we may hope to derive all the physical laws. On the other hand, LQG is based on a canonical approach to QG introduced earlier by Dirac, Bergmann, Wheeler and DeWitt \cite{PhysRev.160.1113}. However, instead of using metrics as the quantized objects \cite{PhysRev.160.1113}, LQG is formulated in terms of connections and is a non-perturbative and background-independent quantization of GR \cite{Thiemann:2007pyv,Rovelli_Vidotto_2014}. The gravitational sector is described by the {SU}(2)-valued 
Ashtekar connection and its associated conjugate momentum, the densitized triad, from which one defines the holonomy of Ashtekar's connection and the flux of the densitized triad \cite{PhysRevLett.57.2244}. Then, one can construct the full kinematical Hilbert space in a rigorous and well-defined way \cite{Thiemann:2007pyv,Rovelli_Vidotto_2014}.

A concrete application of LQG is loop quantum cosmology (LQC) \cite{PhysRevLett.86.5227}, in which  various cosmological models have been studied  (For a recent reviews of LQC, see \cite{Li:2023dwy,Agullo:2023rqq} and references therein). These include the  spatially-flat ($k = 0$),  closed ($k = 1$)  Friedmann-Lema\^itre-Robertson-Walker (FLRW) models, the Bianchi type I, II and IX models, the Kantowski-Sachs cosmological models, and the Gowdy universe. In the last case the model contains an infinite number of degrees of freedom. Based on a rigorous mathematical framework, in all these models a coherent picture of Planck scale physics has emerged: {\em the big bang singularity is replaced by a quantum bounce, purely due to quantum geometric effects} \cite{Ashtekar:2011ni,Li:2023dwy}. More important, its cosmological perturbations are consistent with current experiments \cite{Agullo:2023rqq,Li:2024xxz} and can alleviate some tensions between the classical inflationary models and CMB observations \cite{Ashtekar:2020gec,Ashtekar:2021izi,Agullo:2020fbw,Agullo:2020cvg,Martin-Benito:2023nky}.

Despite a wealth of results on the singularity resolution and consistency of its theoretical predictions and observations, an important issue that has remained open is its connection to LQG (see, for example, \cite{Beetle:2021zsv,Cortez:2021njc,Bojowald:2021kzv} for discussions). In particular, LQC contains two assumptions and none of them is in the spirit of LQG. The first is the minisuperspace assumption, that is, in LQC one first classically reduces the Hamiltonian from infinitely many degrees of freedom to a few gravitational ones by imposing homogeneity and then to quantize the classically reduced Hamiltonian by using the techniques of LQG. However, in LQG the processes of symmetry reduction and quantization do not commute in general, and it is important to understand how well the physics of the full LQG is captured by LQC. 
The second assumption is concerned with the quantization process. In LQG, the gravitational Hamiltonian usually consists of two parts, the Euclidean and Lorentz parts
\bq
\lb{eq1.1}
{\cal{H}}_G = {\cal{H}}_G^{\text{Euclidean}} + {\cal{H}}_G^{\text{Lorentz}}.
\eq
The quantization for each part usually follows quite different regularization schemes \cite{Thiemann:2007pyv,Rovelli_Vidotto_2014}. However, LQC takes advantage of the fact that, in the flat FLRW universe, the two parts are classically proportional to each other. Therefore, one first writes the total gravitational Hamiltonian in terms of ${\cal{H}}_G^{\text{Euclidean}}$, and then quantizes only the Euclidean part.

In the past decade or so, the above issues have been extensively studied by both bottom-up and top-down approaches, and found that  {\em LQC and its major predictions are robust}. In particular, the big bang singularity is resolved in all the models studied so far. However, dramatic changes of the evolution of the universe in the pre-bounce phase are also found \cite{Li:2021mop}.

In the bottom-up approach, symmetries are still imposed before quantization \cite{Yang:2009fp}. However, the Lorentzian term is quantized independently, using Thiemann's regularization from the full theory of LQG \cite{Thiemann:2007pyv}. In doing so, it was found that the resultant wave-function now is described by a fourth-order differential equation \cite{Yang:2009fp}  (For a systematic derivation of the
model, see \cite{Assanioussi:2018hee}), instead of the second-order one obtained in LQC \cite{Ashtekar:2006wn}.

In the top-down approach, a modified LQC model was obtained using complexifier coherent states and by treating the Euclidean and Lorentzian terms in the scalar constraint separately, as done in full LQG  \cite{Dapor:2017gdk}, which is quite similar to the one obtained first in \cite{Yang:2009fp}, referred to as mLQC-I \cite{Li:2021mop}. However, since this approach relies on a fixed graph, it has some important drawbacks, most notably subtle issues with the classical limit \cite{Dapor:2019mil}. To address these shortcomings, the above work was further extended to incorporate graph-changing dynamics in an approach  based on the path-integral reformulation of LQG \cite{Han:2021cwb}, resulting in a consistent model, and this model is precisely mLQC-I.

Therefore, in this paper we shall focus ourselves on mLQC-I. For sharply peaked states, the resulting quantum dynamics are well described by effective Friedman-Raychaudhuri (FR) equations \cite{Yang:2009fp,Li:2018opr}. Then, it was found that not only is the resolution of the big bang singularity generic \cite{Li:2018opr,Li:2018fco,Saini:2018tto,Li:2019ipm,Saeed:2024xhk},
but also a subsequent desired slow-roll inflation typically occurs after the quantum bounce \cite{Li:2019ipm}, with the probability given by
\bqn
\lb{eq1.2}
P^{\;\text{mLQC-I}}(\text{not realized})\lesssim  1.12\times 10^{-5}, 
\eqn
where $P^{\;\text{mLQC-I}}(\text{not realized})$ denotes the probability for the desired slow-roll not to happen.

For the kinetic-energy dominated initial conditions imposed at the quantum bounce, the evolution of the homogeneous and isotropic quantum universe is universal, and the corresponding expansion factor $a(t)$ and scalar field $\phi(t)$ can be  given explicitly  in both of the pre- and post-bounce phases  \cite{Li:2018opr,Li:2018fco,Li:2019ipm,Saeed:2024xhk}. It is this universal behavior that significantly simplifies our analytical investigations of the mode function of the linear perturbations. In fact, finding analytical approximate solutions of the mode function becomes possible precisely because the background is known analytically, a case quite similar to the slow-roll inflation, where the background is quasi-de Sitter. Furthermore, the resulting cosmological perturbations are also consistent with current CMB observations \cite{Agullo:2018wbf,Li:2019qzr,Li:2020mfi}. 

%\begin{figure}[htp]
\begin{figure*}[htbp]
\centering
\resizebox{0.7\textwidth}{!}{%
\begin{tikzpicture}
\tikzstyle{every node}=[font=\LARGE]
\draw [ line width=2pt ] (3.75,18.75) rectangle (27,-1.75);
\node (tikzmaker) [shift={(-0.75, -0.5)}] at (20,7.5) {};
\begin{scope}[rotate around={41.5:(3.75,-1.75)}]
\draw[domain=3.9:34.55,samples=100,smooth, line width=1.5pt] plot (\x,{0.2*sin(4.4*\x r -6 r ) + -1.75});
\end{scope}
\draw [line width=1.3pt, dashed] (3.75,13) .. controls (14.75,14.5) and (17.25,15) .. (27,18.75);
\draw [line width=1.5pt, dashed] (3.75,7) .. controls (11.75,9.25) and (15.5,11.75) .. (27,18.75);
\node [font=\Huge] at (3,18.75) {A};
\node [font=\Huge] at (3,-1.75) {C};
\node [font=\Huge] at (27.8,-1.75) {D};
\node [font=\Huge] at (27.8,18.75) {B};
\node [font=\Huge] at (3,13) {a};
\node [font=\Huge] at (3,7) {b};
\node [font=\Huge] at (15,19.5) {$\hat{t}=+\infty \; (\hat{\eta}=0)$};
\node [font=\Huge, rotate around={25:(0,0)}] at (12,12.5) {$\hat{t}=constant$};
\node [font=\Huge, rotate around={44:(0,2)}] at (14,6.5) {$\hat{t}=-\infty \; (\hat{\eta}=-\infty)$};
\end{tikzpicture}
}%
\caption{The Penrose diagram of the de Sitter spacetime.  
The spacelike horizontal line AB corresponds to $\hat{t} = +\infty$, and the diagonal line BC corresponds to $\hat{t} = -\infty$, which represents a null hypersurface. The curved dashed lines aB and bB correspond to the hypersurfaces of $\hat{t}=$ Constant. }
\lb{fig1:Penrose}
\end{figure*}
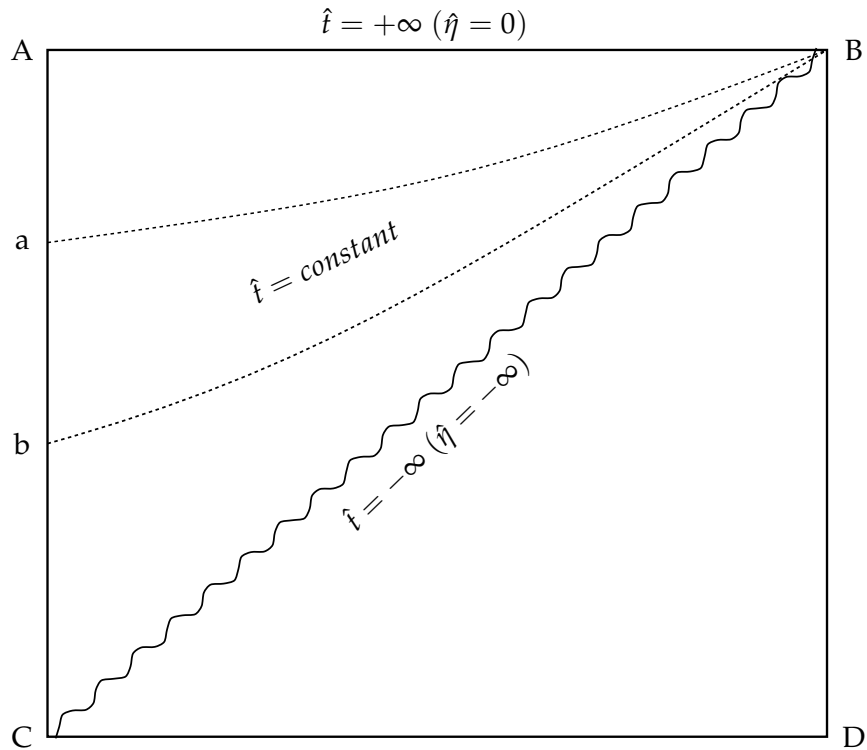

The purposes of this paper are two-folds. First,  the initial conditions of the perturbations. This problem actually involves two aspects: when the initial conditions should be imposed, and what type of initial conditions should be imposed. Naturally, these two aspects are not independent. In fact, the appropriate initial conditions depend on the time at which they are imposed. In mLQC-I, a natural choice for setting initial conditions is a moment in the remote contracting phase, during which the spacetime is approximately de Sitter \cite{Saeed:2024xhk}. 
In particular, using the Birrell-Davies method \cite{Birrell:1982ix}  we find that in this phase the initial condition should be chosen as  
\bq
\lb{eq1.3} 
 \nu^{\text{initial}}_k(\eta) \simeq  \frac{1}{\sqrt{2k}}  e^{-ik\eta}\left(1 - \frac{i}{k\eta}\right), \; (\eta \ll 1).
\eq
It must be noted that this condition is different from that of  the BD vacuum \cite{Bunch:1978yq},
commonly adopted in the classical slow-roll inflation, as now the expansion factor is very large, $a(\eta) = -1/(H\eta) \gg 1$, that is,  $\eta \ll 1$, so that $(k\eta)^{-1} \gg 1$ for sufficiently early time. Then,  the term $i/(k\eta)$ in Eq.(\ref{eq1.3}) is usually not negligible. On the other hand,
the BD vacuum is given by \cite{Birrell:1982ix}
\bq
\lb{eq1.3b} 
 \nu^{\text{initial, BD}}_k(\eta) \simeq  \frac{1}{\sqrt{2k}}  e^{-ik\eta}, \; (\eta \gg 1).
\eq
To distinguish these two vacuums, in \cite{Li:2021mop} the one given by Eq.(\ref{eq1.3}) is called {\em the de Sitter vacuum}. This difference can be further seen from the Penrose diagram of the de Sitter spacetime given by Fig. \ref{fig1:Penrose}. In the classical slow-roll inflation, the BD vacuum is usually imposed in the sufficiently early time $a(\eta_i) \ll 1$, so the current observational modes were all within the Hubble horizon at $\eta_i$, which corresponds to the diagonal line BC and is a null hypersurface. However, in mLQC-I the universe is very large in the deep remote contracting phase where $a(\eta_i) \gg 1$, and this corresponds to the top horizontal line AB, which now is spacelike. During this phase, some observational modes were out of the Hubble horizon, and became non-adiabatic. Even for these modes, the above initial condition is still stable and minimizes particle creations \cite{Penna-Lima:2022dmx}. In addition, the asymptotic Hamiltonian diagonalization prescription also leads to the above initial condition  \cite{ElizagaNavascues:2019itm}.

Second,  we systematically derive  the mode function of the cosmological scalar perturbations using the uniform asymptotic approximation (UAA) \cite{Olver:1997}. The generalization of such studies to the tensor mode function is straightforward. One of the advantages of the UAA method is that the errors can be well controlled by properly choosing the function introduced in the method  
and becomes particularly powerful when the WKB method is not applicable \cite{Joras:2008ck,Ashoorioon:2011eg}. In doing so, we find that in some cases the mode function is a linear combination of the second kind of the cylindrical functions. To our best knowledge, this presents the first example of the use of such special functions in cosmology.

The rest of the paper is organized as follows: In Sec. II, we briefly review the background evolution of the universe, and pay particular attention to the universal properties for the initial conditions of the background that are kinetic-energy dominated at the quantum bounce ($t = t_B$). In Sec. III, we write down the Mukhanov-Sasaki (MS) equation of mLQC-I in the dressed metric approach developed in \cite{Agullo:2012sh,Agullo:2012fc,Agullo:2013ai}. In particular, we adopt the extension of the effective potential to the pre-bounce phases studied in \cite{Li:2020mfi}. In Sec. IV, we show that  the whole evolution of the universe, from the deep remote contracting phase until the end of the slow-roll inflation  into several epochs, can be divided into four different phases - the pre-de Sitter, bouncing, transition and inflation. The mode function can be easily found analytically  in the pre-de Sitter,  the transition, and the inflationary phases [cf. Eqs.  by Eq.(\ref{sol1}) and (\ref{sol3A})].  Then,  in Sec. V, we focus ourselves on the mode function in the  bouncing phase. In particular, dividing the whole range of the comoving mode $k$ into three different sub-ranges: i) $k \gtrsim k_2$, ii) $k_e \lesssim k \lesssim k_2$, and iii) $k \lesssim k_e$,  we find  analytical solutions of the mode function 
in all these sub-ranges analytically by using the UAA method,  where $k_2 \simeq 1.61 \; m_P$ and $k_e \simeq 0.97 \; m_P$, and $m_P$ is the Planck mass \footnote{It should be noted that the observational window of the modes is 
$k/a_0 \in \left(0.1, 10^3\right) k_*$, where $k_*/a_0$ is the pivot mode, and was chosen as $ k_*/a_0 = 5\times 10^{-2}$/Mpc in Planck 2018 \cite{Planck:2018jri}, where $a_0$ is the current value of the expansion factor, given by
$a_0 = a_B e^{N_{\text{tot}}}$ in terms of the expansion factor at the bounce $a_B$ and the total e-fold $N_{\text{tot}}$ since the bounce. For $a_B = 1, \; N_{\text{tot}} \simeq 141$ \cite{Ashtekar:2011ni}, we find that
$k \in  e^{N_{\text{tot}}} \left(0.1, 10^3\right) k_* \simeq  \left(1,  10^4\right)\; m_P$. Therefore,  all the currently observational modes  are at least of order of the Planck scale at the quantum bounce, which is quite reasonable.}. Then, we match them smoothly to the solutions given in other phases, so that the mode function is uniquely determined once the initial condition is imposed in the deep remote contracting phase.  It must be noted that such applications of the UAA method and  their subsequent matching are by no means trivial. In particular, it is the first time in cosmology that the mode function is expressed as the linear combination of the second kind of the cylindrical functions in Case iii). Then, the corresponding matching becomes significantly complicated, and various asymptotical forms of these functions have to be properly used. This also explains why in this section we carry out the calculations step by step all in detail.
 In Sec. VI,  applying the Birrell-Davies trick \cite{Birrell:1982ix} to the remote contracting de Sitter universe, we identify the initial condition (\ref{eq1.3}), while in Sec. VII we summarize our main results and provide some concluding remarks.

Before proceeding to the next section, we would like to note that, once quantum gravitational effects are taken into account,  the corresponding problems of solving the MS equations of mode functions  become very much mathematically involved, and traditional methods, such as the WKB approximations,  are often either not applicable or generate large errors \cite{Joras:2008ck,Ashoorioon:2011eg}, which  are often by far above the accuracy required by current and forthcoming cosmological observations \cite{Abazajian:2013vfg}. Then, one has to turn to  numerical computations \cite{Agullo:2015tca}. Although numerical analyses are very powerful in dealing with such complicated problems, they also have several shortages, for example,  the computations often require the use of high-performance computing, as a result, they are very much time-consuming and expensive and  highly limit  the access of general researchers. In addition, numerical computations can explore only a finite region of the parameter space of the initial conditions, and  other physically interesting regions can be ignored. A concrete example is the calculation of the power spectra of scalar and tensor perturbations in the deformed algebra approach  \cite{Bojowald:2008gz,Cailleteau:2013kqa}, {\em in which it was found numerically that the power spectra are inconsistent with current CMB observations}  \cite{Bolliet:2015raa,Grain:2016jlq}. 
%Hence, the viability of this approach becomes questionable. 
Recently we  first  constructed analytically the general solutions of the mode functions by using the UAA method  and then showed explicitly that {\em there exists one class of initial conditions that does lead to power spectra consistent with current CMB observations} \cite{Li:2018vzr}.

In the past couple of years, various analytical analyses have been carried out \cite{Zhu:2016dkn,Zhu:2017jew,Wu:2018sbr,MenaMarugan:2024vyy,MenaMarugan:2024zcv,Alonso-Serrano:2023xwr}. In particular, we have systematically developed  the UAA method,   so that it has been successfully applied to various theories,  including Horava-Lifshitz cosmology \cite{Wang:2010an},  the  effective theories of quantum cosmology \cite{Zhu:2013fha,Zhu:2013upa,Zhu:2014aea,Zhu:2014wda,Zhu:2016srz,Qiao:2018dpp,Zhu:2018smk,Ding:2019nwu}, $k$-inflation \cite{Zhu:2014wfa}, Einstein-Gauss-Bonnet inflation  \cite{Wu:2017joj}, inflation in parity-violated theories  \cite{Qiao:2019hkz}, and  LQC with both  holonomy and inverse-volume  corrections   \cite{Zhu:2015xsa,Zhu:2015owa,Zhu:2015ata}.   The advantage of this method is that 
 the upper bounds of errors to each order of approximations are known and can be minimized by  minimizing {\em the error control function} introduced in the method.  In particular, to the third-order  approximation,  the upper bounds of errors can be $\lesssim 0.15\%$  \cite{Zhu:2014aea,Zhu:2016srz}, which is sufficient for  current and forthcoming Stage 4 experiments \cite{Abazajian:2013vfg}. To test the method further, recently we \cite {Li:2019cre} applied it to quantum mechanics and studied the energy spectra for five different potentials whose  energy eigenvalues are known exactly  \cite{Dong:2011zzf}.  Among other things, we found that  
 the UAA method yields the  same energy eigenvalue spectra as the exact ones in all five cases, while the WKB method does it
only for the Morse potential. This is one of the most accurate methods existing in the literature \cite{Alinea:2015gpa}.

In addition, to make our current studies as much independent as possible, it is inevitably to have some overlaps with the previous works, although we try our best to limit them to their minima. Finally,  in this paper we use the Planck units $\hbar=c=1$ and keep Newton's constant $G$ implicitly in  $\kappa$ with $\kappa=8 \pi G$. We also use $t_{\text{P}}$ and $m_{\text{P}}$ to denote the Planck time and mass, respectively. In all the numerical plots of the figures and content of this paper we choose units so that $t_{\text{P}} = 1 = m_{\text{P}}$, except in some particular occasions in which they are specified explicitly.

\section{Homogeneous Quantum Universe}  
\lb{sec:analytical_solutions}

In this section, we provide a brief review over the evolution of the homogeneous quantum universe both before and after the quantum bounce in mLQC-I. The evolution is universal for all trajectories, of which the kinetic energy of the inflaton dominates its potential energy at the  quantum bounce \cite{Li:2018opr,Li:2018fco,Li:2019ipm,Saeed:2024xhk}
\bq
\lb{eq2.1}
\frac{1}{2}\dot\phi_B^2 \gg V\left(\phi_B\right),
\eq 
where $\phi_B \equiv \phi(t_B)$ with $t_B$ denoting the moment when the quantum bounce occurs.  This will lay out the foundations for the analytical approach to solving the modified Mukhanov-Sasaki (MS) equations later on.

\subsection{Background Dynamics}

In a spatially-flat FLRW universe filled with a massive scalar field, the effective Hamiltonian of mLQC-I takes the form \cite{Yang:2009fp,Li:2018opr,Li:2018fco,Li:2019ipm,Saeed:2024xhk}
\bq
\lb{ham}
\mathcal {H}^{\scriptscriptstyle{\mathrm{I}}} =\frac{3v}{8\pi G\lambda^2}\left\{\sin^2(\lambda b)-\frac{(\gamma^2+1)\sin^2(2\lambda b)}{4\gamma^2}\right\}+ \frac{p_{\phi}^2}{2v} + vV(\phi), 
\eq
where $v$ is the expectation value of the volume operator, $b$ its  conjugate moment, and $\phi$ and $p_{\phi}$ are 
respectively the scalar field and the corresponding conjugate moment
with the potential $U(\phi)$, satisfying the relations
\bq
\lb{eq2.1aa}
\{b,v\}=4\pi G\gamma, \quad \{\phi,p_\phi\}=1. 
\eq
The parameter $\gamma$ is  the Barbero-Immirzi parameter, and $\lambda \equiv \sqrt{\Delta}$, where $\Delta \;(\equiv 4\sqrt{3}\pi\gamma \ell_{\mathrm{P}}^2)$ denotes the minimal non-zero area gap obtained in LQG, and $\ell_{\mathrm{P}}$ is the Planck length.  Then, it is straightforward to find the Hamilton's equations
\bqn
\lb{eomA}
\dot v&=&\Big\{v, \mathcal H^{\scriptscriptstyle{\mathrm{I}}}\Big\}=\frac{3v\sin(2\lambda b)}{2\gamma \lambda}\Big\{(\gamma^2+1)\cos(2\lambda b)-\gamma^2\Big\},\\
\lb{eomB}
\dot b&=&\Big\{b, \mathcal H^{\scriptscriptstyle{\mathrm{I}}}\Big\}=-4\pi G  \gamma \left(\rho+P\right),\\
\lb{eomC}
\dot \phi&=&\Big\{\phi, \mathcal H^{\scriptscriptstyle{\mathrm{I}}}\Big\}=\frac{p_{\phi}}{v},\\
\lb{eomD}
\dot p_{\phi}&=&\Big\{p_{\phi}, \mathcal H^{\scriptscriptstyle{\mathrm{I}}}\Big\}= - v V_{,\phi},
\eqn
where $V_{,\phi} \equiv \partial V(\phi)/\partial \phi$, and the energy density and the pressure of the scalar field are defined as usual 
\bq
\rho=\frac{1}{2}\dot \phi^2+V(\phi),~~~~P=\frac{1}{2}\dot \phi^2 -V(\phi).
\eq

In terms of $H \; (\equiv \dot{v}/3v)$, $\rho$ and $P$, the corresponding modified FR equations  take different forms before and after the bounce, although they are smoothly connected across the bounce \cite{Li:2018opr}. In particular, in the pre-bounce phase ($t \leq t_B$), we have
%\begin{widetext}
\bqn
\lb{eq3.1a}
H^2 &=&\frac{8\pi \alpha G  \rho_\Lambda}{3}\left(1-\frac{\rho}{\rho_c^{\text{I}}}\right)\left[1+\left(\frac{1-2\gamma^2+\sqrt{1-\rho/\rho_c^{\text{I}}}}{4\gamma^2
\left(1+\sqrt{1-\rho/\rho_c^{\text{I}}}\right)}\right)\frac{\rho}{\rho_c^{\text{I}}}\right], \;\;\;\; \left(t \le t_B\right), \\
\lb{eq3.1b}
\frac{\ddot a}{a} &=&- \frac{4 \pi  \alpha G}{3}\left(\rho + 3P - 2\rho_\Lambda \right)  +4\pi G\alpha P\left(\frac{2-3\gamma^2 +2\sqrt{1-\rho/\rho_c^{\text{I}}}}
{(1-5\gamma^2)\left(1+\sqrt{1-\rho/\rho_c^{\text{I}}}\right)}\right)\frac{\rho}{\rho_c^{\text{I}}}\nb\\
&& - \frac{4\pi G\alpha \rho}{3}\left[\frac{2\gamma^2+5\gamma^2\left(1+\sqrt{1-\rho/\rho_c^{\text{I}}}\right)-4\left(1+\sqrt{1- \rho/\rho_c^{\text{I}}}\right)^2}
{(1-5\gamma^2)\left(1+\sqrt{1-\rho/\rho_c^{\text{I}}}\right)^2}\right]\frac{\rho}{\rho_c^{\text{I}}},
 \; \left(t \le t_B\right), \;\;\;\;\;
\eqn
%\end{widetext}
where  
\bqn
\alpha \equiv \frac{1-5\gamma^2}{\gamma^2+1}, \quad
 \rho_\Lambda \equiv \frac{3}{8 \pi \alpha G \Delta (1+\gamma^2)^2}, \quad  
 \rho_c^{\text{I}} \equiv \frac{\rho_c}{4(\gamma^2+1)}, \quad
 \rho_c \equiv \frac{3}{8\pi \Delta\gamma^2 G}.
 \eqn
 From Eqs.(\ref{eq3.1a}) and (\ref{eq3.1b}) we find that
\bq
\lb{eq3.3}
\dot\rho + 3H\left(\rho + P\right) = 0  \quad \Rightarrow \quad \ddot{\phi} + 3H\dot{\phi} + V_{,\phi} = 0,
\eq
which can be also obtained from Eqs.(\ref{eomC}) and (\ref{eomD}) and is precisely the Klein-Gordon equation for the scalar field (Also called the conservation law of energy in terms of $\rho$ and $P$). It is remarkable to note that this is the same  as  in LQC (as well as in GR). 
When  $\rho \ll \rho_c^{\text{I}}$,  Eqs. (\ref{eq3.1a}) and (\ref{eq3.1b}) reduce, respectively, to
\bqn
\lb{eq3.3a}
H^2&\approx&\frac{8\pi G_\alpha}{3}\left(\rho+\rho_\Lambda\right) \simeq 
\frac{8\pi G_\alpha}{3}\rho_\Lambda, \\
\lb{eq3.3b}
\frac{\ddot a}{a}&\approx&-\frac{4\pi G_\alpha}{3}\left(\rho+3P - 2\rho_\Lambda\right) \simeq \frac{8\pi G_\alpha}{3}\rho_\Lambda,
\eqn
where $G_{\alpha} \equiv \alpha G$. Note that $\rho_{\Lambda}/\rho_c^{\text{I}} = 4\gamma^2/(1-5\gamma^2) \simeq 0.31$ for $\gamma=0.2735$ \cite{Meissner:2004ju}. Thus,  the effective Planck-size cosmological constant $\rho_\Lambda$ dominates the evolution of the universe 
almost during the entire  contracting phase. Only very near to the bounce ($t \simeq t_{B}$), the scalar field takes over, whereby a scalar-field-dominated bounce is achieved, $w (t \simeq t_B) \simeq 1$, so that a natural inflationary phase can be resulted generically after the universe expands for about $\Delta t \simeq 10^{5}t_{P}$,  whose exact value is usually model-dependent \cite{Li:2018opr,Li:2018fco,Li:2019ipm,Saeed:2024xhk}.  This is quite different from LQC. In order to distinguish this phase with the inflationary one, we refer it to as {\it the pre-de Sitter phase}.

After the bounce ($t \ge t_B$), the corresponding FR equations take the forms \cite{Li:2018opr}
\begin{widetext}
\bqn
\lb{eq3.4a}
H^2 &=&\frac{8\pi G \rho}{3}\left(1-\frac{\rho}{\rho_c^\text{I}}\right)\Bigg[1  +\frac{\gamma^2}{\gamma^2+1}\left(\frac{\sqrt{\rho/\rho_c^\text{I}}}{1 +\sqrt{1-\rho/\rho_c^\text{I}}}\right)^2\Bigg],
\;\;\;\; \left(t \ge t_B\right),\\
\lb{eq3.4b}
\frac{\ddot a}{a} &=&-\frac{4\pi G}{3}\left(\rho + 3P\right)
  + \frac{4\pi G \rho}{3}\left[\frac{\left(7\gamma^2+ 8\right) -4\rho/\rho_c^{\text{I}} +\left(5\gamma^2 +8\right)\sqrt{1-\rho/\rho_c^{\text{I}}}}{(\gamma^2 +1)\left(1+\sqrt{1-\rho/\rho_c^{\text{I}}}\right)^2}\right]\frac{\rho}{\rho_c^{\text{I}}}\nb\\
  &&  + 4\pi G P \left[\frac{3\gamma^2+2+2\sqrt{1-\rho/\rho_c^{\text{I}}}}{(\gamma^2+1)\left(1+\sqrt{1-\rho/\rho_c^{\text{I}}}\right)}\right]\frac{\rho}{\rho_c^{\text{I}}}, \;\;\;\; \left(t \ge t_B\right).
\eqn
\end{widetext}
From these two equations, it can be shown that the conservation law of energy given by Eq.(\ref{eq3.3}) holds also in the post-bounce phase. In addition, when 
$\rho \ll \rho_c^{\text{I}}$,  Eqs. (\ref{eq3.4a}) and (\ref{eq3.4b}) reduce, respectively, to
\bqn
\lb{eq3.5a}
H^2 &\approx& \frac{8\pi G}{3}\rho, \\
\lb{eq3.5b}
\frac{\ddot a}{a}&\approx&-\frac{4\pi G }{3}\left(\rho+3P\right),  
\eqn
whereby  the standard relativistic  cosmology is recovered. It is also remarkable to note that {\it the Planck-size cosmological constant $\rho_{\Lambda}$ no longer appears  in the post-bounce phase, and the Newtonian coupling constant $G$ is automatically restored}. All these features of mLQC-I are unique and different from other models, including LQC and mLQC-II \cite{Li:2021mop}.

\begin{figure*}[htbp]
\resizebox{\linewidth}{!}
{\begin{tabular}{cc}
\includegraphics[height=3cm,width=6cm]{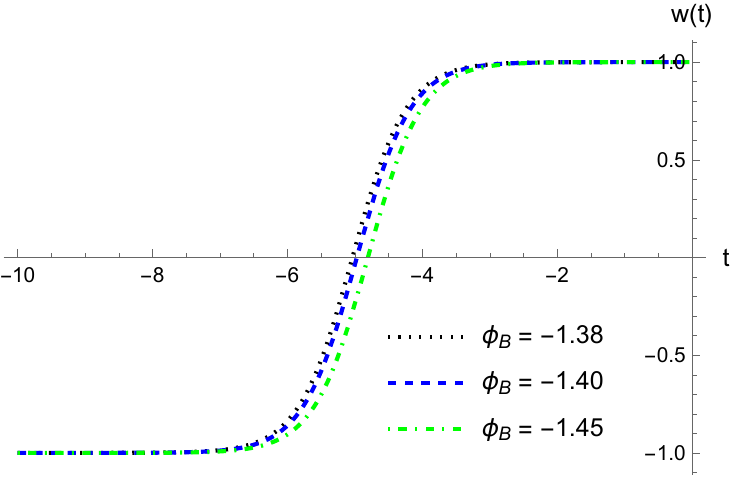}
\includegraphics[height=3cm,width=6cm]{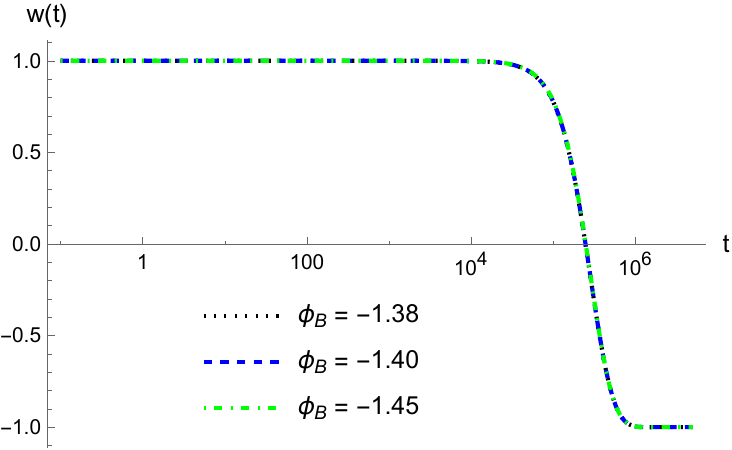}
\end{tabular}}
\caption{The equation of state for the Starobinsky potential, $V(\phi) = \frac{3m^2}{32\pi G}\left(1 - e^{-\phi/\phi_s}\right)^2$ with $\phi_s \equiv \sqrt{3/(16\pi G)}$, for different initial values of $\phi_B$ that result in greater than 50 e-Folds during the inflationary period.   For other potentials, similar conclusions are obtained \cite{Saeed:2024xhk}, as long as the initial conditions satisfy Eq. (\ref{eq2.1}).}
\label{fig2:eos-w}
\end{figure*}

\subsection{Universal Properties of the Background Evolution}

As shown explicitly in \cite{Li:2018opr,Li:2018fco,Li:2019ipm,Saeed:2024xhk}, if the initial conditions satisfy Eq. (\ref{eq2.1}), the evolution of the homogeneous universe is unique and can be well described by the analytical solutions. In particular, in the 
post-bounce region ($t \geq t_B$), three different phases are identified
\bqn
\lb{eq2.18}
&& (i) \; \text{bouncing} \; (w(t) \simeq 1),\nb\\
&& (ii)\; \text{transition}\;  (-1 < w(t) < 1), \nb\\
&& (iii) \; \text{inflationary},\; (w(t) \simeq -1),\; (t \geq t_B), 
\eqn 
where $w(t)$ denotes the equation  of the state of the inflaton field, defined by  
\bq
\lb{eq2.19}
w(t) \equiv \frac{P}{\rho} = \frac{\dot\phi^2/2 - V(\phi)}{\dot\phi^2/2 + V(\phi)}.
\eq
Moreover, during the bouncing phase, the expansion factor $a(t)$ and the scalar field $\phi(t)$ can be well described by the analytical solution \cite{Li:2019ipm}
\bqn
\lb{eq2.21}
a(t) &=& \left(1+24\pi\rho^I_c\left( 1+\frac{A_0\gamma}{1+B_0 t} \right)t^2   \right)^\frac{1}{6},\nb\\
\phi(t) &=& \phi_B +\text{sgn}\left(\dot\phi_B\right) \frac{\text{sinh}^{-1}{\sqrt{24\pi\rho^I_c\left(1+\frac{C_0\gamma^2}{1+D_0t}\right)t}}}{\sqrt{12\pi\left(1+\frac{C_0\gamma^2}{1+D_0t}\right)}},\; (t \geq t_B),
\eqn 
where 
\bq
\lb{eq2.21a}
A_0 = 1.2, \;\;\; B_0 = 6,\;\;\;  C_0 = 1.2, \;\;\; D_0 = 2,
\eq
with relative errors no larger than $0.3\%$.  The analytical solutions are universal and independent of the choice of inflationary potential $V(\phi)$ as long as the initial kinetic-energy dominated condition (\ref{eq2.1}) holds.

It is remarkable that before the bounce ($t\leq t_B$), three different phases are also identified \cite{Saeed:2024xhk}, as illustrated in Fig. \ref{fig2:eos-w},
\bqn
\lb{eq2.22}
&& (i) \; {\text pre-bouncing} \; (w(t) \simeq 1), \nb\\
&& (ii)\; {\text pre-transition}\;  (- 1 < w(t) < 1), \nb\\
&& (iii) \; {\text pre-de \; Sitter} \; (w(t) \simeq - 1),\; (t\leq t_B).
\eqn
Again such a  
division is universal and holds for all potentials as long as the initial conditions satisfy
(\ref{eq2.1}). In the pre-bounce regime, the expansion factor $a(t)$  can be well approximated by the analytical solutions
\begin{equation}
\label{eq2.23}
a(t) =   \begin{cases}
\left(1+d_0\rho^I_ct^2  \right)^\frac{1}{6}\sum_{n=2}^4{d_n t^n}, &
t_{m} \leq t \leq 0, \cr
a(t_{m})\exp\left\{-H_{\Lambda}\left(t-t_{m}\right)\right\}, & t \leq t_m ,\cr
\end{cases}
\end{equation}
where we have set $t_B = 0$, $H_{\Lambda} \equiv \sqrt{8\pi \alpha G\rho_{\Lambda}/3}$, $t_m $ is determined by $\dot{H}(t_m ) \simeq 0$, and $d_n\; (n = 0, ..., 4)$ are fitting constants, which must satisfy the junction conditions at both the bounce ($t = 0$) and the point ($t = t_m $). In particular, we require that {\em $a(t)$ and its first derivative be contiguous across these two points},
where $a(t)$ is given by Eq.(\ref{eq2.21}) in the post-bounce regime ($t \ge 0$).  
Then, it was found that \cite{Saeed:2024xhk}
\bqn
\lb{eq2.24}
d_0 &\simeq& 74.057, \quad
d_2 \simeq 0.104, \quad
d_3 \simeq 0.001, \quad
d_4 \simeq 0.002,
\eqn
for which the relative errors $\delta a(t)$ are less than $1\%$ at any given moment $t \le t_B$,  where the relative errors are defined by
\bq
\lb{eq2.25}
\delta{A} \equiv \left|\frac{A_n - A_a}{A_n}\right|,
\eq 
where $A_n$ and $A_a$ denote the numerical and analytical values of the physical quantity $A$.

In addition, to model the scalar field $\phi(t)$, it was found that in the whole pre-bounce regime ($t \le t_B$), the scalar field can be cast in the form \cite{Saeed:2024xhk}
\begin{equation}
\lb{eq2.26}
\phi(t) - \phi_B = e_3+\text{sgn}\left(\dot\phi_B\right) (e_0+e_2 t) e^{e_1t},\; (t \leq t_B),
\end{equation}
where $e_m\; (m = 0, ..., 3)$ are the fitting constants. Again, {\em $\phi(t)$ and its first derivative are required to be continuous across the bounce}, where $\phi(t)$ is given by Eq.(\ref{eq2.21}) in the post-bounce regime ($t \ge t_B$).
Then, fitting the above with the numerical data, 
it was found that, with the choices of fitting constants as
\bqn
\lb{eq3.16}
e_0 &\simeq& 0.3526, \quad
e_1 \simeq 0.9808,\quad
e_2 \simeq 0.0943, \quad
e_3 \simeq -0.3526,\;\; (\dot\phi_B > 0),
\eqn
for $\dot\phi_B > 0$, the relative errors $\delta\phi(t)$ are less than $1\%$ at any given moment $t \le t_B$. On the other hand, for $\dot\phi_B < 0$, the coefficients $e_n$ are given by 
\bqn
\lb{eq3.17}
e_0 &\simeq& 0.3526, \quad
e_1 \simeq 0.981,\quad
e_2 \simeq 0.094, \quad
e_3 \simeq 0.3526,\;\; (\dot\phi_B < 0).
\eqn

\section{Cosmological perturbations and Modified Mukhanov-Sasaki Equation}
\lb{sec:effective_mass_function}

In this section, we first review the modified Mukhanov-Sasaki (MS) equation of mLQC-I developed in the dressed metric approach \cite{Li:2019qzr,Agullo:2023rqq}, which
 takes the form 
\bq
\lb{MS_equation}
\nu^{\prime\prime}_{k}(\eta) +\left(k^2+m^2\right)\nu_{k}(\eta)=0,
\eq
here $\nu_{k}(\eta)$ is the MS variable, related to the comoving curvature perturbation $\mathcal{R}_{k}$ via $\nu_{ k}=z_s\mathcal{R}_{ k}$ with $z_s=a\dot{\phi}/H$ and a prime denotes the derivative with respect to the conformal time $\eta$.  The mode function $\nu_{k}(\eta)$ must satisfy the Wronskian
\bq
\lb{eq3.2}
\nu^{*}_{k}\nu^{\prime}_{k} - \nu^{*\prime }_{k}\nu_{k} = -i.
\eq
In terms of $\nu_{k}(\eta)$, the primordial scalar power spectrum is given by 
\bq
\lb{scalarpower}
P_\mathcal R=\frac{k^3}{2\pi^2}\frac{|\nu_{ k}|^2}{z^2_s}.
\eq
In the MS equation, the specific form of the mass function $m^2$ is model-dependent. In the classical perturbation theory of cosmology, depending on the gauge choices and use of the different MS variables, the classical mass function develops different forms, which turn out to be equivalent on the classical dynamical trajectories. In particular, the form related with the effective mass function used in the dressed metric approach is \cite{Li:2022evi}
\bq
\lb{dressed_mass_classical}
m^2 =\frac{3\kappa p^2_\phi}{ v^{4/3}}\left(1- \frac{3p^2_\phi}{4\pi G a^2 \pi_a^2}\right)+v^{2/3}\left(V_{, \phi\phi}-\frac{12 p_\phi }{a\pi_a}V_{,\phi}\right)-\frac{a^{\prime\prime}}{a}.
 \eq

\subsection{Effective Mass Functions in Dressed Metric Approach}

\begin{figure}
{
\includegraphics[width=8cm]{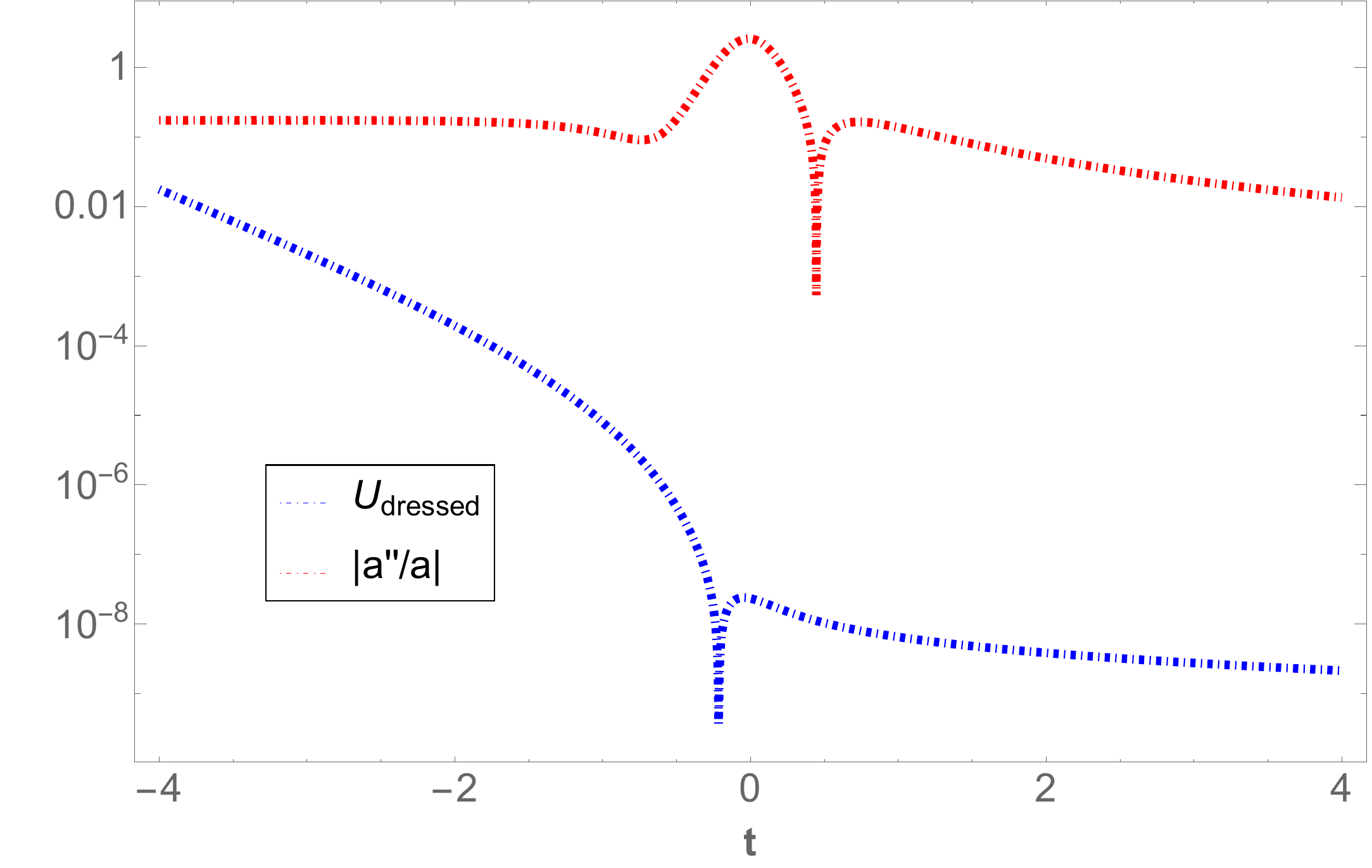}
\includegraphics[width=8cm]{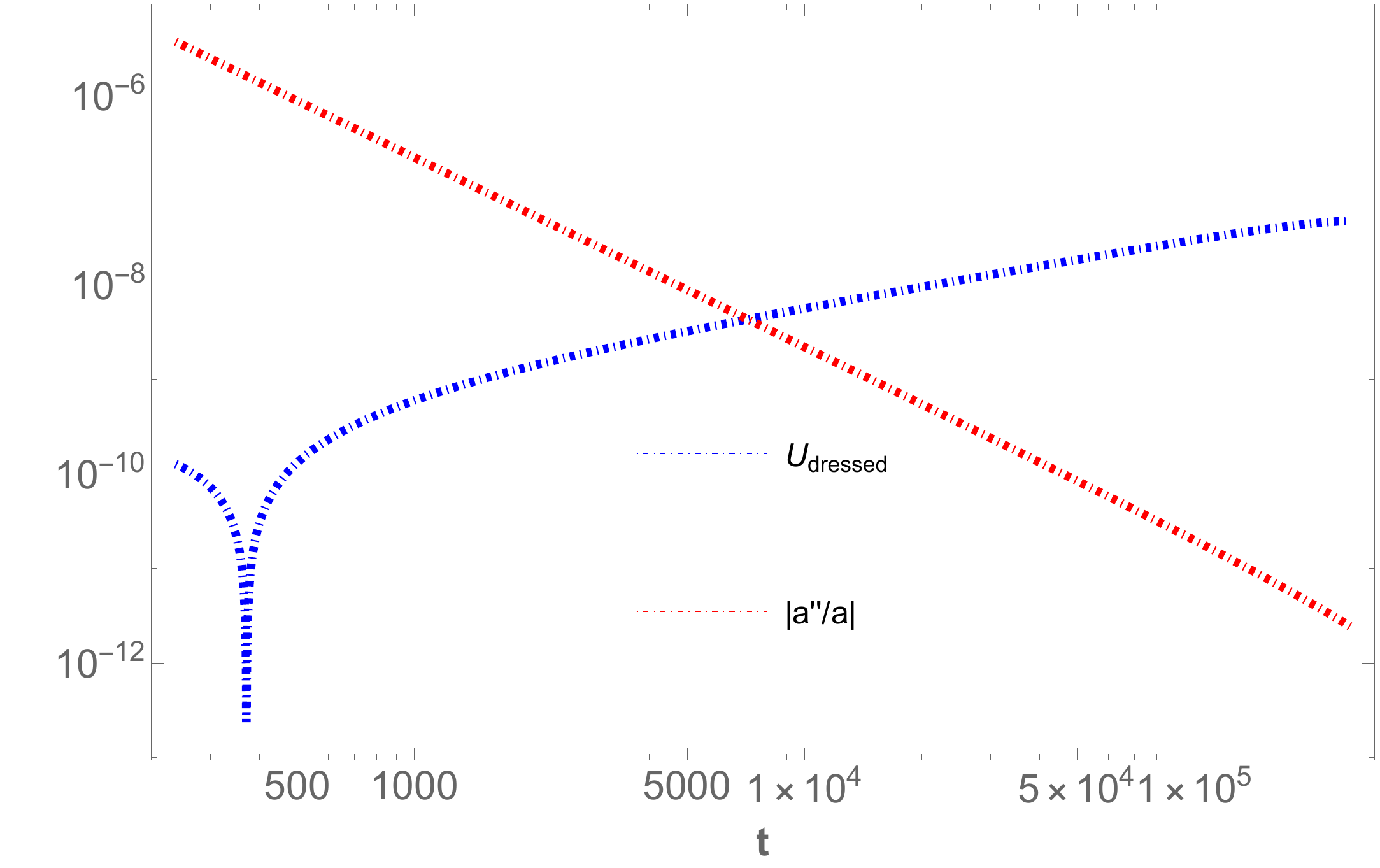}
}
\caption{In this figure, we compare the relative magnitudes of the effective potential and the acceleration term in the effective mass function in the dressed metric approach for mLQC-I. In the left panel, we focus on the region near the quantum bounce ($t = t_B = 0$), while in the right panel the two parts become compared in the post-bounce phase until the beginning of inflation.} 
\label{fig3:p-vs-a}
\end{figure}

\begin{figure}
{
\includegraphics[width=8cm]{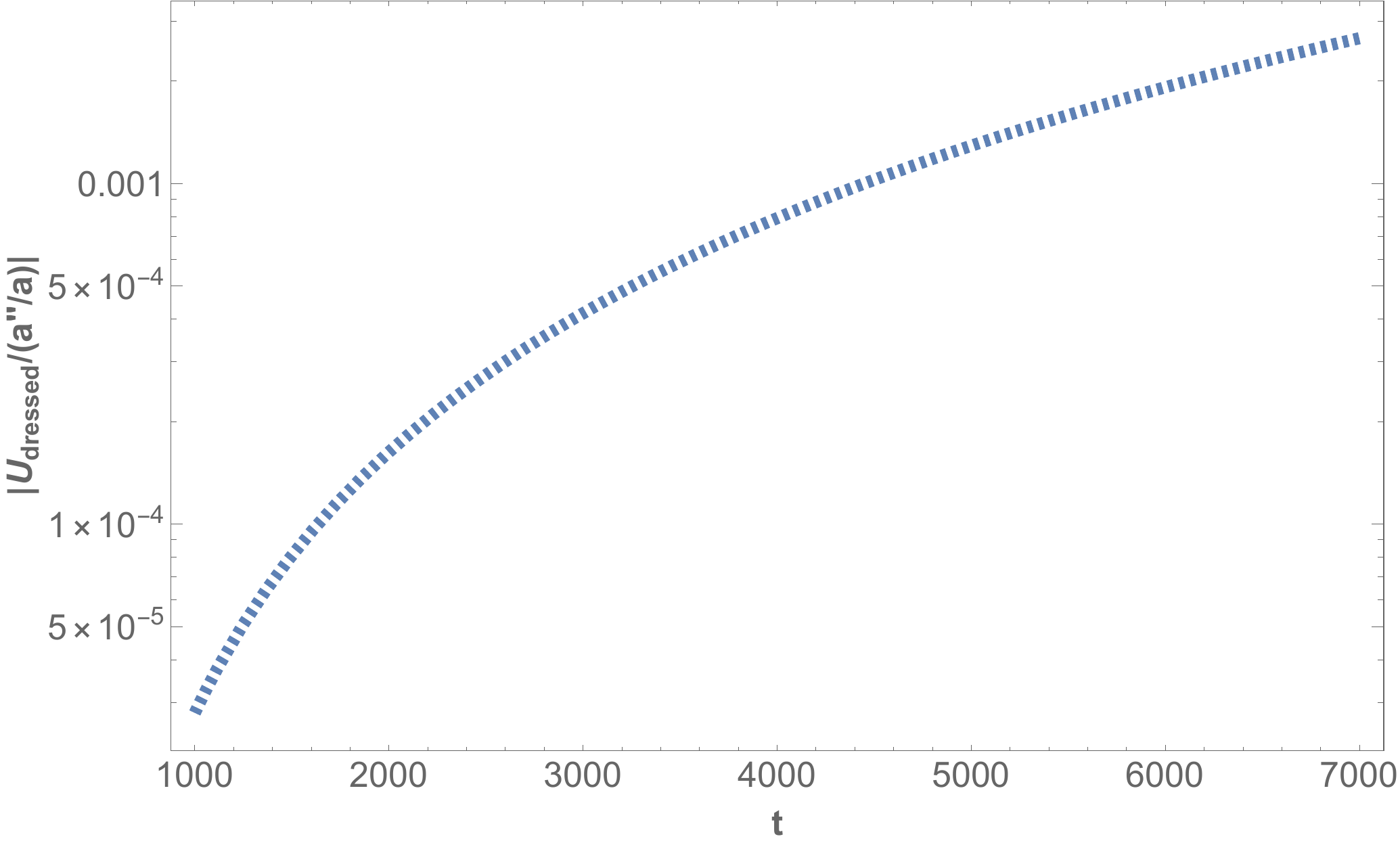}
\includegraphics[width=8cm]{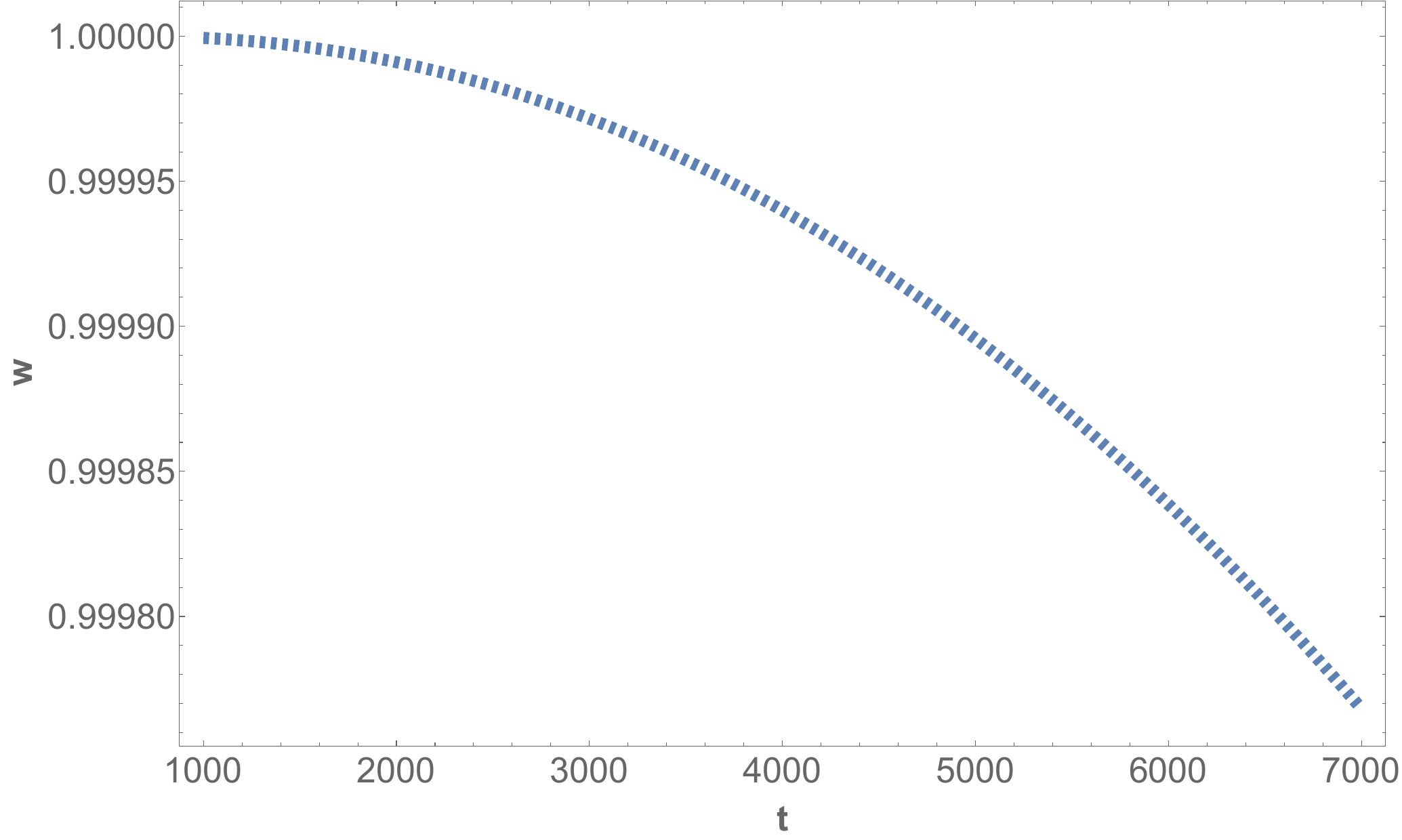}
}
\caption{In this figure, we show the range near the bounce in which the magnitude of the effective potential terms is less than $1\%$ of that of the acceleration term so that the former can be safely ignored when approximating the effective mass analytically.}
\label{fig4:bounce-regime}
\end{figure}

In the dressed metric approach, the effective mass function is obtained from the polymerization of the classical mass function given above. 
In particular, a viable ansatz of the polymerization has to deal with two terms, $1/\pi_a$ and its square $1/\pi^2_a$, since classically we have \cite{Ashtekar:2011ni}
\bq
\lb{eq3.4}
\pi_a = -\frac{3a\dot a}{4\pi G},
\eq
which now vanishes at the quantum bounce. Thus, an extension across the bounce to the pre-bounce is needed. Such extension must be  consistent with the polymerization of the background dynamics. Specifically, recall that the classical Hamiltonian constraint of the background dynamics takes the form
\bq
\mathcal H^{(0)}=-\frac{3\Omega^2}{8\pi G v\gamma^2}+\frac{p^2_\phi}{2v}+v V,
\eq
where $\Omega \equiv -4\pi G \gamma a \pi_a/3$. The effective Hamiltonian constraint is obtained by polymerizing the above classical counterpart with the replacement \cite{Li:2020mfi}
\bq
\Omega^2\rightarrow\Omega^2_{{\scriptscriptstyle{\mathrm{I}}}}=-\frac{v^2\gamma^2}{\lambda^2}\Big\{\sin^2\left(\lambda b\right)-\frac{\gamma^2+1}{4\gamma^2} \sin^2\left(2 \lambda b\right) \Big\}. 
\eq
As a result, when polymerizing the classical mass function (\ref{dressed_mass_classical}), we adopt the following ansatz
\bq
\frac{1}{\Omega^2}\rightarrow \frac{1}{\Omega^2_{{\scriptscriptstyle{\mathrm{I}}}}}, \quad \quad  \frac{1}{\Omega}\rightarrow \frac{\Lambda_{{\scriptscriptstyle{\mathrm{I}}}}}{\Omega^2_{{\scriptscriptstyle{\mathrm{I}}}}},\quad \Lambda_{{\scriptscriptstyle{\mathrm{I}}}} \equiv \frac{v}{2\lambda}\sin\left(2\lambda b\right).
\eq 
The first replacement is to render the polymerization of the classical mass function consistent with that of the background, while the second replacement is aimed to respect the superselection rule of the background dynamics \footnote{We refer the readers to the  papers \cite{Li:2020mfi,Kowalczyk:2025qbi} and references therein for more details.}. With the above polymerization ansatz, the effective mass function in the dressed metric  is given by
\bq
\lb{eq3.9aaa}
m^2_{\mathrm{eff}}=-\frac{a^{\prime\prime}}{a}+\mathfrak{U}_{\mathrm{dressed}},
\eq
where the effective potential is given by
\bq
\lb{eq3.9bbb}
\mathfrak{U}_{\mathrm{dressed}}=a^2\left(V_{,\phi \phi}+48 \pi G V+\frac{6H}{\Theta(b)}\frac{\dot {\phi}}{\rho}V_{, \phi}-\frac{48 \pi G}{\rho}V^2\right),
\eq
with $\Theta(b)\equiv (\gamma^2+1)\cos(2\lambda b)-\gamma^2$. Therefore, similar to the effective mass function in the standard LQC, the effective mass function in mLQC-I is also composed of two parts,  the effective potential $\mathfrak{U}_{\mathrm{dressed}}$ and the acceleration  term $-a^{\prime\prime}/a$. The relative magnitude of these two parts are compared in Figs. \ref{fig3:p-vs-a} and \ref{fig4:bounce-regime} where numerical results are used to show the differences between these two terms. From these figures we can see that 
$\left|\mathfrak{U}_{\mathrm{dressed}}/(a^{\prime\prime}/a)\right| \ll 1$
holds for $t \in(t_B, t_E)$ where $t_E \simeq 7.157\times10^3$.

\begin{figure}
{
\includegraphics[width=8cm]{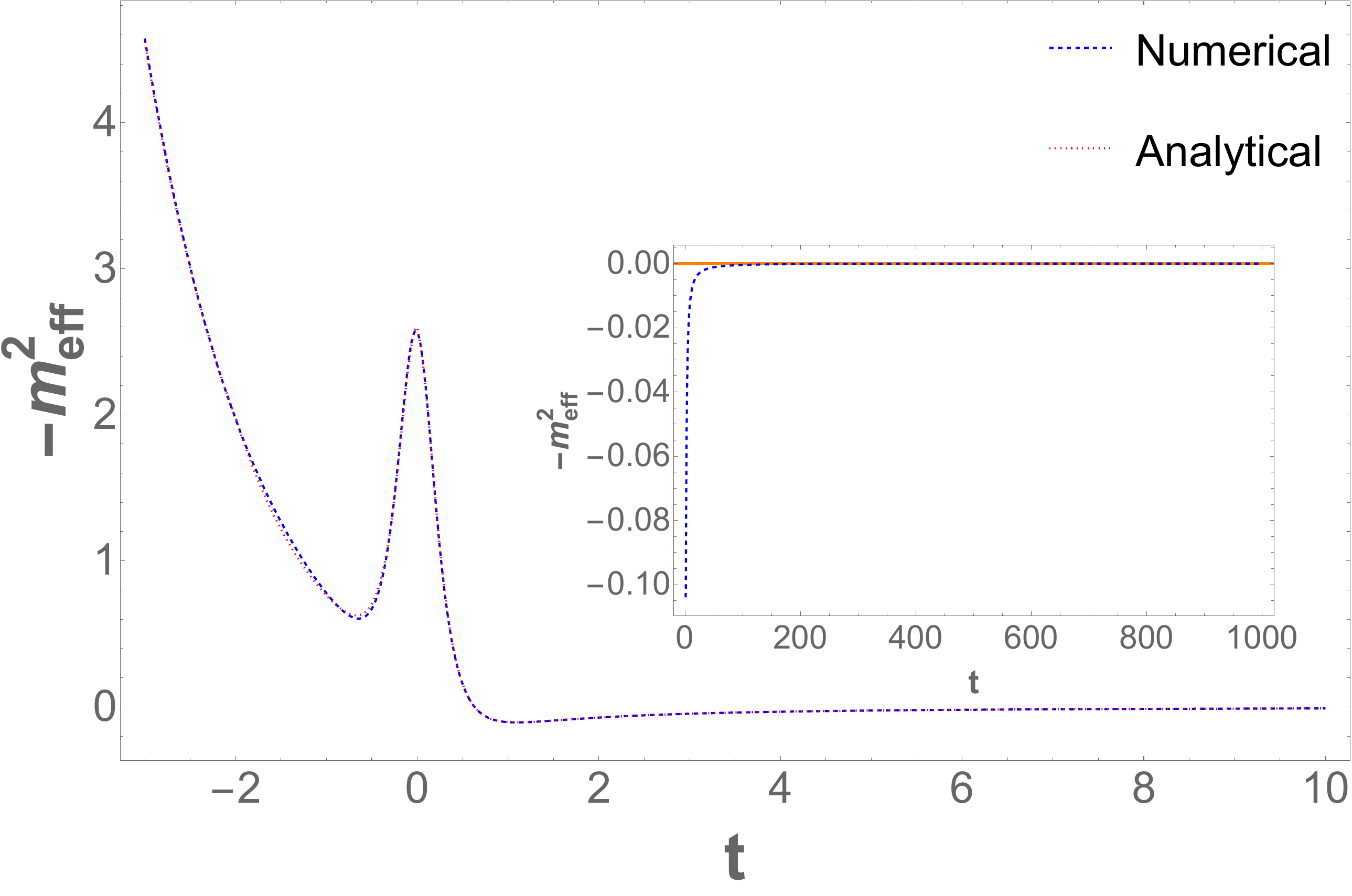}
\includegraphics[width=8cm]{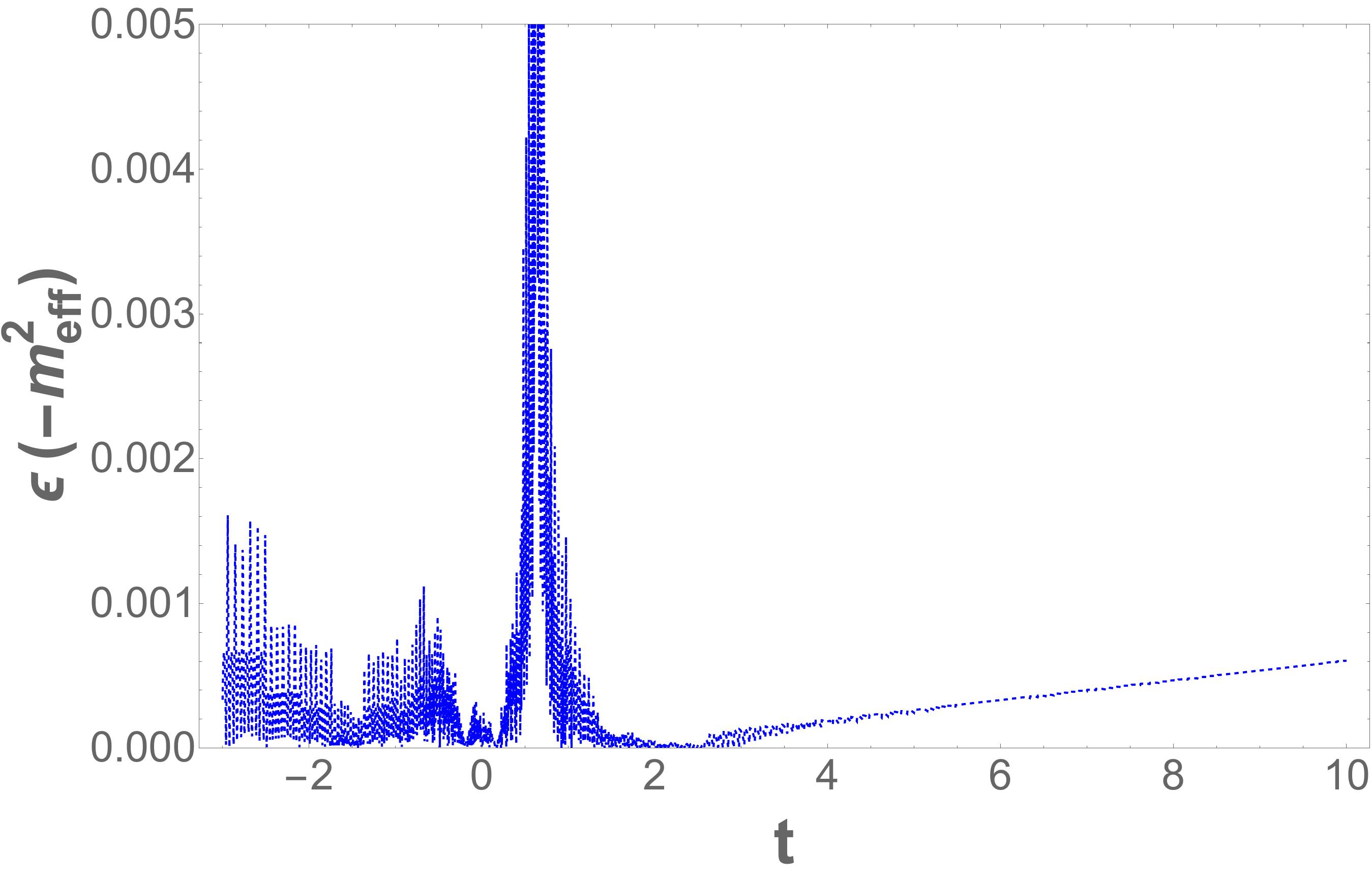}
}
\caption{{\bf Left Panel:} The  effective mass function $(-m^2_{\text{eff}})$ in the dressed metric approach for mLQC-I near the bounce and in the bouncing phase ($t > 0$). The effective mass approaches zero quickly from below as $t$ increases, as shown in the inserted plot. {\bf Right Panel:} The relative errors $\epsilon(-m^2_{\text{eff}})$ defined by Eq.(\ref{eq2.26}) between the numerical and analytical solutions of
the effective mass function. The spark point is where $m^2_{\text{eff}} \simeq 0$, apart from which the relative errors are negligible. }
\label{fig5:meff}
\end{figure}

In order to solve the modified MS equation analytically, it is important to understand the behavior of the effective mass function in the entire evolution of the universe. In Fig. \ref{fig5:meff}, we numerically plot the (negative) effective mass function near the bounce point (the left-hand panel) and in the post bounce phase
(the right-hand panel). We find that in the post bounce phase (up to the onset of the inflationary phase) the effective mass functions in mLQC-I and the standard LQC have similar qualitative behavior. In particular, we find that
\bq
\lb{eq3.10a}
\left|m^2{\text{eff}}\right| < 10^{-5}\; m_P ^2, \quad (t \gtrsim 10^3 t_P ).
\eq
Hence, we have
\bq
\lb{eq3.10b}
k^2 + m^2_{\text{eff}} \simeq k^2, \quad (t \gtrsim 10^3 t_P ).
\eq

Note that  the observational window of the modes is $k/a_0 \in (0.1 k_*, 10^3 k_*)$ \cite{Ashtekar:2011ni}, where $k_*/a_0\; (\equiv 5\times 10^{-2}\; \text{Mpc}^{-1})$ denotes the pivot mode used by Planck observation \cite{Planck:2018jri},  $a_0 \left(\equiv e^{N_{\text{tot}}} a_B\right)$ is the current value of the expansion factor  \footnote{In this paper, we shall set the value of the expansion factor $a(t)$ at the bounce equal to one, $a_B = 1$, instead of $a_0 = 1$ as used by Planck \cite{Planck:2018jri}.}, and $N_{\text{tot}}$ denotes the total e-fold of the expansion of the universe from the quantum bounce to the current time. Thus, for $a_B = 1, \; N_{\text{tot}} \simeq 141$ \cite{Ashtekar:2011ni,Zhu:2017jew} we find that
\footnote{It is also worth pointing out that, when the number of the inflationary e-folds changes from 50 to 70, the comoving wavenumber of the pivot mode $k_*$ increases from $0.00025 m_P$ to $98.7969 m_P$. When the number of the inflationary e-folds is $N_{\text{inf}}=60$, we find that $k^{60}_*=0.00455342 m_P$. For more details, see \cite{Agullo:2013ai}.
}
\bq
\lb{eq3.11a}
k \in  e^{N_{\text{tot}}} \left(0.1, 10^3\right)k_* \simeq \left(1, 10^4\right) m_P , \; \left(a_B = 1, \;\; N_{\text{tot}} \simeq 141\right).
\eq
However, 
 near the bounce distinct features arise, as can be seen from the left-hand panel of Fig. \ref{fig5:meff}. 
Particularly, near the bounce the acceleration term $\left|a''/a\right|$ is much larger than the effective potential $\left|\mathfrak{U}_{\mathrm{dressed}}\right|$ as shown in Fig. \ref{fig3:p-vs-a}, so the latter can be safely ignored. Then, we can introduce  a characteristic wavenumber at the bounce, namely
\bq
k_\mathrm{B}\equiv\left.{\sqrt{\frac{a^{\prime\prime}}{a}}}\right|_{t=t_B}=
\sqrt{8 \pi G \left(1+A_1\gamma^2\right)\rho_{\text{c}}^{\scriptscriptstyle{\mathrm{I}}}}\approx 1.6015\; m_P,
\eq
which determines the order of magnitude of the wavenumber of the perturbation modes that are expected to be influenced  by the quantum gravity effects in the Planck regime. 

\begin{figure}[htp]
{
\includegraphics[width=8cm]{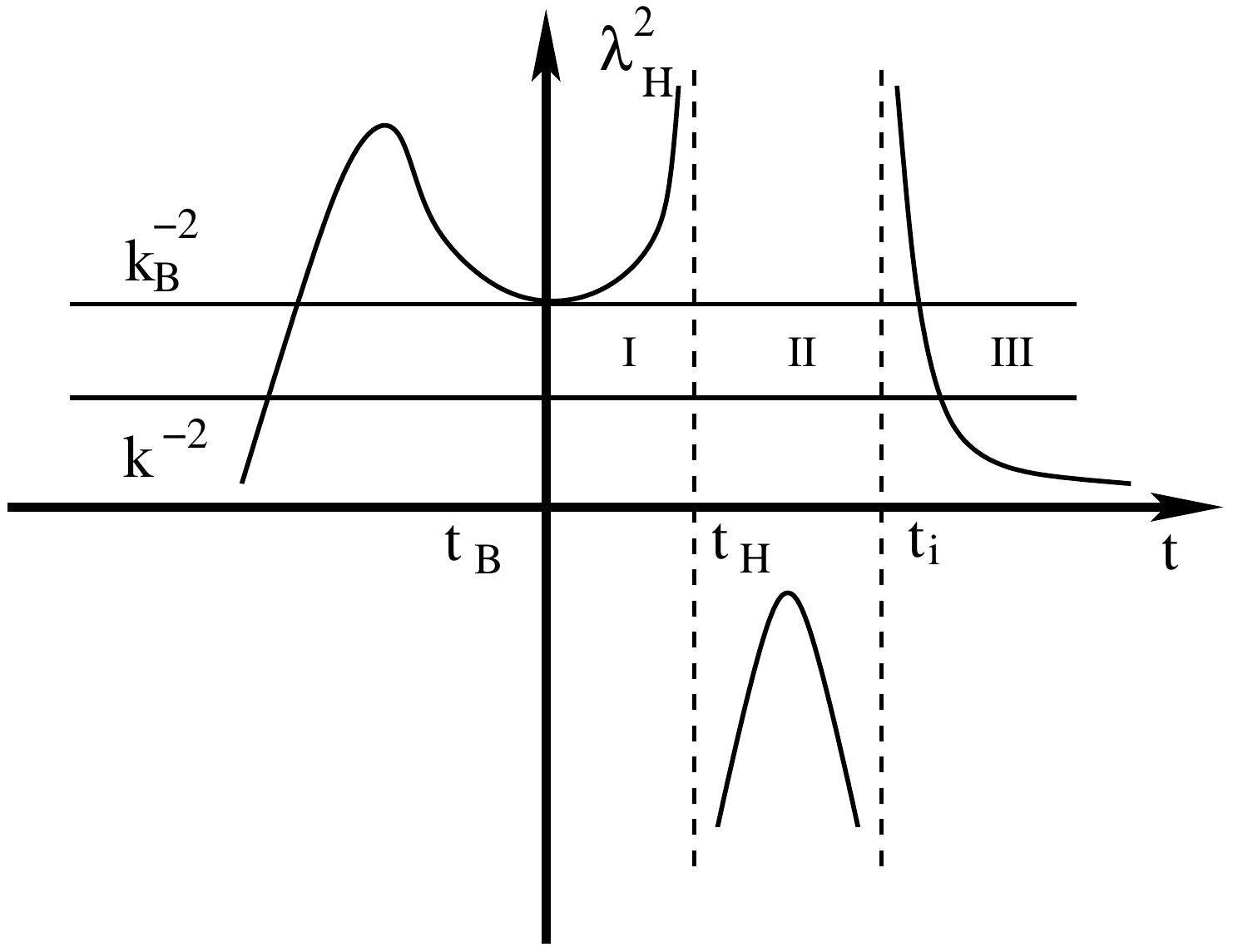}
\includegraphics[width=8cm]{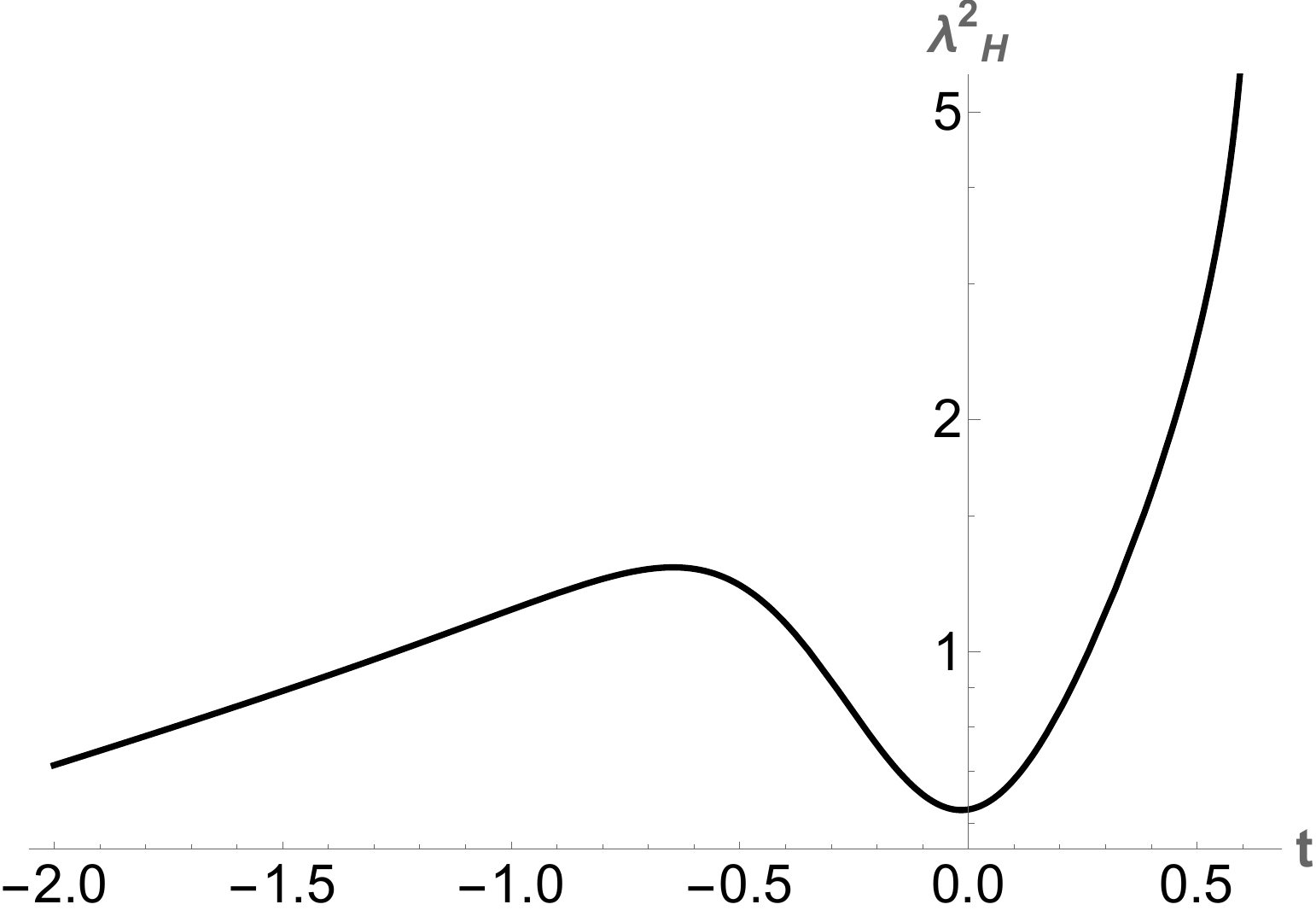}\\
(a) {\mbox{\hspace{8.3cm}}} (b)\\
}
\caption{(a) Schematic plot of the quantity $\lambda^2_H$ defined in Eq.(\ref{eq3.11}) in the whole range of $t$ \cite{Li:2021mop}. (b) The numerical plot of $\lambda^2_H$ near the bounce. }
\label{fig6:lamdaH}
\end{figure}

In addition, from the left-hand panel of Fig. \ref{fig5:meff}, one can also see that $- m^2_{\text{eff}}$ has one minimum and one maximum point near the bounce, which is found to be located respectively at $t_{\text{min}} \simeq -0.01525 t_P$ and  $t_{\text{max}}\simeq -0.6457 t_P$. The corresponding comoving wavenumber is $k_\mathrm{min} \simeq 0.7777 m_P$ and $k_\mathrm{max}\simeq 1.6041 m_P$. Then, it is expected that the modes with $k \in (k_\mathrm{min}, k_\mathrm{max})$ will feel strong quantum gravitational effects, while the ones with $k \gg k_\mathrm{max}$ enter the Hubble horizon at a very early time $t_{\text{enter}} \ll -2 t_P$ and does not feel the influence of the background near the bounce region, where $t_{\text{enter}}$ is defined as $k^2 = - m^2_{\text{eff}}(t_{\text{enter}})$. On the other hand, the modes with $k \ll k_\mathrm{min}$ stay outside of the Hubble horizon during the whole contracting phase, but soon enter the Hubble horizon after the bounce. For more details, we refer readers to \cite{Li:2021mop}.

\section{Mode Functions in Different Phases}
\lb{ModeFunctions}

In review of the above analysis, to find the approximate analytical solutions of the modified MS equation, it is convenient to divide the history of the evolution of the mode function into several different phases.  Before studying the modified MS equation in each of these phases, let us first introduce the quantity $\lambda_H$
\bq
\lb{eq3.11}
\lambda^2_H \equiv - \frac{1}{m^2_{\text{eff}}},
\eq
so that 
\bq
\lb{eq3.12}
\Omega^2_{\text{tot}} \equiv k^2 + m^2_{\text{eff}}   = \frac{1}{\lambda^2} -  \frac{1}{\lambda_H^2} =
\begin{cases}
> 0, & \lambda_H^2 > \lambda^2,\cr
< 0, & 0 < \lambda_H^2 < \lambda^2,\cr
> 0, &  \lambda_H^2 < 0,\cr
\end{cases}
\eq
where $\lambda\; (\equiv k^{-1})$ denotes the comoving wavelength of the mode $k$. Note that such a defined quantity $\lambda_H^2$ becomes negative when  the effective mass
is positive.  
In Fig. \ref{fig6:lamdaH},  we plot $\lambda_H^2$  schematically for the   Starobinsky potential, where     
the moments $t_H$ and $t_i$ are defined, respectively, 
by $\ddot a(t_{H}) = \ddot a(t_i)=0$, so
$t_i$ represents the beginning  of the inflationary phase,
and during the slow-roll inflation (Region III), we have $\lambda_H^2\approx L_\text{H}^2/2 \simeq 1/(2a^2H^2)$, which is exponentially decreasing, and all the  modes observed today were inside the comoving Hubble radius at $t=t_i$. Between the times $t_H$ and $t_i$, $\lambda_H^2$ is negative, and $\Omega^2_{\text{tot}}$   is strictly positive. Therefore,   during this period the mode functions are oscillating, while
during the epoch  between $t_B$ and $t_H$,  some modes ($k^{-2} > k_B^{-2}$) are inside the comoving Hubble radius,  and others ($k^{-2} < k_B^{-2}$) are outside it right after the bounce, where
$k_B\equiv \lambda_B^{-1}(t_B)$. 
 In the contracting phase,  {when $t \ll  t_B$, the universe is quasi-de Sitter  and  $\lambda_H^2 \simeq 1/(2a^2H^2)$ increases exponentially  toward  the bounce $t \rightarrow t_B$, as now $a(t)$ is decreasing exponentially. However, several Planck seconds before the bounce, the universe enters a non-de Sitter state, during which $\lambda_H^2$ starts to decrease until the bounce, at which  a characteristic Planck
 scale $k_B\; (\equiv 1/\lambda_H)$ can be well defined.
 Therefore,  for $t \ll t_B$, some modes are outside the comoving Hubble radius, and become non-adiabatic. Then, it is impossible to impose the BD vacuum.
 To study the perturbation further, let us consider the modified MS equation in each of these phases separately.

\begin{figure}[htp]
{
\includegraphics[width=8cm]{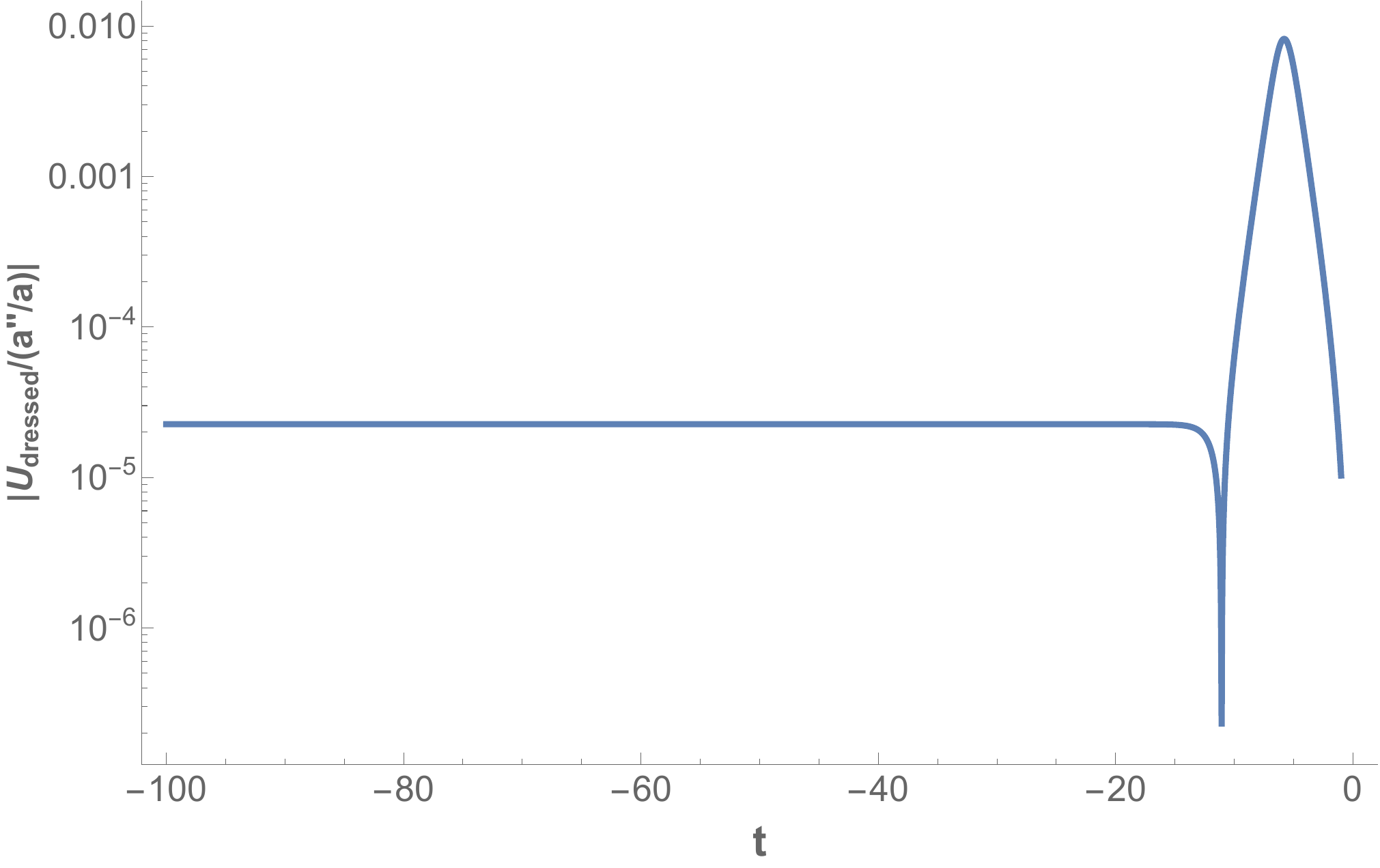}
\includegraphics[width=8cm]{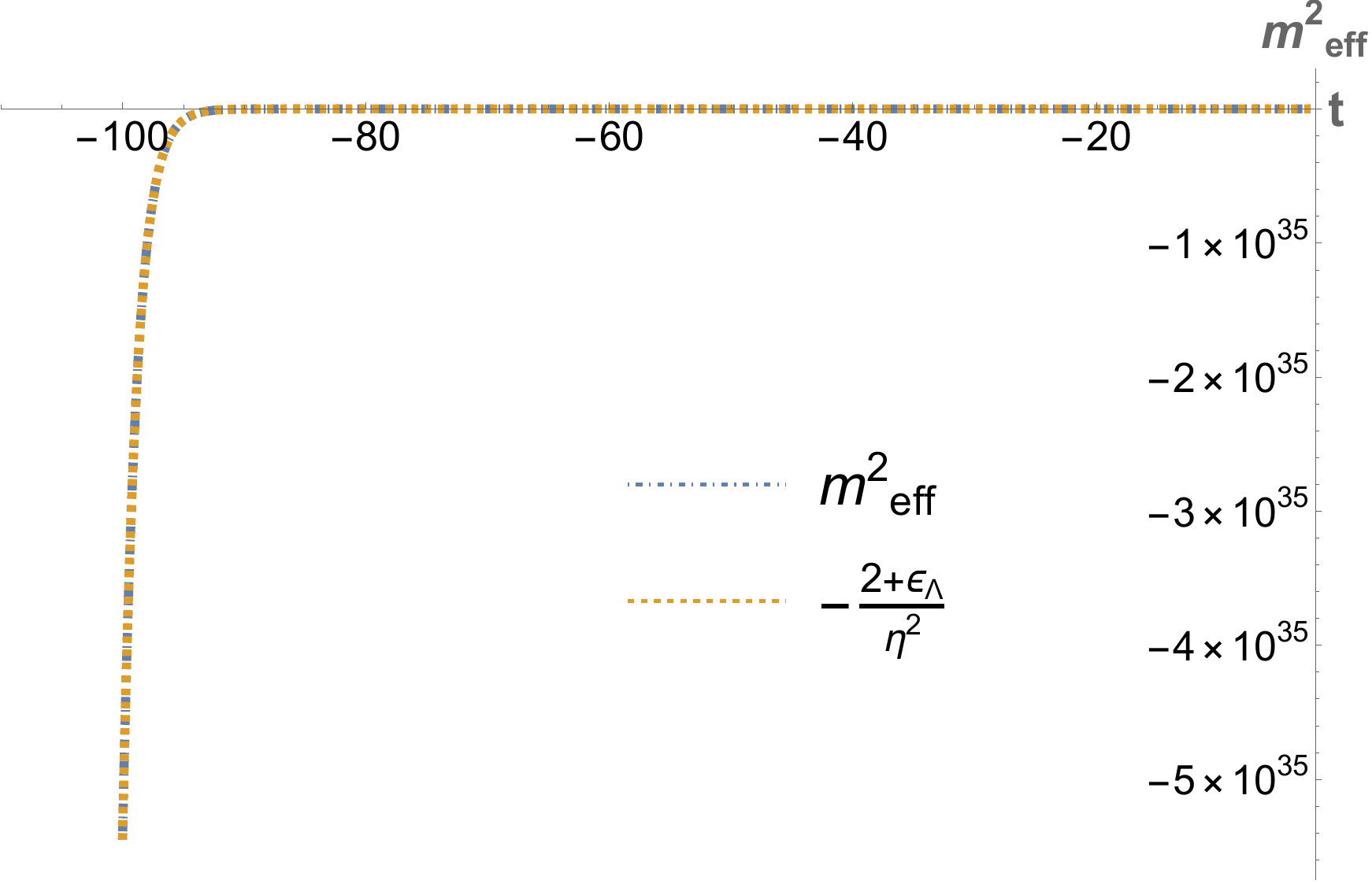}
}
\caption{{\bf Left Panel:} The ratio between $\mathfrak{U}_{\mathrm{dressed}}$ and $a''/a$ in the pre-bounce phase $t < 0$.    {\bf Right Panel:} The numerical effective mass function $m^2_{\text{eff}}(t)$ and its analytical approximation $-(2+\epsilon_{\Lambda})/\eta^2$.}
\label{fig7:U-and-meff}
\end{figure}
 
\subsection{The Pre-de Sitter Phase}

In this phase, the expansion factor $a(t)$ is given by Eq.(\ref{eq2.23}), from which we find 
\bq
\lb{eq3.13}
\eta(t)=\int^t_{-\infty}\frac{dt}{a}=\frac{1}{a_m H_{\Lambda}} e^{H_{\Lambda}\left(t-t_m \right)},
\eq
where $a_m \equiv a(t_m )$. Then, we find that
\bq
\lb{eq3.13aa}
\mathfrak{U}_{\mathrm{dressed}} \simeq - \frac{\epsilon_{\Lambda}}{\eta^2}, \quad m^2_{\text{eff}} \simeq -\frac{2+\epsilon_{\Lambda}}{\eta^2}, \quad \epsilon_{\Lambda} \equiv  - \Delta\left(1+\gamma^2\right)^2 V''(\phi_{\text{initial}}),
\eq
with  $\phi_{\text{initial}} \equiv \phi_B + e_3$ [cf. Eq.(\ref{eq2.26})].
In the left-hand side of Fig. \ref{fig7:U-and-meff} we show the ratio between $\mathfrak{U}_{\mathrm{dressed}}$ and $a''/a$, from which it can be seen that
it indeed approaches a constant, which is of order of $10^{-5}$. 
Hence, the  MS equation (\ref{MS_equation})   reduces to 
\bq
\lb{MS_de_Sitter}
\nu^{\prime \prime}_k+\left(k^2-\frac{2 + \epsilon_{\Lambda}}{\eta^2}\right)\nu_k=0.
\eq 
The above equation allows the following exact solutions,   
\bq
\lb{sol1}
\nu_k=\sqrt{\frac{\pi k\eta}{2}}\left[a_k H_{\nu}^{(2)}(k\eta)  +  b_k H_{\nu}^{(1)}(k\eta)\right],  \quad \nu \equiv \left(\frac{9}{4} + \epsilon_{\Lambda}\right)^{1/2},
\eq
where $H_{\nu}^{(1)}$ and $H_{\nu}^{(2)}$ are the 
Hankel functions of the first and second kinds, respectively, and satisfies the   Wronskian
\bq
\lb{sol1aa}
{\cal{W}}\left(H_{\nu}^{(1)}(y), H_{\nu}^{(2)}(y)\right) \equiv H_{\nu}^{(1)}(y) H_{\nu}^{(2)}{}'(y)  - H_{\nu}^{(1)}{}'(y) H_{\nu}^{(2)}(y)  = - \frac{4i}{\pi y},
\eq
where $H_{\nu}^{(2)}{}'(y) \equiv dH_{\nu}^{(2)}(y)/dy$, etc. Then,  the  Wronskian of  Eq.(\ref{eq3.2}) leads to  
\bq
\lb{Wcd}
\left|a_k\right|^2-\left|b_k\right|^2= \frac{1}{2k}.  
\eq
The coefficients $a_k$ and $b_k$ will be determined by  the initial conditions and the above Wronskian.

It should be noted that, since $\epsilon_{\Lambda} \simeq 10^{-5}$, we have $\nu \equiv \sqrt{9/4 + \epsilon_{\Lambda}} \simeq 3/2$. Then, we find
\bq
\lb{eq3.18} 
 H_{3/2}^{(1)}(z) = - \sqrt{\frac{2}{\pi z}}e^{iz}\left(1 + \frac{i}{z}\right), \quad  
 H_{3/2}^{(2)}(z) = - \sqrt{\frac{2}{\pi z}}e^{-iz}\left(1 - \frac{i}{z}\right).
\eq
Inserting the above expressions into Eq.(\ref{sol1}) we find that
\bq
\lb{eq3.19} 
 \nu_k(\eta) = - \left\{a_k e^{-ik\eta}\left(1 - \frac{i}{k\eta}\right) + b_k e^{ik\eta}\left(1 + \frac{i}{k\eta}\right)\right\}.
\eq
\subsection{The Extended Bouncing Phase}

This phase consists of the pre-bouncing ($t_m  \le t \le 0$) and bouncing ($0 \le t \le t_E$) phases. Without causing any confusion, in the rest of this paper, we simply call it the  bouncing phase. During this phase, 
some modes stay outside of the Hubble horizon all the time, while others enter, exit and re-enter a couple of times, before all exiting from the Hubble horizons, as shown by Fig. \ref{fig6:lamdaH} (a). It is in this phase that quantum gravitational effects  on some modes may become important, whereby lead to observational evidence for current and forthcoming observations. We shall dedicate  the next section to calculate the mode function analytically during this phase, using the UAA method.

\subsection{The Transition Phase}

During this phase, the effective potential starts to pick up and soon becomes larger than the kinetic term $|a''/a|$, as shown in Fig. \ref{fig3:p-vs-a}. However, as shown by the right-hand panel of Fig. \ref{fig5:meff} and Eqs.(\ref{eq3.10a})-(\ref{eq3.11a}), $\left|m^2_{\text{eff}}\right|$ remains very small in the whole post-bounce phase. In particular, for $t > 10^3 t_P $,
we find $\left|m^2_{\text{eff}}\right| < 10^{-5}$. Therefore, for the observational window given by Eq.(\ref{eq3.11a})
we have $k^2 + m^2_{\text{eff}} \simeq k^2$ for $t \gtrsim 10^{3}t_P$. Then, we find that  
\bq
\lb{sol3A}
\nu_k=\frac{1}{\sqrt{2k}}\left(\alpha_ke^{-ik\eta}+ \beta_k e^{ik\eta}\right),
\eq
where $\alpha_k$ and $\beta_k$ are two integration constants and will be determined uniquely by the initial conditions imposed in the remote contracting phase.

\subsection{ The Inflationary Phase}

During this phase, marked as Region III in Fig. \ref{fig6:lamdaH} (a), the wavenumber of the relevant modes are much larger than the effective mass function. Therefore, the mode function is just a linear combination of the plane waves given also by
Eq.(\ref{sol3A}).

It should be noted that with the initial conditions (\ref{eq3.17}) imposed in the pre-de Sitter phase ($t_{\text{initial}} \ll t_B$), in general we have $\beta_k \not= 0$, that is, at the onset of inflation, the Universe is no longer in the BD vacuum.  
In particular,  the resulting primordial power now is given by  \cite{Zhu:2017jew}
\bq
P_\mathcal R=|\alpha_k+\beta_k|^2 P^{GR}_\mathcal R,
\eq
where $P^{GR}_\mathcal R$ stands for the primordial power spectrum predicted in the classical GR.

\section{Analytical Solutions of the Mode Function}
\lb{sec:Analytical_Solutions}

In this section, we shall use the well-developed   UAA method to solve the MS equation (\ref{MS_equation}) in the (extended) bouncing phase. To this goal,  we first rewrite it in the standard form \cite{Olver:1997,Zhu:2013upa,Zhu:2014aea,Zhu:2016srz,Pan:2023aiv}
\bq
\lb{standard_eom}
\frac{d^2\nu_k(y)}{dy^2}=-\left(1+\frac{m^2_\mathrm{eff}}{k^2}\right)\nu_k(y)=\left(\lambda^2g(y)+q(y)\right)\nu_k(y),
\eq
where $y=k\eta$.   It should be noted that the parameter $\lambda$ introduced above serves as a bookmarker, and the solutions $\nu_k$ can be expanded as 
\bq
\nu_k(y) = \sum\frac{\nu_k^{\text{(n)}}(y)}{\lambda^n}. 
\eq
But, finally we shall set $\lambda = 1$. In this paper we shall consider only the first-order approximations, $n = 1$.

In addition, the limiting case when $k\gg k_B$ is one of the simplest situations in which approximate solution can be obtained without turning to the UAA method. In particular, when $k\gg k_B$, the solution of the equation of motion (\ref{standard_eom}) is the plane wave solution. Moreover, the de Sitter initial state (\ref{sol1}) in the contracting phase also reduces to the plane wave solution. As a result, for short-wavelength modes, $\alpha_k=1$ and $\beta_k=0$, the classical results are recovered in the short-wavelength limit. 

\begin{figure}[htp]
\centering
\includegraphics[width=\textwidth]{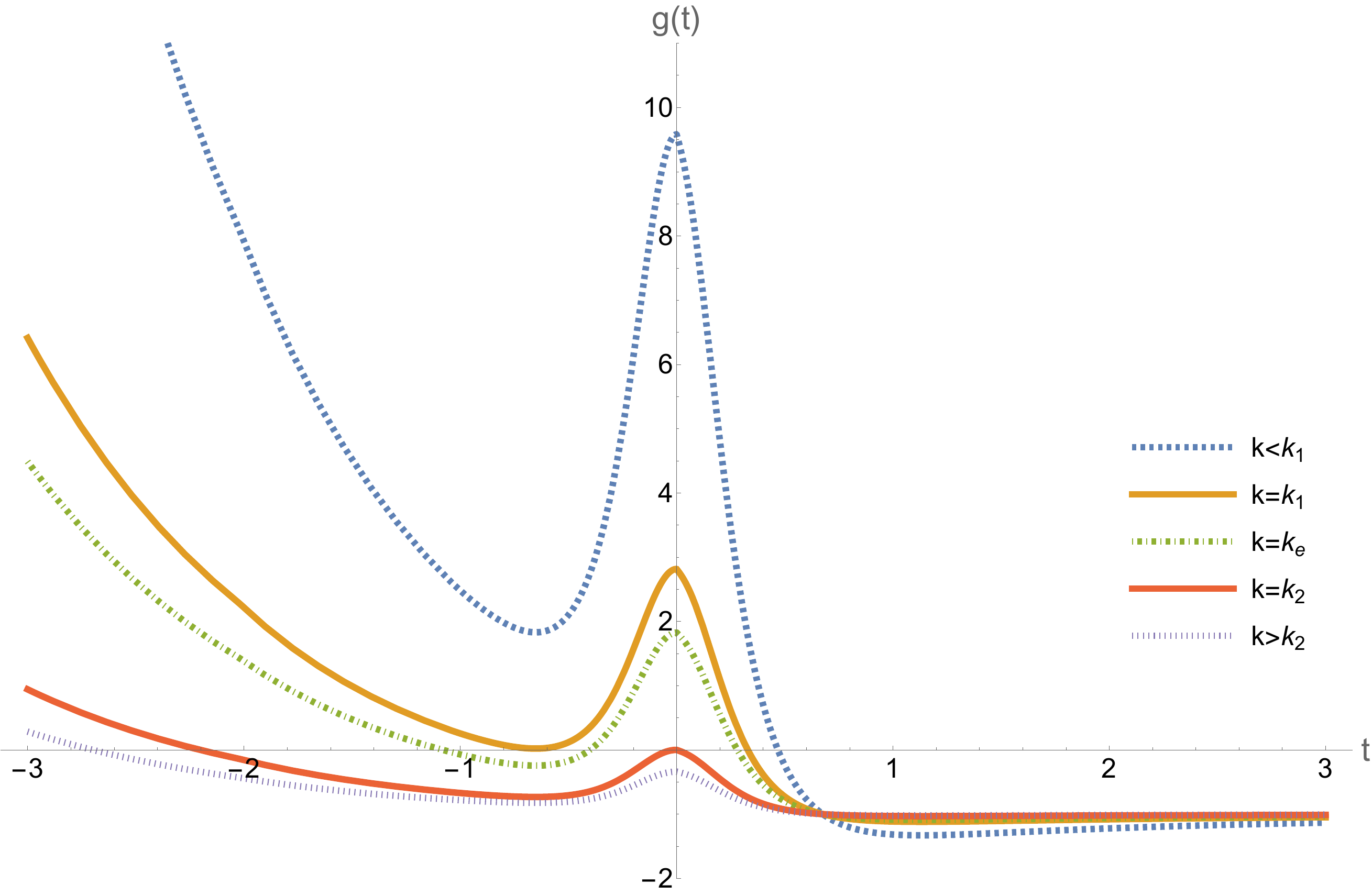}
\caption{The function $g(y(t))$ for several representative cases of different $k$, where $k_1 \simeq 0.83255 \;m_P$, $k_e \simeq 0.96754\; m_P$ and $k_2 \simeq 1.61230 \;m_P$. The roots of $g(y(t)) = 0$ are called turning points.}
\label{fig8:g5k}
\end{figure}

On the other hand, we apply the UAA method to other modes. In particular, in the pre-de Sitter phase, the  MS equation in terms of $y$ reduces to 
\bq
\lb{eq4.2}
\frac{d^2\nu_k(y)}{dy^2}=\left(-1+\frac{2}{y^2}\right)\nu_k(y)=\left(\lambda^2g(y)+q(y)\right)\nu_k(y), \; (t \ll t_m ).
\eq
Then, in order to minimize the error at the singular point $y=0$, we must choose \cite{Zhu:2013upa}
\bq
\lb{eq4.3}
q(y)=-\frac{1}{4y^2}.
\eq
As a result, in other regions we  have
\bq
\lb{eq4.4}
\lambda^2g(y)= -\left(1 +\frac{m^2_\mathrm{eff}}{k^2} + q(y)\right)  = \frac{1}{4y^2} -1 -\frac{m^2_\mathrm{eff}}{k^2}. 
\eq

In Fig. \ref{fig8:g5k} we plot $g(t) \left[\equiv g(y(t))\right]$ for various modes (with $\lambda = 1$), from which we find that we need to consider the three cases: A) $k \gtrsim k_2$;
B) $k_e \lesssim k \lesssim k_2$; and C) $ k \lesssim k_e$, separately. Denoting the three real roots of $g(y,k) = 0$ by $y_0(k)$, $y_1(k)$ and $y_2(k)$ with $y_0 \ge y_1 \ge y_2$, then $k_1,\; k_e, \; k_2$ correspond to $y_0(k_1) = y_1(k_1)$, $\left|y_0(k_e) - y_1(k_e)\right| = \left|y_1(k_e) - y_2(k_e)\right|$, and $y_1(k_2) = y_2(k_2)$, respectively, as shown in Fig. \ref{fig8:g5k}. They are numerically determined as
\bq
\lb{eq4.4aa}
 k_1 \simeq 0.83255 \;m_P, \quad k_e \simeq 0.96754\; m_P, \quad k_2 \simeq 1.61230 \;m_P.
\eq

In this section, our main task is to solve Eqs.(\ref{eq4.2}) - (\ref{eq4.4})  using the UAA method for various intervals of $k$, as illustrated in Fig. \ref{fig8:g5k}. To this goal, let us first introduce a new variable $\xi$ and function $U(\xi)$ via the relations 
\cite{Zhu:2013upa,Olver:1997}
\bqn
\lb{eq4.5}
\nu_k(y) = \dot{y}^{1/2} U(\xi(y)), \quad \dot{y}^2 g(y) = {\cal{F}}(\xi),
\eqn
where $\dot{y} \equiv dy/d\xi$ \footnote{It should be noted that in this paper we also use an over dot to denote the derivative with respect to the cosmic time $t$, for example, $\dot a \equiv da/dt, \dot\rho \equiv d\rho/dt, \dot\phi = d\phi/dt$, etc. However, the dot over $y$ in this paper always represents the derivative of $y$ with respect to $\xi$, $\dot{y} \equiv dy/d\xi$. This is the only exception, and in all other cases an over dot always denote the derivative with respect to the cosmic time $t$, without causing any confusion.}, and ${\cal{F}}(\xi)$ is a function of $\xi$ only, the choice of which depends on the properties of the turning points (or roots)
of $g(y) = 0$ \cite{Zhu:2013upa,Olver:1997}. In terms of $U(\xi)$ Eq.(\ref{eq4.2}) reads
\bq
\frac{d^2U(\xi)}{d\xi^2} = \left(\lambda^2 \dot{y}^2 g + \psi(y)\right)U(\xi), 
\eq
where 
\bq
\lb{eq4.7}
\psi(y) \equiv \dot{y}^2 q(y) + \dot{y}^{1 /2} \frac{d^2}{d\xi^2}\left(\dot{y}^{-1/2}\right). 
\eq
By properly choosing $\dot{y}$, we can make $\left|\lambda^2 \dot{y}^2 g\right| \gg \left|\psi(y)\right|$, so that, to the first-order approximations, we have 
\bq\label{UAA-equation}
\frac{d^2U^{(1)}(\xi)}{d\xi^2} = \lambda^2 \dot{y}^2 g\; U^{(1)}(\xi).
\eq
With the above brief introduction of the UAA method, in the following let us consider  different values of $k$, as illustrated in Fig. \ref{fig8:g5k}.

 \begin{figure}[htp]
\centering
\includegraphics[width=8cm]{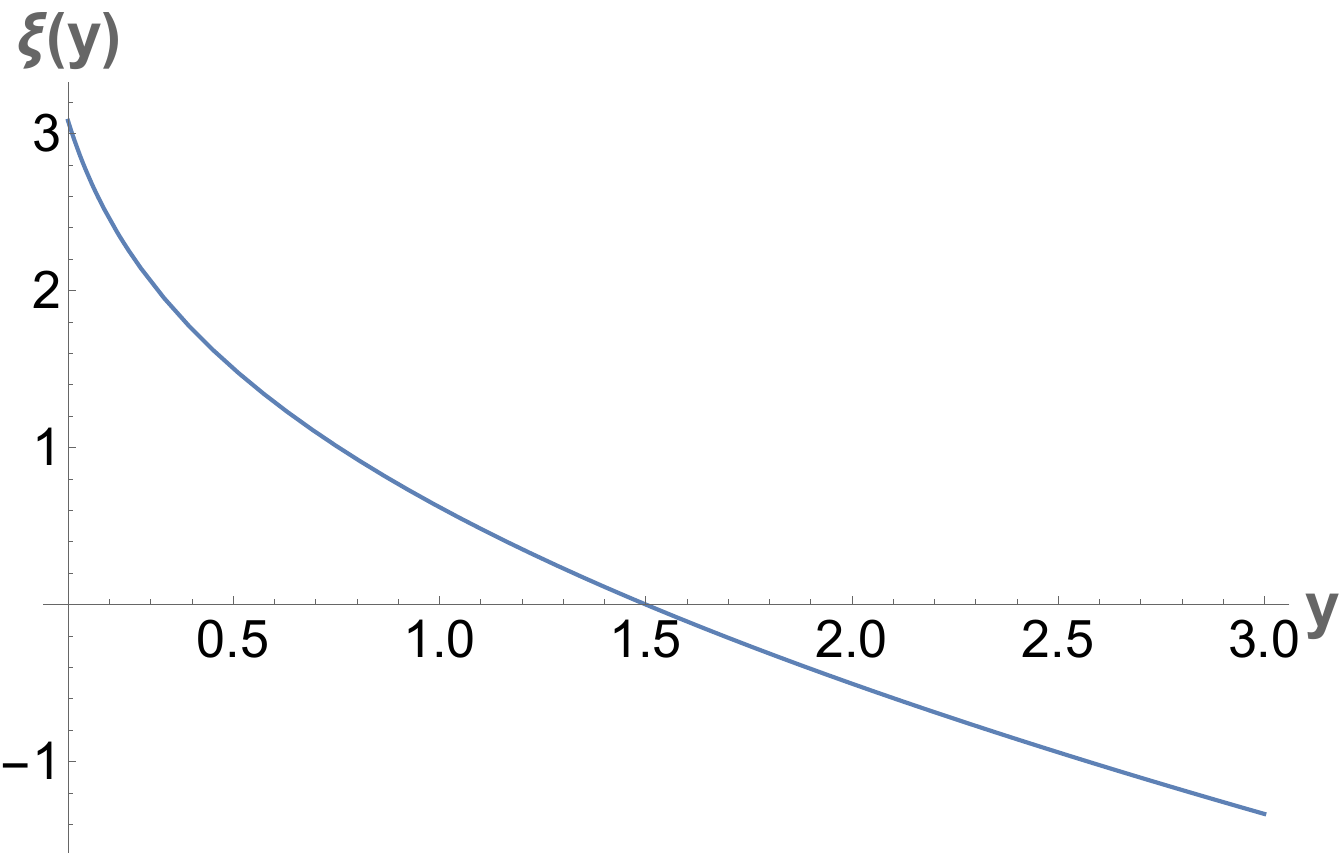}
\caption{The function $\xi(y)$ defined by Eq.(\ref{eq4.6b}) in the neighborhood of the turning point $y_0=1.5$ for $k=28$. It is clear that $\xi(y)$ is monotonically decreasing and vanishes at
the turning point.}
\label{fig9:xi-k28}
\end{figure}

\subsection{$ k \gtrsim k_2$} 

In this case, as shown in Fig. \ref{fig8:g5k}, $g(y) = 0$ has only one turning point (or zero), which will be denoted as $y_0$. Then, $\dot{y}$ can be chosen as
\cite{Olver:1975a,Zhu:2013upa}
\bq
\dot{y}^2 g =  \xi, 
\eq
where $\xi$ and $g(y)$ have the same sign and $\xi(y_0) = 0$. Then, Eq.(\ref{UAA-equation}) becomes the Airy differential equation. Therefore,  the mode function $\nu_k$ can be well approximated by the linear combination of the 
Airy functions of the first and second kinds 
\bq
\lb{analytical_sol1}
\nu_k(t) = \left(\frac{\xi}{ g}\right)^{1/4} \Big\{\tilde a_k {\rm{Ai}}\left(\xi\right) + \tilde b_k{\rm{Bi}}\left(\xi\right)\Big\},
\eq
here $\tilde a_k$ and $\tilde b_k$ are two integration constants to be determined by the initial conditions. 
In addition, $\xi$ and $y$ are related via the relation \cite{Olver:1975a,Zhu:2013upa}
\bq
\lb{eq5.6}
\sqrt{\left|\xi\right|}d\xi = - \sqrt{\left|g(y)\right|}dy.
\eq
Note that the ``-" sign appearing in the right-hand side of the above equation is consistent with the fact that $\xi(y)$ now is chosen to be a monotonically decreasing function of $y$, as shown in Fig. \ref{fig9:xi-k28}.  Then,
we find  
\bqn
\lb{eq4.6}
 \int_{y_0}^y\sqrt{|g(y)|} dy= 
\begin{cases}
- \frac{2}{3} \xi^{3/2}, & y \leq  y_0,\cr
 \frac{2}{3} \left(- \xi\right)^{3/2}, & y\geq y_0, \cr
\end{cases}
\eqn
or inversely 
\bqn
\lb{eq4.6b}
\xi(y)= 
\begin{cases}
 \left(\frac{3}{2} \int^{y_{0}}_y \sqrt{g(y)} dy\right)^{2/3}, ~~y\le y_0, \\
 -\left(\frac{3}{2} \int_{y_{0}}^y \sqrt{- g(y)} dy \right)^{2/3}, ~~y\ge  y_0,
\end{cases}
\eqn
here $y_0$ stands for the location of the turning point, that is, $g(y_0) = 0$. 
It should be noted that
\bq
\lb{eq4.6aa}
\xi\left(y_0\right) = 0, \quad \left.\frac{\xi(y)}{g(y)}\right|_{y= y_0} =\; \text{finite and non-zero}.
\eq
Once the mode function is given in the neighborhood of the turning point,  
we need to match it together with the one given by Eq.(\ref{sol1}) in the pre-de Sitter phase and the one given by Eq.(\ref{sol3A}) in the transition and inflationary phases. In the following, let us first consider the matching between the solutions given by Eqs.(\ref{sol1}) and (\ref{analytical_sol1}). 

Let us first note that for $\eta \ll \eta_1\; (t \ll t_1)$ we have 
\bqn
\xi(y)&=&\left(\frac{3}{2} \int_{y}^{3/2} \sqrt{-1+\frac{9}{4y^2}} d y \right)^{2/3}\simeq \left(-\frac{9}{4}\ln y \right)^{2/3}, \; (y \equiv k \eta  \ll 1), 
\eqn
and
\bqn
\lb{eq4.9}
{\rm{Ai}}(\xi) &\simeq& \frac{e^{-x}}{2(\pi^2 \xi)^{1/4}}, \quad {\rm{Bi}}(\xi) \simeq \frac{e^x}{(\pi^2 \xi)^{1/4}},
\eqn
for $\xi \gg 0$, where $x \equiv (2/3)\xi^{3/2}$. Inserting the above expressions into Eq.(\ref{analytical_sol1}) we find 
\bqn
\lb{eq4.10}
\nu_k(\eta) &\simeq& \left(\frac{\xi}{g}\right)^{\frac14} \left(\tilde a_k \frac{e^{-x}}{2(\pi^2 \xi)^{1/4}} + \tilde b_k \frac{e^{x}}{(\pi^2 \xi)^{1/4}}\right)
\simeq \frac{\tilde a_k}{\sqrt{6 \pi}}  (k\eta)^2 + \sqrt{\frac{2}{3 \pi}}\;  \frac{\tilde b_k}{k\eta},\; \left(k\eta \ll 1\right).
\eqn %}
On the other hand, expanding $e^{\pm i k \eta}$ in terms of $k\eta$, we find that \footnote{Note that here we neglect the contribution from $\epsilon_{\Lambda} \simeq {\cal{O}}(10^{-5})$, so that $\nu^2 = 9/4 + \epsilon_{\Lambda} \simeq 9/4$, and the mode function given by Eq.(\ref{sol1}) can be well approximated by the one given by  Eq.(\ref{eq3.19}).}
\bqn 
\lb{eq4.11}
\nu_k(\eta) & = & - \left\{a_k e^{-ik\eta}\left(1 - \frac{i}{k\eta}\right) + b_k e^{ik\eta}\left(1 + \frac{i}{k\eta}\right)\right\}\nb\\
&=& i(a_k-b_k)\left(\frac{1}{k \eta}+ {\cal{O}}\left(k\eta\right)\right)%^6 \frac12 (k \eta)-\frac18 (k \eta)^3+O((k\eta)^4)\right)
+(a_k+b_k)\left(\frac13 (k \eta)^2 + {\cal{O}}\left(\left(k\eta\right)^4\right)\right). %-\frac{1}{30} (k \eta)^4 + O((k\eta))^6\right) \\
\eqn
Comparing the leading order of Eqs.(\ref{eq4.11}) and (\ref{eq4.10}), we obtain 
\bqn
\lb{eq4.12}
\tilde a_k &=&  \sqrt{\frac{2\pi}{3}} \; (a_k+b_k) , \quad \tilde b_k  = i\sqrt{\frac{3\pi}{2}} \; (a_k-b_k).
\eqn

 \begin{figure}[htp]
{
\includegraphics[width=8cm]{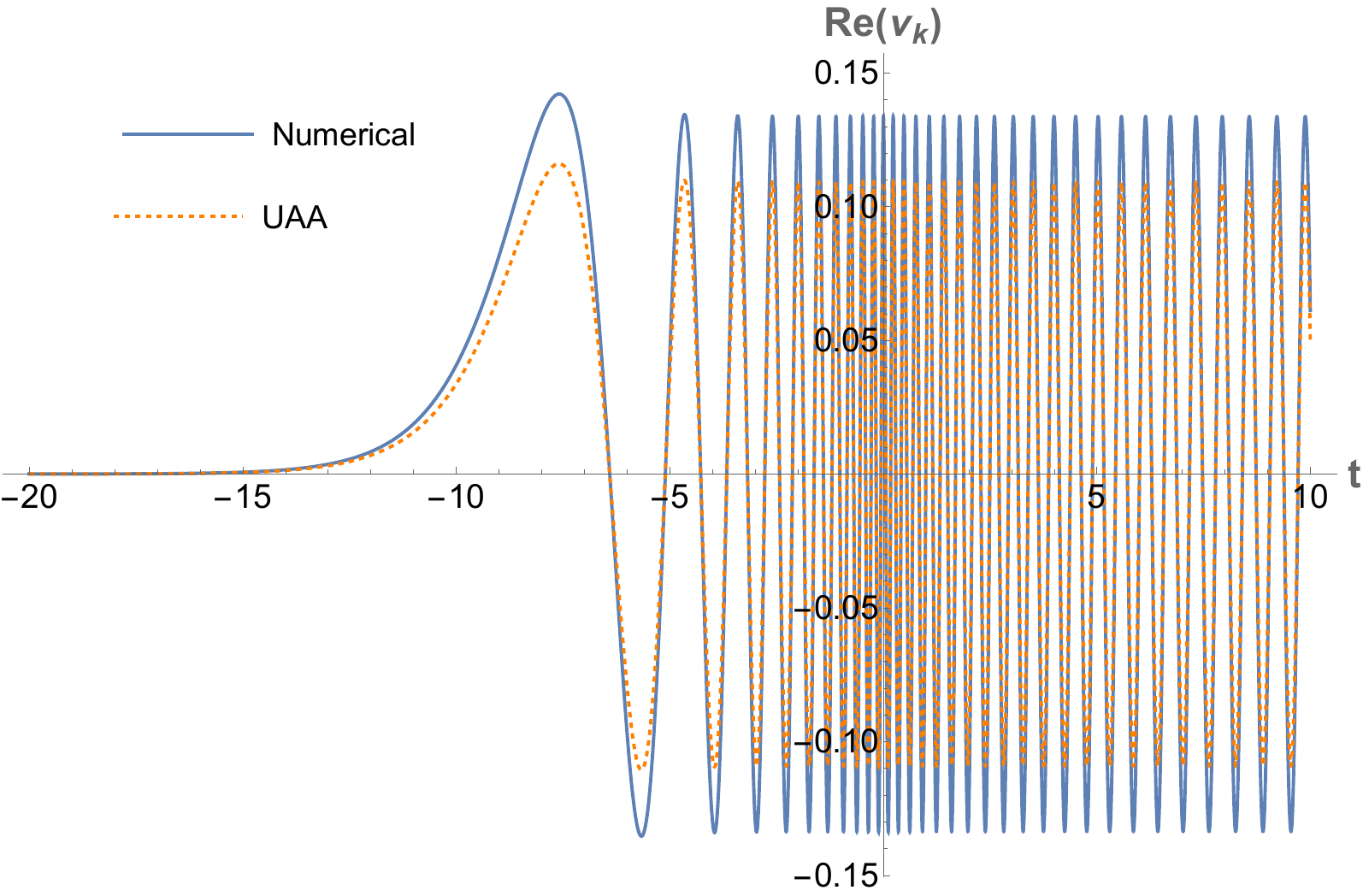}
\includegraphics[width=8cm]{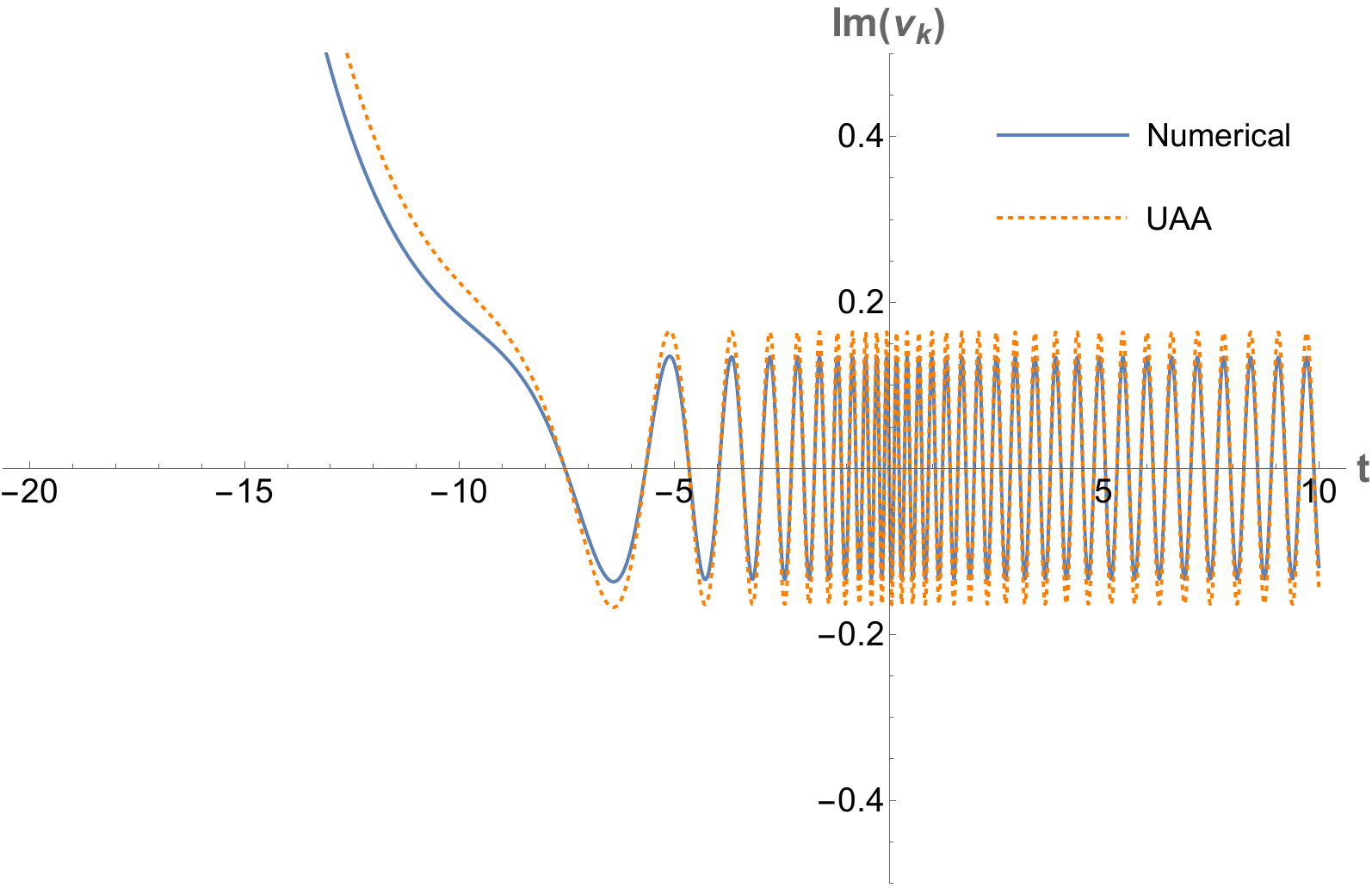}
}
\caption{The numerical and UAA solutions  with $k = 28 m_P$ and the corresponding turning point defined by $g(t_0) = 0$ is $t_0 \simeq -9.0383$. The numerical solutions are obtained by integrating the MS Eq.(\ref{MS_equation}) with the initial conditions given by the analytical solutions of Eq.(\ref{MS_equation}) with the initial time taken at $t_i=-20$. The UAA solutions are given by Eq.(\ref{analytical_sol1}).}
\label{fig10:num-vs-uaa-k28}
\end{figure}

\begin{figure}[htp]
{
\includegraphics[width=8cm]{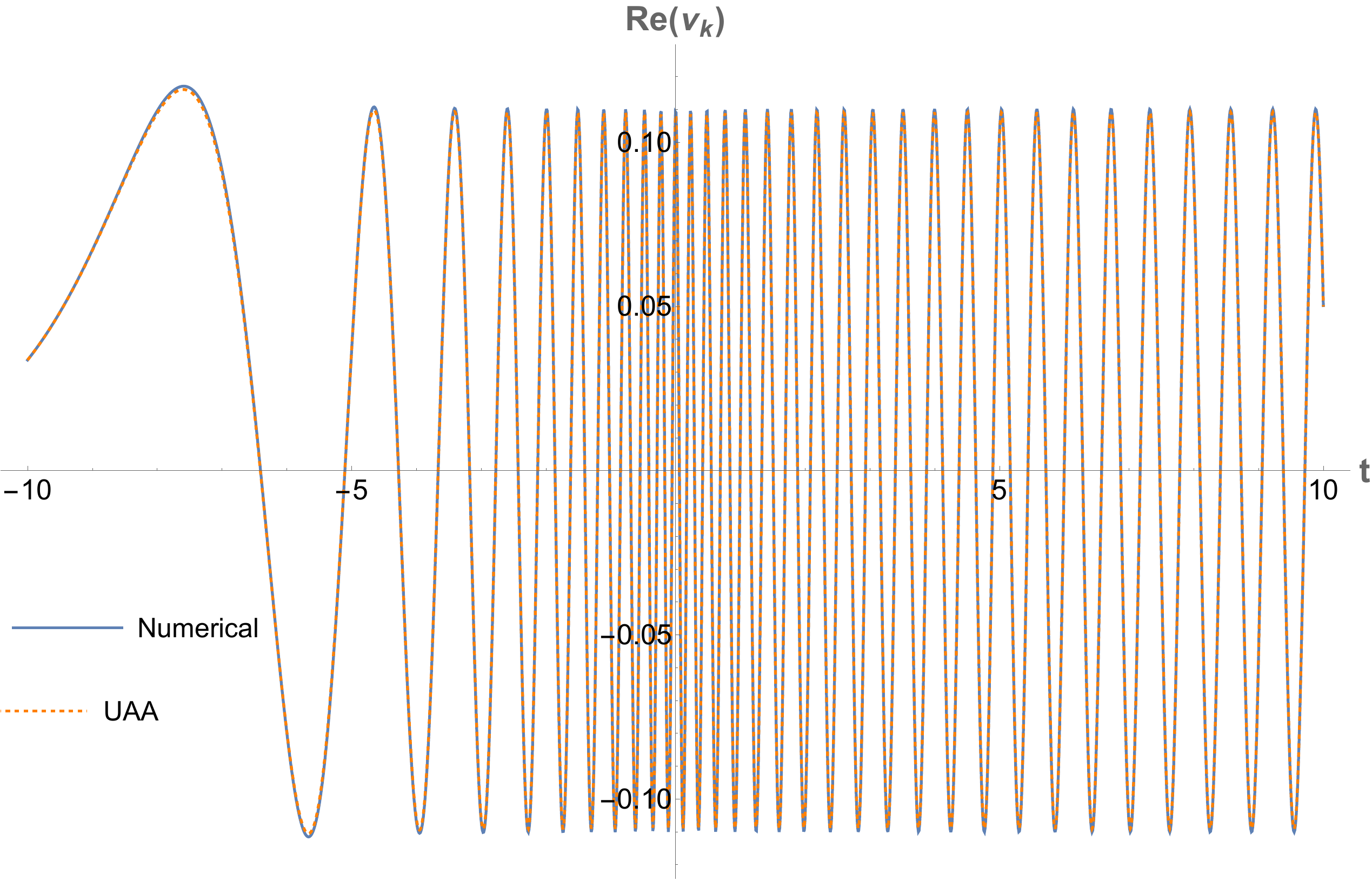}
\includegraphics[width=8cm]{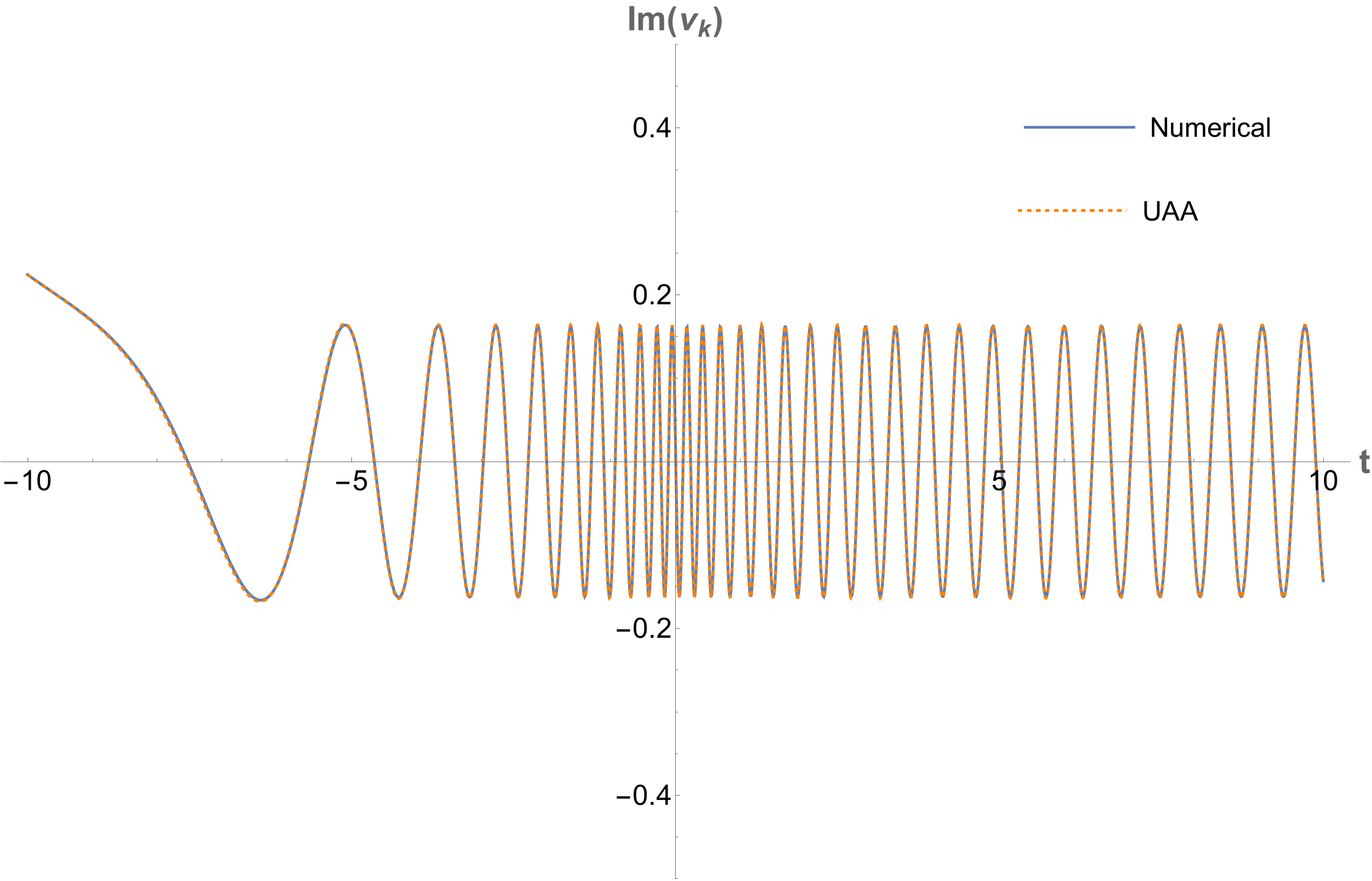}
}
\caption{The numerical and  UAA solutions  with $k = 28 m_P$ and the corresponding turning point is $t_0 \simeq -9.0383$. The numerical solutions are obtained by integrating the MS Eq.(\ref{MS_equation}) but now with the initial conditions at $t_i = -10\; t_P$ given by the UAA solutions of Eq.(\ref{analytical_sol1}).}
\label{fig11:num-vs-uaa-k28-initial-uaa}
\end{figure}

On the other hand, to match the solution of Eq.(\ref{analytical_sol1}) to the one given by Eq.(\ref{sol3A}), we first note that now $\xi$ is given by 
\bqn
\lb{eq4.14}
\xi(y)&=& - \left(\frac{3}{2} \int^y \sqrt{1-\frac{9}{4y^2}} dy\right)^{2/3} \simeq -\left(\frac32 y\right)^{2/3},
\quad (y \equiv k \eta  \gg 1).
\eqn
Then, we get
\bqn
\lb{eq4.15}
\rm{Ai}(-x) &\simeq&  \frac{1}{(\pi^2 x)^{1/4}}\cos\left(\frac{2}{3}x^{3/2}-\frac{\pi}{4}\right), \quad \rm{Bi}(-x) \simeq  -\frac{1}{(\pi^2 x)^{1/4}}\sin\left(\frac{2}{3}x^{3/2}-\frac{\pi}{4}\right),
\eqn
for $x \gg 1$.   
Inserting Eqs.(\ref{eq4.14}) and (\ref{eq4.15}) into Eq.(\ref{analytical_sol1}), we find
\bq
\lb{eq4.16}
\nu_k(y) = \frac{e^{-i\pi/4}}{2\sqrt{\pi}}\left(\left(i\tilde a_k+\tilde b_k\right)e^{-ik\eta}+\left(\tilde a_k+i\tilde b_k\right)e^{ik\eta}\right), \; (y \gg 1).
\eq
Comparing it with Eq.(\ref{sol3A}), we obtain
\bqn
\lb{eq4.17}
\alpha_k &=&    e^{-i\pi/4}\sqrt{\frac{k}{2\pi}}(i\tilde a_k+\tilde b_k) = i e^{-i\pi/4}\sqrt{\frac{k}{12}}\left(5a_k-b_k\right), \nb\\
\beta_k &=&    e^{-i\pi/4}\sqrt{\frac{k}{2\pi}}(\tilde a_k+i\tilde b_k) =  - e^{-i\pi/4}\sqrt{\frac{k}{12}}\left(a_k-5b_k\right).
\eqn

In Fig. \ref{fig10:num-vs-uaa-k28} we plot the numerical and UAA solutions with $k = 28 m_P$, for which the corresponding turning point defined by $g(t_0) = 0$ is $t_0 \simeq -9.0383$. It must not be confused with the current cosmic time and $t_0$ used here. The numerical solutions are obtained by integrating directly the MS equation (\ref{MS_equation}) with the initial conditions given by the analytical solutions  at the initial time $t_i=-20$.
From the figures it can be seen that the UAA solutions trace the numerical ones well
with typical errors about $15\%$. This is consistent with the general analysis of errors for the first-order approximation of the UAA method \cite{Zhu:2014aea}, and it is expected that the upper bound of the errors will be no larger than $0.15\%$ up to the third-order approximation. 
To study the UAA solutions further, in Fig. \ref{fig11:num-vs-uaa-k28-initial-uaa} we show the numerical solutions but now with the UAA solutions at $t_i = -10 t_P$ as the initial conditions. It can be seen that now the numerical and UAA solutions match very well in terms of not only  their frequencies but also their amplitudes. 

On the other hand, in Fig. \ref{fig12:num-vs-uaa-k2} we compare our UAA solution given by Eq.(\ref{analytical_sol1}) with the corresponding numerical one for $k = 2 m_p$, which is a little bit greater than  $k_2 \simeq 1.61230 m_P$. From the figure it can be seen clearly that our UAA solution traces the numerical one well even in this case, considering the fact that this represents only the first-order approximations. To higher order, we expect that our UAA solution will trace much better to the numerical one.

\begin{figure}[htp]
{
\includegraphics[width=8cm]{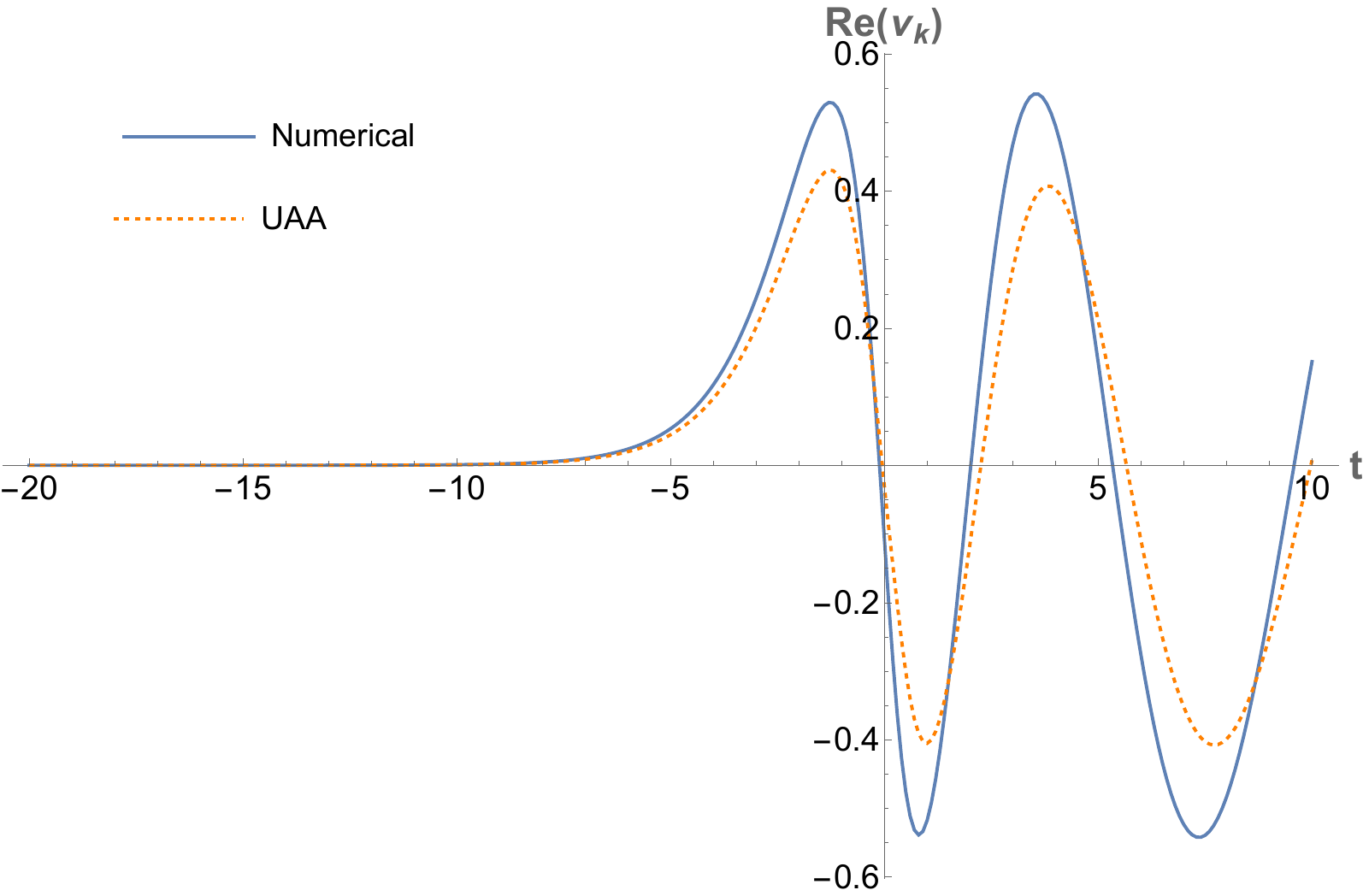}
\includegraphics[width=8cm]{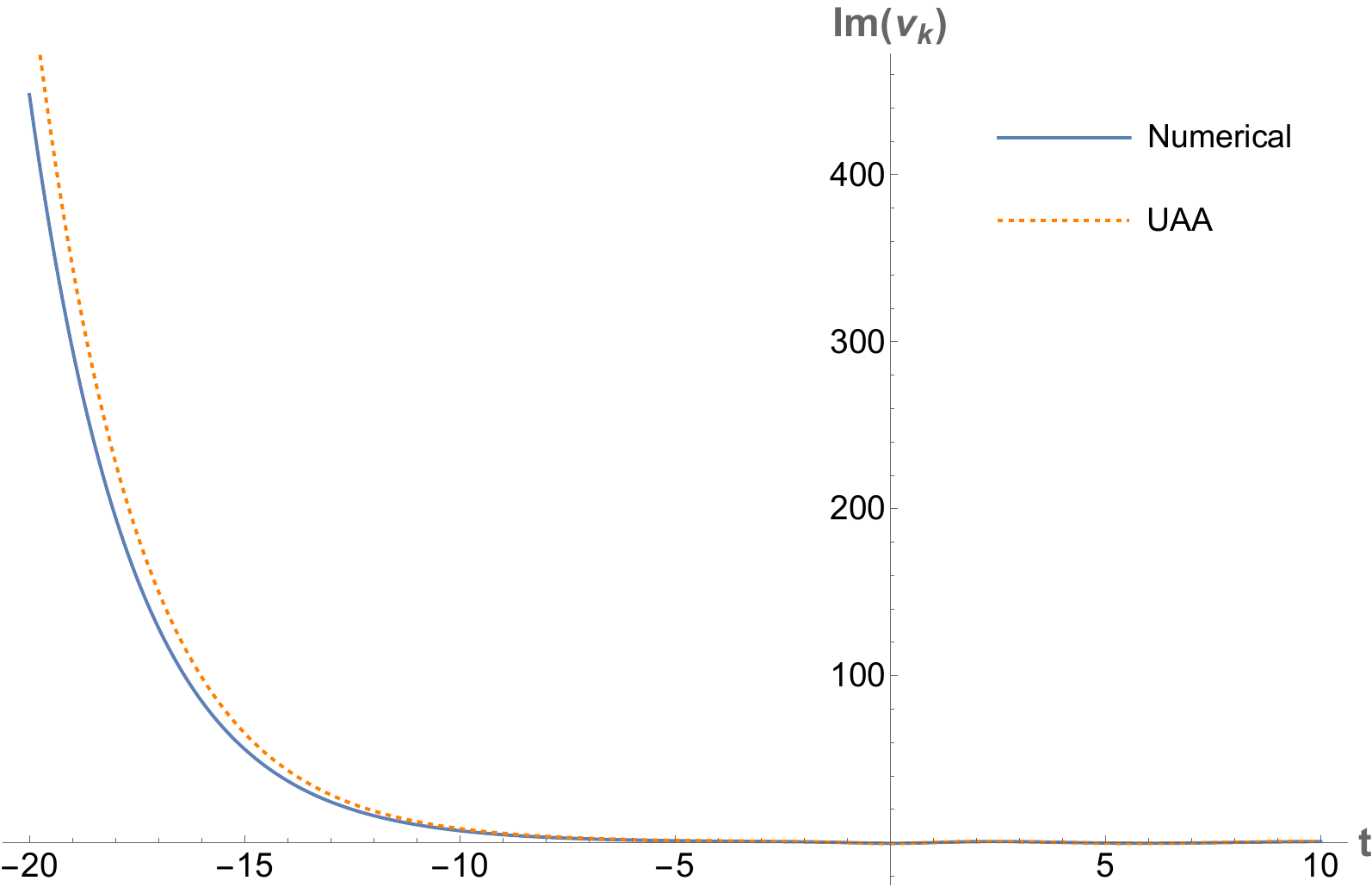}
}
\caption{The numerical and UAA solutions  with $k = 2 m_P$ and the corresponding turning point defined by $g(t_0) = 0$ is $t_0 \simeq -2.09$. The numerical solutions are obtained by integrating the MS Eq.(\ref{MS_equation}) with the initial conditions given by the analytical solutions of Eq.(\ref{MS_equation}) with the initial time taken at $t_i = -20$. The UAA solutions are given by Eq.(\ref{analytical_sol1}). }
\label{fig12:num-vs-uaa-k2}
\end{figure}

\subsection{$k_{e} \lesssim  k \lesssim k_2$} 

In this case, $g(y) = 0$ has three real roots, as shown in Fig. \ref{fig13:g-k155}, which will be referred to as $y_0, \; y_1$ and $y_2$, respectively \cite{Zhu:2013upa}. Without loss of generality, we assume that $y_2 \ge y_1 \ge y_0$. In particular, from the figure it can be seen that $y_0,\;  y_1 < 0$, $y_2 > 0$,  $|y_0| \gg |y_1|$ and $|y_1| \simeq |y_2|$ for $k \gg k_e$. Recall that $k_e$ is defined by
\bq
\lb{eq5.21a}
\left|y_0(k_e) - y_1(k_e)\right| = \left|y_1(k_e) - y_2(k_e)\right|,
\eq
from which we find that $k_e \simeq 0.96754\; m_P$. So, following \cite{Zhu:2014aea}, we can treat $y_0$ as a single-turning point, and $y_1$ and $y_2$ as two-turning points. To discuss the mode function in each of the regions divided by these roots, it is convenient to introduce three constants $y_{a, b, c}$ by assuming that
\bq
\lb{eq5.20}
y_a \ll y_0, \quad y_0 < y_b < y_1, \quad y_c \gg y_2.
\eq
Then, we divide the whole $y$-axis into four different regions [cf. Fig. \ref{fig13:g-k155}], I: $y \leq y_a$; II: $y_a \leq y \leq y_b$; III: $ y_b \leq y \leq y_c$; and IV: $y \geq y_c$.

\begin{figure}[htp]
\centering
\includegraphics[width=8cm]{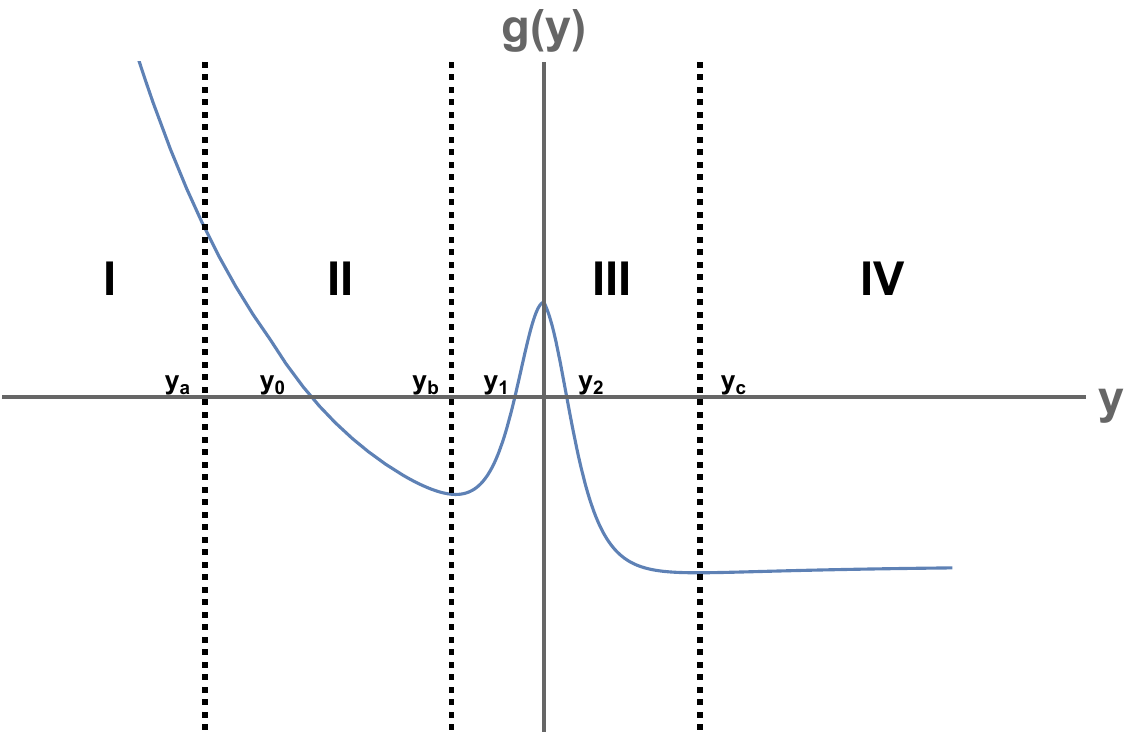}
\caption{The four regions, I - IV for $k=1.55 m_P$, defined, respectively, as I: $y \leq y_a$; II: $y_a \leq y \leq y_b$; III: $ y_b \leq y \leq y_c$; and IV: $y \geq y_c$.}
\label{fig13:g-k155}
\end{figure}

\begin{figure}[htp]
\centering
\includegraphics[width=6cm]{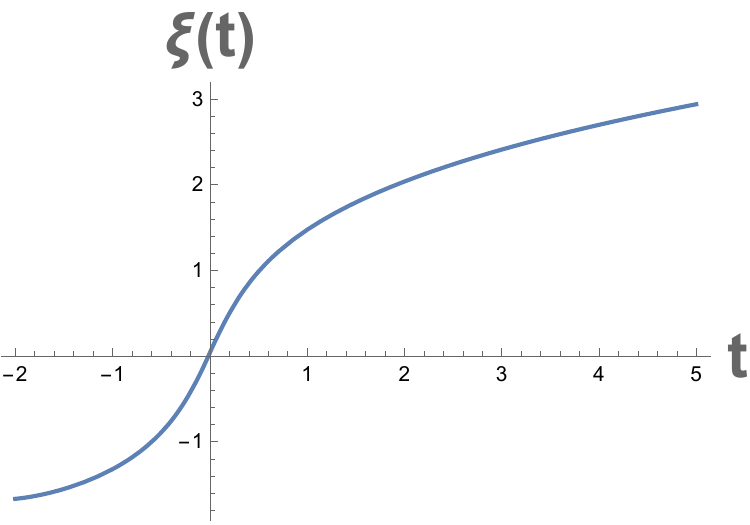}
\caption{The function $\xi(t)$ defined by Eqs.(\ref{eq4.19a})-(\ref{eq4.19c}) for $t_a<t<t_c$.}
\label{fig14:xi-k155}
\end{figure}
 
Thus, it is clear that in Region I the mode function $\nu_k$ is given by Eq.(\ref{eq3.19}) with two arbitrary constants $(a_k, b_k)$, and in Region IV, the mode function $\nu_k$ is given by Eq.(\ref{sol3A})  with  two arbitrary constants $(\alpha_k, \beta_k)$, which includes the transition and inflationary phases.
 In Region II, the function  $g(y)$ has a single-turning point, $y_0$, and the corresponding   mode function  is given by Eq.(\ref{analytical_sol1})  with  two arbitrary constants $(\tilde a_k, \tilde b_k)$, where the relation between the variables  $\xi$ and $y$ is given by Eq.(\ref{eq4.6}) or inversely by Eq.(\ref{eq4.6b}).
Therefore, we have 
\bqn
\lb{eq5.22}
\nu_k(y) = \begin{cases}
- \left\{a_k e^{-ik\eta}\left(1 - \frac{i}{k\eta}\right) + b_k e^{ik\eta}\left(1 + \frac{i}{k\eta}\right)\right\}, & \text{Region I},\cr
\left(\frac{ \zeta}{ g}\right)^{1/4} \Big\{\tilde a_k \rm{Ai}\left(\zeta\right) + \tilde b_k\rm{Bi}\left( \zeta \right)\Big\}, & \text{Region II}, \cr
    \frac{1}{\sqrt{2k}}\left(\alpha_ke^{-ik\eta}+ \beta_k e^{ik\eta}\right), & \text{Region IV}. \cr
\end{cases}
\eqn
Note that in writing the solution in Region II we had used the argument $\zeta$ instead of $\xi$ in order to distinguish the argument of the solution in Region III to be considered below. In particular,   now $\zeta$ is defined by
[cf. Eq.(\ref{eq5.6})]
\bq
\lb{eq5.22aa}
\dot{y}^2 g(y) = \zeta,
\eq
with $\zeta(y_0) = 0$, where $\dot{y} \equiv dy/d\zeta$. Then, we can see that $\zeta(y)$ has the same sign as that of $g(y)$. Without loss of generality, similar to the last case we shall choose  
\bq
\lb{eq5.22bb}
\dot{y} = - \sqrt{\frac{\zeta}{g}},
\eq
so that $\zeta(y)$ is a monotonically decreasing function of $y$ in the neighbourhood of $y_0$ [cf. Fig. \ref{fig9:xi-k28}].

In Region III, the mode function is given by the two-turning point function \cite{Zhu:2013upa}
\bqn
\lb{eq4.18}
\nu^{\text{III}}_k(y)&=&\left(\frac{\xi^2-\xi_0^2}{-g(y)}\right)^{1/4} \left\{\tilde\alpha_k  W\left(\frac{1}{2}\xi_0^2, \sqrt{2}\xi \right)+\tilde\beta_k  W\left(\frac{1}{2}\xi_0^2, -\sqrt{2}\xi \right)\right\},
\eqn
where $(\tilde\alpha_k, \tilde\beta_k)$ are two integration constants, $W\left(\frac{1}{2}\xi_0^2, \pm \sqrt{2}\xi \right)$ denote the parabolic cylindrical functions, and now the relation between $\xi$ and $y$ is  given by \cite{Olver:1975a,Zhu:2013upa}
\bq
\lb{eq4.18aa}
\sqrt{\left|\xi^2 - \xi_0^2\right|}d\xi = \sqrt{\left|g(y)\right|} dy,
\eq
so that $\xi(y)$ is a monotonically increasing function in the region $y \in (y_b, y_c)$, as shown in Fig. \ref{fig14:xi-k155}. 
Then, we find 
\bqn
\lb{eq4.19a}
 \int_{y}^{y_1} \sqrt{-g(y)} dy &=& \int_{\xi}^{-\xi_0}\sqrt{\xi^2 - \xi_0^2} d\xi\nb\\
&=&   - \frac{1}{2} \xi 
\;  \sqrt{\xi^2-\xi_0^2} + \frac{1}{2} \xi_0^2 \; \ln\left(\frac{\xi + \sqrt{\xi^2 - \xi_0^2}}{-\xi_0}\right)\nb\\
&=& - \frac{1}{2} \xi 
\;  \sqrt{\xi^2-\xi_0^2} - \frac{1}{2} \xi_0^2 \; \ln\left(\frac{\sqrt{\xi^2 - \xi_0^2} - \xi}{\xi_0}\right), \;\;\;\; y_b \leq y \leq y_1,\\ 
\lb{eq4.19b}
 \int_{y_1}^{y} \sqrt{g(y)} dy&=& \int^{\xi}_{-\xi_0}\sqrt{\xi_0^2 - \xi^2} d\xi \nb\\
 &=&\frac{1}{2} \xi 
    \sqrt{\xi_0^2 - \xi^2} + \frac{1}{2} \xi_0^2 \left[\sin^{-1}\left(\frac{\xi}{\xi_0}\right) +  \frac{\pi}{2}\right],  \;\;\;\; y_1 \leq y \leq y_2,\\
    \lb{eq4.19c}
 \int_{y_2}^{y} \sqrt{-g(y)} dy &=& \int^{\xi}_{\xi_0}\sqrt{\xi^2 - \xi_0^2} d\xi \nb\\
 &=&   \frac{1}{2} \xi 
\;  \sqrt{\xi^2-\xi_0^2} - \frac{1}{2} \xi_0^2 \; \ln\left(\frac{\xi + \sqrt{\xi^2 - \xi_0^2}}{\xi_0}\right),  \;\;\;\; y_2 \leq y \leq y_c,
\eqn
where 
\bqn
\lb{eq4.22}
\xi_0^2 \equiv \frac{2}{\pi} 
    \int_{y_1}^{y_2} \sqrt{-g(y)} dy = \int_{-\xi_0}^{\xi_0} \sqrt{\xi_0^2 - \xi^2} d\xi. 
\eqn
Note that in writing down the above expressions we assumed that
\bq
\lb{eq4.23}
\xi\left(y_1\right) = - \xi_0, \quad \xi\left(y_2\right) =  \xi_0, \quad \xi_0 \ge 0.
\eq 
It is clear that  $\xi_0 = 0$, when the two roots $y_{1}$ and $y_2$ are identical. 

It should be noted that the above treatments can be extended to the case  where $k \gtrsim k_2$. From Fig. \ref{fig8:g5k} we can see that this corresponds to the case where $y_1$ and $y_2$ are complex conjugate \cite{Olver:1975a,Zhu:2013upa}. 
In particular, we assume that  $y_1$ and $y_2$ are purely imaginary \footnote{Otherwise, we can first make the coordinate transformation, $\tilde y = y - \text{Re}(y_1)$, then in terms of $\tilde y$, we find that the two complex turning points becomes purely imaginary.}, that is, $y_1 = - y_1^*$ and $y_2 = y_1^*$, and that  
\bq
\lb{eq4.23aa}
\xi\left(y_1\right) = - i\left|\xi_0\right|, \quad \xi\left(y_2\right) =  i\left|\xi_0\right|, 
\eq
 with $\xi_0^2 < 0$. Then, we find that   
\bqn
\lb{eq4.22aa}
\xi_0^2 \equiv - \frac{2}{\pi} \left|\int_{y_1}^{y_2} \sqrt{-g(y)} \; dy\right| = -  \frac{2}{\pi} \left|\int_{-i|\xi_0|}^{i|\xi_0|} \sqrt{\xi^2 + |\xi_0^2|} \; d\xi\right|,
\eqn
where the integrations are performed along the imaginary axes \cite{Olver:1975a}. Therefore, in the rest of this subsection, we shall treat all these cases together \cite{Zhu:2013upa}.

Once the solutions are given  in each of the above four regions, we need to match them together across their common regions. In particular, the matching between 
the solutions of Eqs.(\ref{eq3.19}) and (\ref{analytical_sol1}) given in Regions I and II, respectively, is still given by Eq.(\ref{eq4.12}), that is
\bqn
\lb{eq5.31}
\tilde a_k &=&  \sqrt{\frac{2\pi}{3}} \; (a_k+b_k) , \quad \tilde b_k  = i\sqrt{\frac{3\pi}{2}} \; (a_k-b_k).
\eqn
On the other hand, to  match  
the solutions of Eq.(\ref{analytical_sol1}) with those given by Eq.(\ref{eq4.18}), we first note that  for a large and positive $x$ the parabolic cylinder functions take the asymptotical forms \cite{Olver:1975a,Gil:2004,Zhu:2013upa},
\bqn\lb{Wasymp}
W\left(b, x\right)  &\simeq& \left(\frac{ j^2(b)}{\frac14x^2-b}\right)^{1/4} \cos{\mathfrak{D}(x)} \nb\\
W\left(b,- x\right) &\simeq& \left(\frac{ j^{-2}(b)}{\frac14x^2-b}\right)^{1/4} \sin{\mathfrak{D}(x)} \;\;\; (x \gg 1), 
\eqn
where $j(b)\equiv \sqrt{1+e^{2\pi b}}-e^{\pi b}$ and
\bqn
\lb{eq4.25}
&&\mathfrak{D}(b,x) \equiv \frac12x\sqrt{\frac14 x^2-b} - b\ln \left(\frac{\frac12 x+ \sqrt{\frac14 x^2-b}}{\sqrt{b}}\right) +\frac{\pi}{4} + \phi (b),\nb\\  
&&  \phi(b) \equiv  \frac{b}{2} - \frac{b}{4} \ln{b^2} +\frac{1}{2} \text{ph}\Gamma\left(\frac{1}{2}+i b\right),
\eqn
with the phase  $\text{ph}\Gamma\left(\frac{1}{2}+i b\right)$ being zero when $b=0$, and determined by continuity, otherwise. For more details, see \cite{Olver:1975a,Gil:2004,Zhu:2013upa}.
Setting 
$x = -\sqrt{2}\xi$, $b =\xi_0^2/2$,
for $\xi \ll 0$, we find that Eq.(\ref{Wasymp}) yields
\bqn\lb{eq5.38}
&W\left(\frac{1}{2}\xi_0^2, \sqrt{2}\xi\right)  \simeq
\left(\frac{2 j^{-2}\left(\frac12 \xi_0^2\right)}{\xi^2-\xi_0^2}\right)^{1/4} \sin{\mathfrak{D}(-\xi)}, \\
&W\left(\frac{1}{2}\xi_0^2, -\sqrt{2}\xi\right)  \simeq
\left(\frac{2 j^{2}\left(\frac12 \xi_0^2\right)
}{\xi^2-\xi_0^2}\right)^{1/4} \cos{\mathfrak{D}(-\xi)}, %\;\;\; (\xi \ll 0).
\eqn
where
\bqn
\lb{eq5.39}
\mathfrak{D}(-\xi)
= -\frac{1}{2}\xi \sqrt{\xi^2-\xi_0^2} +\frac{1}{2}\xi_0^2\ln\left(\frac{\xi+\sqrt{\xi^2-\xi_0^2}}{-\xi_0}\right) + \phi\left(\frac12 \xi_0^2\right), \;(\xi \ll 0).
\eqn
Note that from \eqref{eq4.19a}
\bqn
\lb{eq5.40}
\int_{y}^{y_1}\sqrt{-g(y)} dy 
= \left(\int_{y_0}^{y_1} - \int_{y_0}^{y}\right)\sqrt{-g(y)} dy 
= -\frac{1}{2}\xi \sqrt{\xi^2-\xi_0^2} +\frac{1}{2}\xi_0^2\ln\left(\frac{\xi+\sqrt{\xi^2-\xi_0^2}}{-\xi_0}\right).\nb\\
\eqn
Thus, we find 
\bqn
\lb{eq5.41}
\mathfrak{D} &=&  {\cal{B}} - {\cal{A}}, \quad 
{\cal{A}} \equiv   \int_{y_0}^{y}\sqrt{-g(y)} dy , \quad
{\cal{B}} \equiv \int_{y_0}^{y_1}\sqrt{|g(y)|}dy + \phi \left(\frac12 \xi_0^2\right).
\eqn
Inserting the above expressions into Eq.(\ref{eq4.18})
we obtain
\bqn
\lb{eq5.42}
\nu^{\text{III}}_k(y)&\simeq& \frac{1}{\left(-g\right)^{1/4}}\left[ \left(\hat\alpha_k\sin{\cal{B}} + \hat\beta_k\cos{\cal{B}}\right)\cos{\cal{A}}
- \left(\hat\alpha_k\cos{\cal{B}} - \hat\beta_k\sin{\cal{B}}\right)\sin{\cal{A}}\right],
\eqn
where 
\bqn
\lb{eq5.43}
 \hat\alpha_k \equiv \left(\frac{\sqrt{2}}{j(b)}\right)^{1/2} \tilde\alpha_k,  \quad
  \hat\beta_k \equiv \left(\sqrt{2}\; j(b)\right)^{1/2} \tilde\beta_k.
\eqn

On the other hand, for $\xi \ll 0$, from Eq.(\ref{eq4.15})  we find 
\bq
\lb{eq5.44}
\text{Ai}(\xi) \simeq \frac{1}{\left(- \pi^2\xi\right)^{1/4}}\cos{\cal{A}}, \quad
\text{Bi}(\xi) \simeq - \frac{1}{\left(- \pi^2\xi\right)^{1/4}}\sin{\cal{A}},
\eq
where  ${\cal{A}}$ is defined in Eq.(\ref{eq5.41}). Thus, the solution given by Eq.(\ref{analytical_sol1}) becomes
\bq
\lb{eq5.45}
\nu^{\text{II}}_k(\xi) \simeq \frac{1}{\left(- \pi^2 g\right)^{1/4}} \left(\tilde{a}_k \cos{\cal{A}} 
- \tilde{b}_k\sin{\cal{A}}\right), \;\;(\xi \ll 0).
\eq
Comparing Eqs.(\ref{eq5.45}) and (\ref{eq5.42}) we find 
\bqn
\lb{eq5.46}
\tilde{\alpha}_k &=& \left(\frac{j(b)}{\sqrt{2}\; \pi}\right)^{1/2}\left(\sin{\cal{B}} \; \tilde{a}_k
+ \cos{\cal{B}} \; \tilde{b}_k\right), \nb\\
\tilde{\beta}_k &=& \frac{1}{\left(\sqrt{2}\; \pi j(b)\right)^{1/2}} \left(\cos{\cal{B}} \; \tilde{a}_k
- \sin{}{\cal{B}} \; \tilde{b}_k\right).~~~
\eqn

\begin{figure}[htp]
{
\includegraphics[width=8cm]{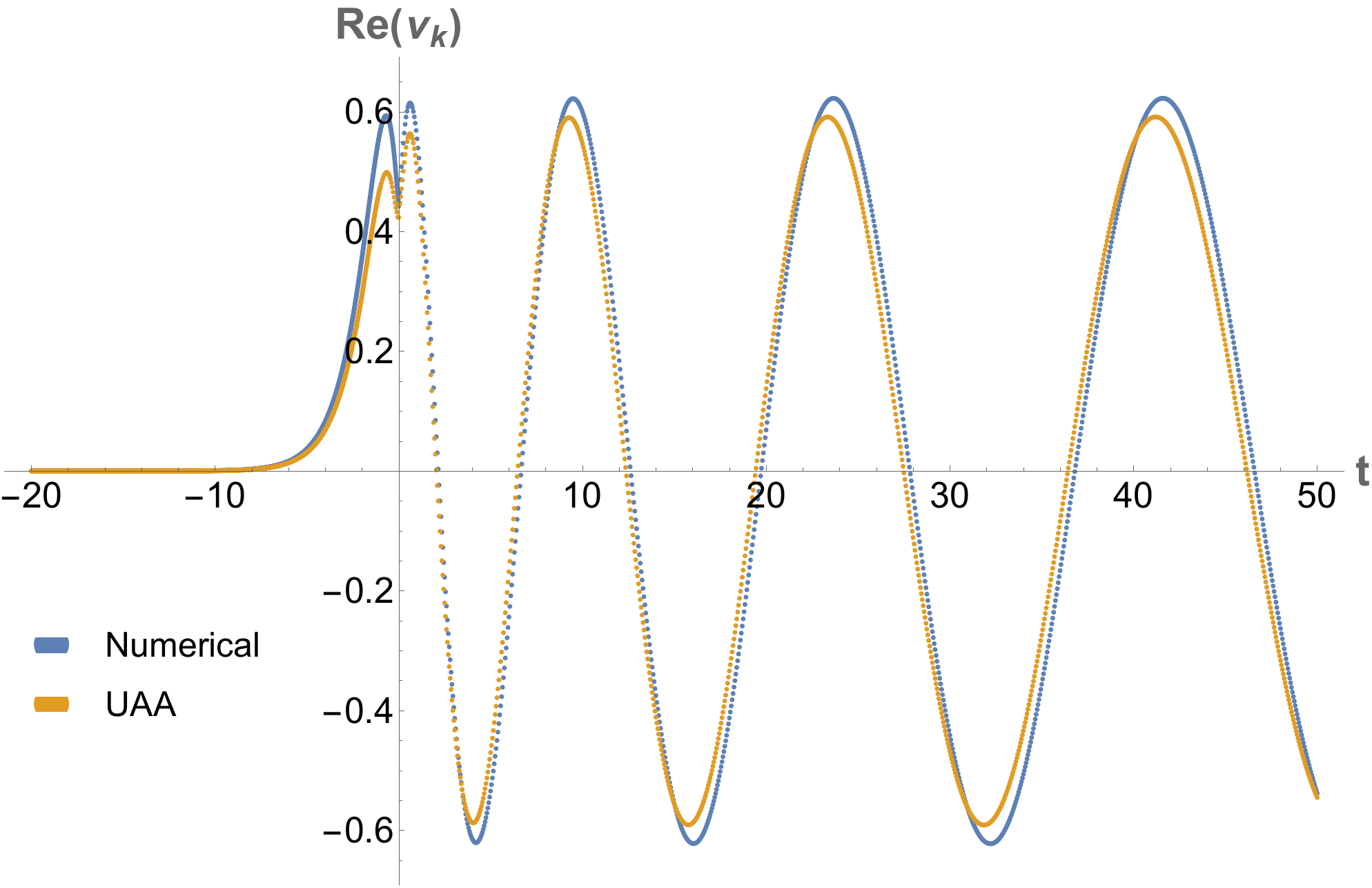}
\includegraphics[width=8cm]{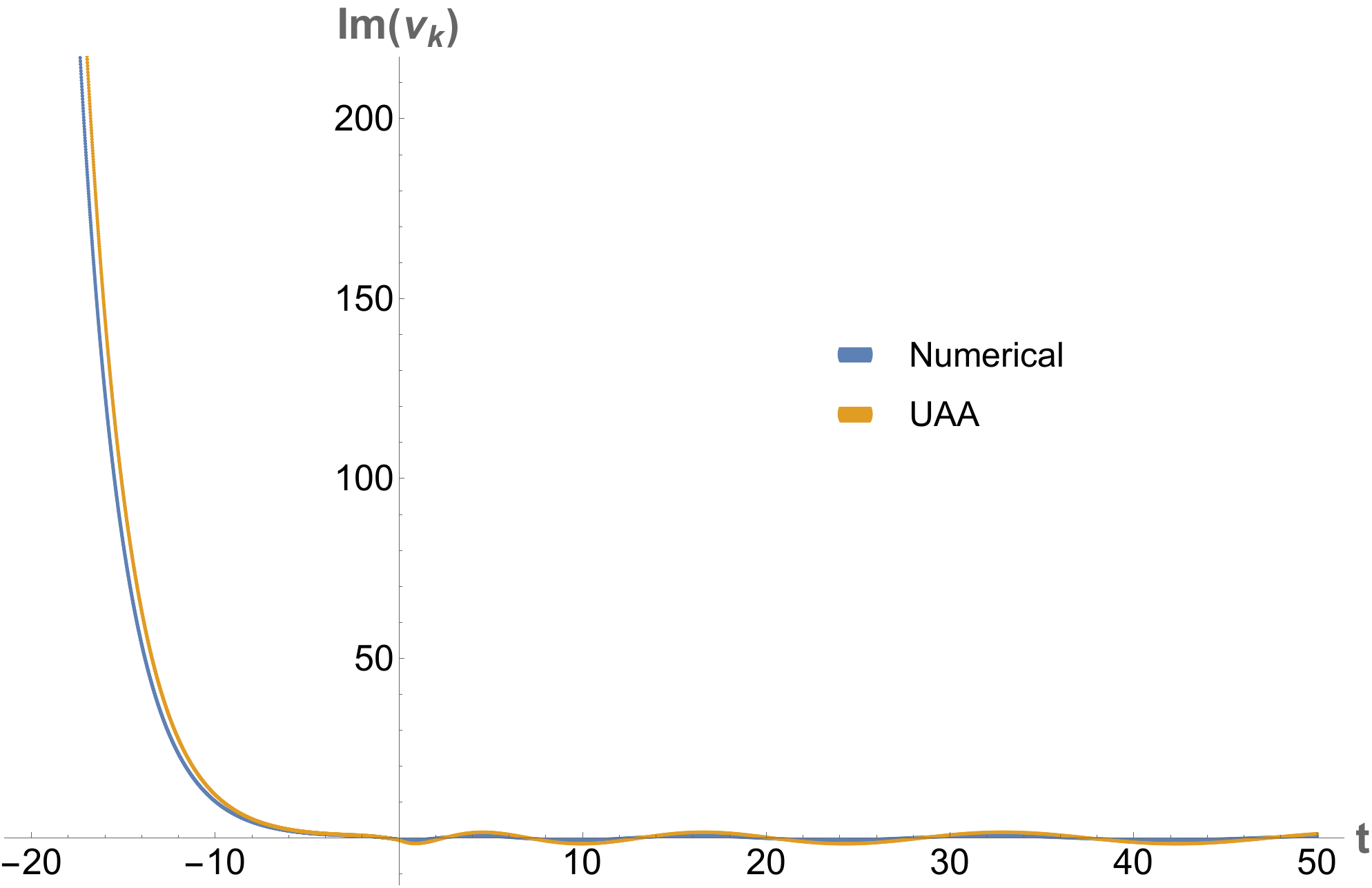}
}
\caption{The numerical and  UAA solutions with $k = 1.55 m_P$ and $b_k=0$. The corresponding turning points are $y_0=1.5 \;(t_0 \simeq -2.0874)$, $ y_1=3.51683 \;(t_1\simeq -0.0917) $ and $ y_2=3.75345 \;(t_2 \simeq 0.06131)$. The numerical solutions are obtained by integrating the MS with the initial conditions at $t_i = -20\; t_P$. the UAA solutions of Eq.(\ref{eq4.18}).}
\label{fig15:num-vs-uaa-k155}
\end{figure}

To match the solution (\ref{eq4.18}) with that given by Eq.(\ref{sol3A}), let us first note that for $\xi \gg \xi_0$ from Eq.(\ref{Wasymp}) we find
\bqn\lb{eq5.47}
W\left(\frac{1}{2}\xi_0^2, \sqrt{2}\xi\right) \simeq \left(\frac{\sqrt{2} j(b)}{\xi}\right)^{1/2} \cos{y},  \;\;
W\left(\frac{1}{2}\xi_0^2, -\sqrt{2}\xi\right)=\left(\frac{\sqrt{2}}{j(b) \xi}\right)^{1/2} \sin{y}, \;\;\; (\xi \gg \xi_0).\nb\\
\eqn
Note that in writing down the above expressions, we used the fact ${\cal{D}} \simeq y$. Then, we find
that
\bqn
\lb{eq5.48}
\nu^{\text{III}}_k(y)&\simeq& \left(\sqrt{2}\; j(b)\right)^{1/2}  \tilde\alpha_k\cos{y}  + \left(\frac{\sqrt{2}}{j(b)}\right)^{1/2} \tilde\beta_k\sin{y},
\;\; (\xi \gg \xi_0).
\eqn
Comparing the above expression with that of Eq.(\ref{eq4.18}) we obtain
\bqn
\lb{eq5.49a}
{\alpha}_k &=& \sqrt{\frac{k}{2}}\left[\left(\sqrt{2}\; j(b)\right)^{1/2}  \tilde{{\alpha}}_k + i \left(\frac{\sqrt{2}}{j(b)}\right)^{1/2}  \tilde{\beta}_k\right],\nb\\
{\beta}_k &=& \sqrt{\frac{k}{2}}\left[\left(\sqrt{2}\; j(b)\right)^{1/2}  \tilde{{\alpha}}_k - i \left(\frac{\sqrt{2}}{j(b)}\right)^{1/2}  \tilde{\beta}_k\right].
\eqn
Inserting Eqs.(\ref{eq5.31}) and (\ref{eq5.43}) into the above expressions, we finally get
\bqn
\lb{eq5.49}
{\alpha}_k &=& \sqrt{\frac{k}{12}}\; \Bigg\{\Big[\left(2j + 3j^{-1}\right)\sin{\cal{B}}+ i \left(3j + 2j^{-1}\right)\cos{\cal{B}}\Big]a_k\nb\\
&& ~~~~~~~~~~~ + \Big[\left(2j - 3j^{-1}\right)\sin{\cal{B}} - i \left(3j - 2j^{-1}\right)\cos{\cal{B}}\Big]b_k\Bigg\},\nb\\
{\beta}_k &=& \sqrt{\frac{k}{12}}\; \Bigg\{\Big[\left(2j - 3j^{-1}\right)\sin{\cal{B}} + i \left(3j - 2j^{-1}\right)\cos{\cal{B}}\Big]a_k\nb\\
&& ~~~~~~~~~~~ + \Big[\left(2j + 3j^{-1}\right)\sin{\cal{B}} - i \left(3j + 2j^{-1}\right)\cos{\cal{B}}\Big]b_k\Bigg\}.
\eqn
It should be noted that the junction conditions \eqref{eq5.49a} hold only for $y_1 \gg y_0$. Otherwise we request
\bqn
\lb{eq5.50}
\mu_k^{\text{II}}(y_b) = {\mu_k^{\text{III}}(y_b)}, \quad
\frac{d\mu_k^{\text{II}}(y_b)}{dy} = \frac{d\mu_k^{\text{III}}(y_b)}{dy} .
\eqn
In Fig.\ref{fig15:num-vs-uaa-k155}, we plot the numerical and UAA solutions for $k=1.55\; m_P$, for which  $y_0=1.5$, $y_1=3.51683$ and $y_2=3.75345$. Since $|y_1-y_0|$ is not much greater than zero, we use the junction conditions \eqref{eq5.50} instead of \eqref{eq5.49a}. The initial conditions are chosen to be $b_k=0$. From the figure, we can see the UAA solution traces the numerical solution very well.

In Fig.\ref{fig16:num-vs-uaa-k1}, we plot the numerical and UAA solutions for $k=1.0 m_P$, which is a little bit greater than $k_e \simeq 0.96754 m_p$, for which $y_0= 1.4285$, $y_1= 1.99364$ and $y_2=2.61394$. Since $|y_1-y_0|$ is also not much greater than zero, so we use the junction conditions \eqref{eq5.50} instead of \eqref{eq5.49a}. The initial conditions are chosen to be $b_k=0$. From the figure, we can see even in this case ($k \simeq k_e$) the UAA solution still traces the numerical one very well.

\begin{figure}[htp]
{
\includegraphics[width=8cm]{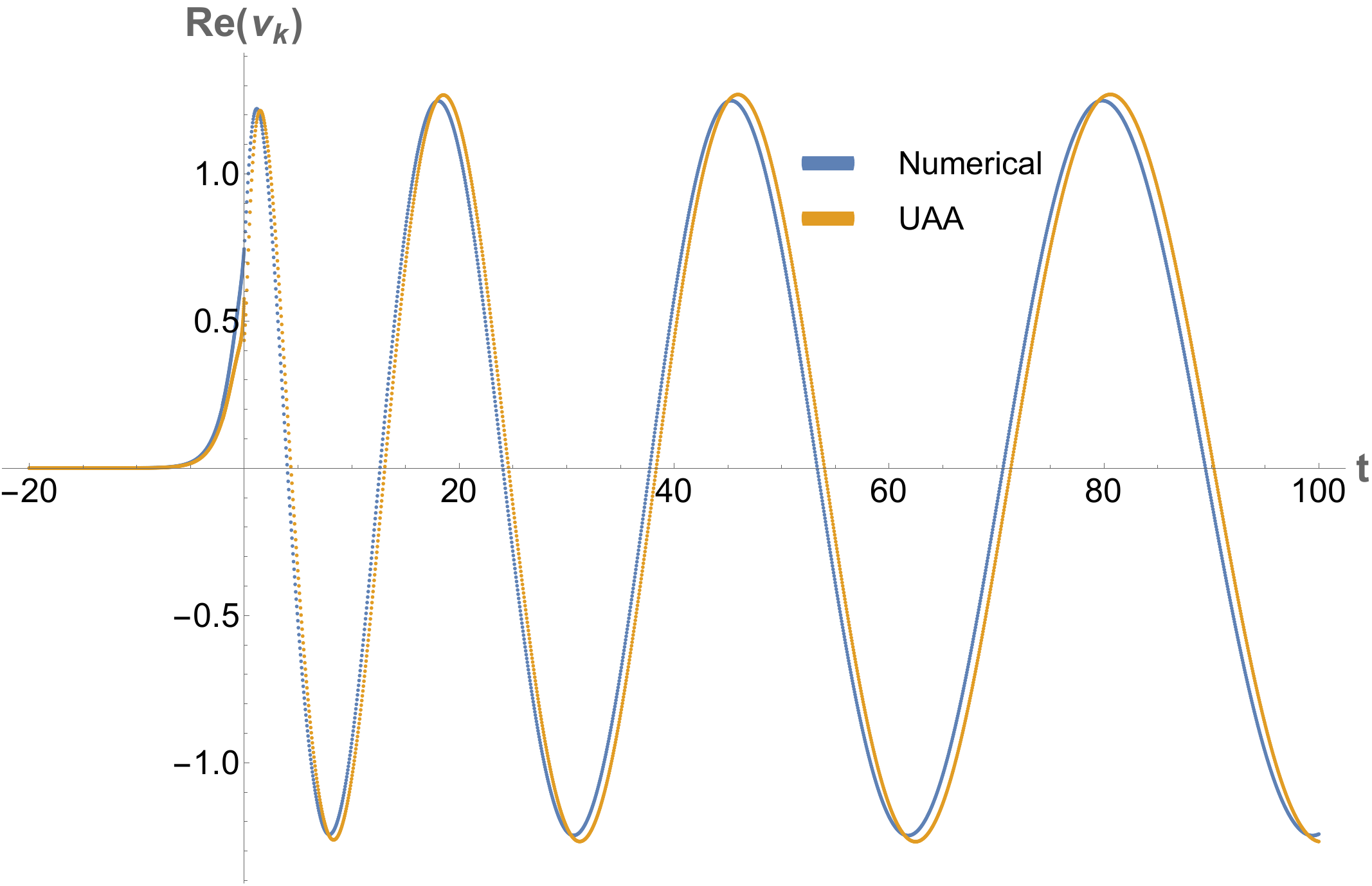}
\includegraphics[width=8cm]{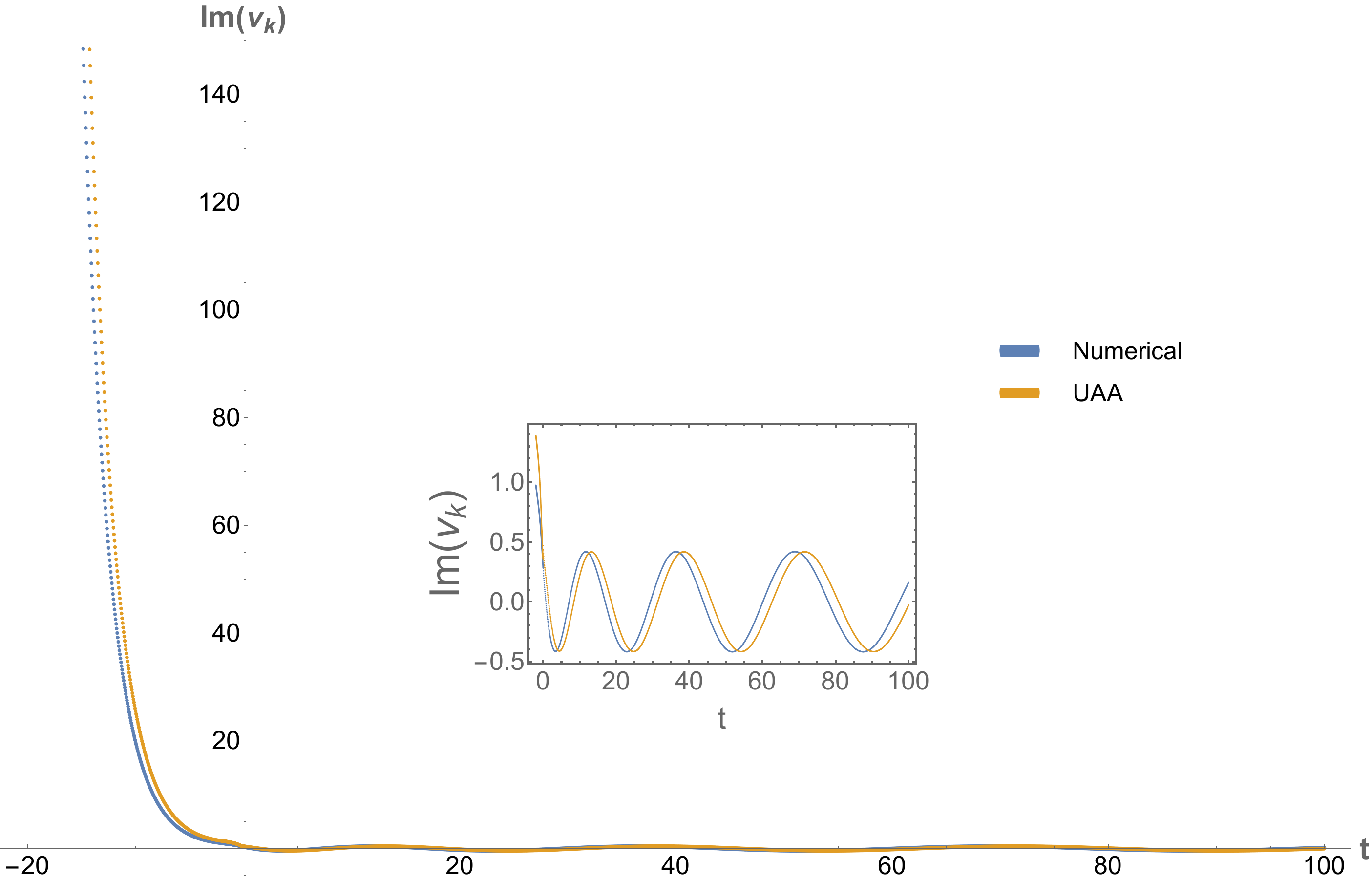}
}
\caption{The numerical and  UAA solutions with $k = 1.0 m_P$ and $b_k=0$. The corresponding turning points are $y_0= 1.4285 \;(t_0 \simeq -1.15458)$, $ y_1= 1.99364 \;(t_1\simeq -0.375304) $ and $ y_2= 2.61394 \;(t_2 \simeq 0.269674)$. The numerical solutions are obtained by integrating the MS with the initial conditions at $t_i = -20\; t_P$. The UAA solution is given by Eq.(\ref{eq4.18}). }
\label{fig16:num-vs-uaa-k1}
\end{figure}

\subsection{$k \lesssim k_e$} 

In this case, we find that $y_0 \simeq y_1 \ll y_2$, as can be seen form Fig. \ref{fig17:g-k087}. So, now we must treat the two points $y_0$ and $y_1$ as double-turning points and $y_2$ as a single-turning point. Then, we find that the whole $y$-axis  can be still divided into four regions, I: $y \leq y_a$; II: $y_a \leq y \leq y_b$; III: $ y_b \leq y \leq y_c$; and IV: $y \geq y_c$, but now with $y_a < y_0 < y_1 < y_b$. In Regions I, III and IV, the mode function is given by
\bqn
\lb{eq5.51}
\nu_k(y) = \begin{cases}
- \left\{a_k e^{-ik\eta}\left(1 - \frac{i}{k\eta}\right) + b_k e^{ik\eta}\left(1 + \frac{i}{k\eta}\right)\right\}, & \text{Region I},\cr
\left(\frac{ \zeta}{ g}\right)^{1/4} \Big\{\tilde\alpha_k {\rm{Ai}}\left( \zeta \right) + \tilde\beta_k{\rm{Bi}}\left( \zeta \right)\Big\}, & \text{Region III}, \cr
    \frac{1}{\sqrt{2k}}\left(\alpha_ke^{-ik\eta}+ \beta_k e^{ik\eta}\right), & \text{Region IV}. \cr
\end{cases}
\eqn

\begin{figure}[htp]
\includegraphics[width=10cm]{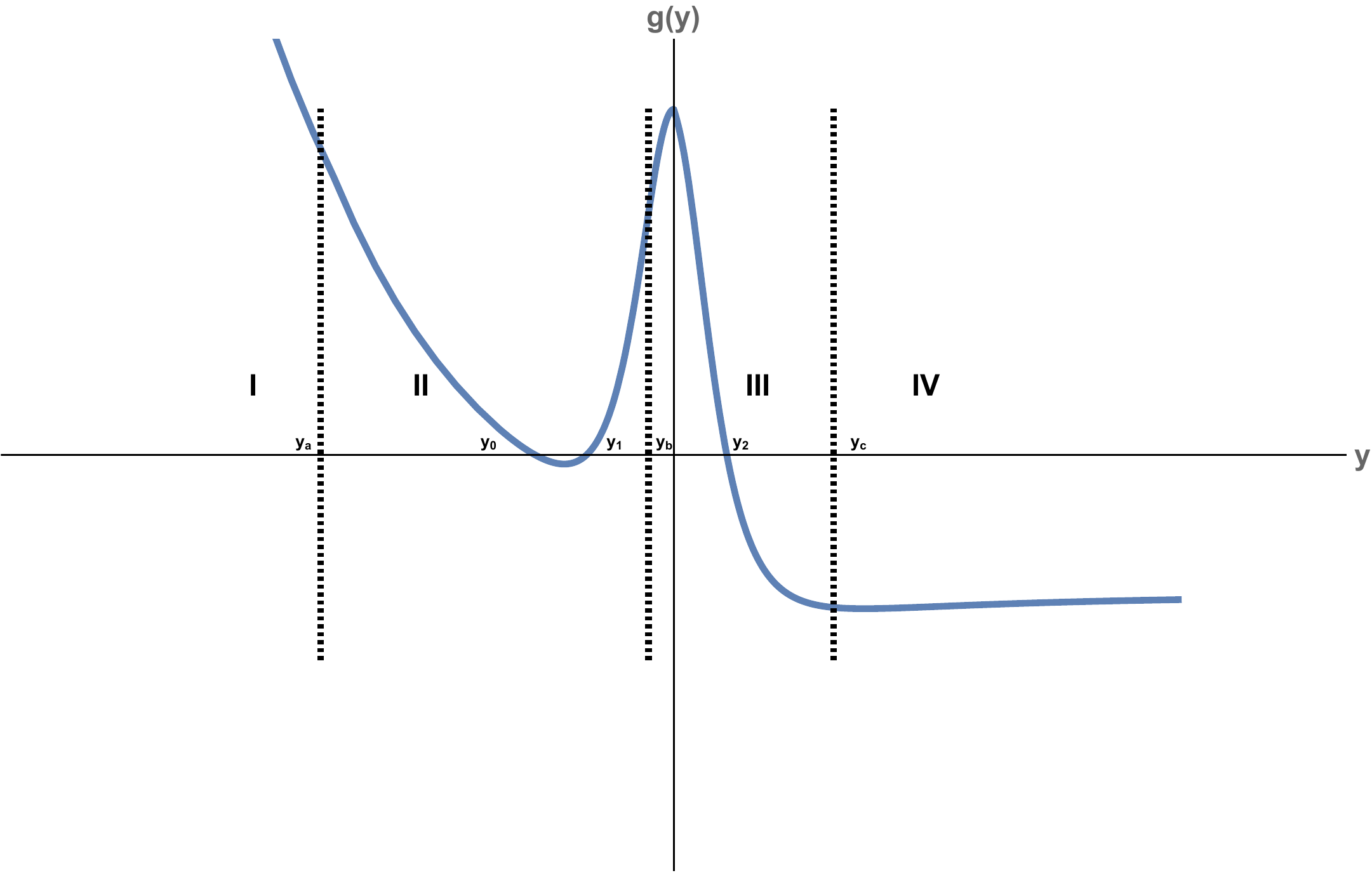}
\caption{The four regions, I - IV for $k=0.87 m_P$, defined, respectively, as I: $y \leq y_a$; II: $y_a \leq y \leq y_b$; III: $ y_b \leq y \leq y_c$; and IV: $y \geq y_c$.}
\label{fig17:g-k087}
\end{figure}

Similar to the last case, now the argument $\zeta$ of the Airy functions ${\rm{Ai}}\left(\zeta\right)$ and ${\rm{Bi}}\left(\zeta\right)$ is defined as that given by Eqs.(\ref{eq5.22aa}) and (\ref{eq5.22bb}), but with the only difference that now we have 
\bqn
\lb{eq5.51aa}
\zeta(y) = \begin{cases}
    > 0, & y < y_2, \cr
      = 0, & y = y_2, \cr
        < 0, & y > y_2, \cr
\end{cases}
\eqn
so that $\zeta$ has the same sign as that of $g(y)$ in the neighborhood of $y_2$.

On the other hand, in the region $ y_0 < y < y_1$ now we have $g(y) < 0$. This is opposite to the last case, in which we have 
$g(y) > 0$ in the region between the two turning points $y_1 < y < y_2$, as can be seen from Fig. \ref{fig13:g-k155}. Thus, in Region II now we choose \cite{Olver:1975a}
\bq
\lb{eq5.52}
\dot{y}^2 g(y) = \xi^2 - \xi_0^2,
\eq
with $\xi(y_0) = - \xi_0$ and $\xi(y_1) =  \xi_0$, where $\dot{y} \equiv dy/d\xi$ and $\xi_0 \geq 0$. Again, choosing $\xi(y)$ as a monotonically increasing function, that is, $\dot{y} = \sqrt{(\xi^2 - \xi_0^2)/g}$,  we find that
\bqn
\lb{eq5.53}
\int_{y_0}^y\sqrt{-g(y)} dy &=& \int_{-\xi_0}^{\xi}\sqrt{\xi_0^2 - \xi^2} d\xi \nb\\
&=&\frac{1}{2}\xi_0^2\cos^{-1}\left(-\frac{\xi}{\xi_0}\right)
+ \frac{1}{2}\xi\left(\xi_0^2 - \xi^2\right)^{1/2},  \; (y_0 \leq y \leq y_1),
\eqn
from which we obtain
\bq
\lb{eq5.54}
\xi_0^2 = \frac{2}{\pi}\int_{y_0}^{y_1}\sqrt{-g(y)} dy.
\eq
Similarly, in other regions we have
\bqn
\lb{eq5.55}
\int_y^{y_0}\sqrt{g(y)} dy &=& \int_{\xi}^{-\xi_0}\sqrt{\xi^2 - \xi_0^2} d\xi \nb\\
&=& - \frac{1}{2}\xi_0^2\ln\left(\frac{-\xi + \sqrt{\xi^2-\xi_0^2}}{\xi_0}\right)
- \frac12\xi\left(\xi^2 - \xi_0^2\right)^{1/2}, \; (y_a < y \leq y_0), ~~~~~~~~
\eqn
and 
\bqn
\lb{eq5.56}
\int_{y_1}^{y}\sqrt{g(y)} dy &=& \int_{\xi_0}^{\xi}\sqrt{\xi^2 - \xi_0^2} d\xi \nb\\
&=& - \frac{1}{2}\xi_0^2\ln\left(\frac{\xi + \sqrt{\xi^2-\xi_0^2}}{\xi_0}\right)
+ \frac{1}{2}\xi\left(\xi^2 - \xi_0^2\right)^{1/2}, \;  (y_1 \leq y < y_b).
\eqn

Then, the general solution in Region II can be written as \cite{Olver:1975a}
\bqn
\lb{eq5.57}
\nu_k^{\text{II}}(\xi) = \left(\frac{\xi^2 - \xi_0^2}{g}\right)^{1/4}\left[\tilde{a}_k U\left(- \frac{1}{2} \xi_0^2, \sqrt{2}\xi\right)
+ \tilde{b}_k \bar{U}\left(- \frac{1}{2} \xi_0^2, \sqrt{2}\xi\right)\right],
\eqn
where in terms of the confluent hypergeometric function ${}_1 F_{1}(a; b; x)$ the functions $U\left(b, \pm x\right)$ and 
$\bar{U}\left(b, \pm x\right)$ are defined as
\bqn
\lb{eq5.58}
U\left(b, \pm x\right) &=& \frac{\sqrt{\pi}\;  e^{-x^2/4}}{2^{(2b+1)/4}\Gamma\left(\frac{3}{4} + \frac{1}{2}b\right)}\;
{}_1 F_{1}\left(\frac{1}{2}b + \frac{1}{4}; \frac{1}{2}; \frac{1}{2}x^2\right) \nb\\
&&  \mp \frac{\sqrt{\pi}\; x e^{-x^2/4}}{2^{(2b-1)/4}\Gamma\left(\frac{1}{4} + \frac{1}{2}b\right)}\;
{}_1 F_{1}\left(\frac{1}{2}b + \frac{3}{4}; \frac{3}{2}; \frac{1}{2}x^2\right),\\
\lb{eq5.59}
\bar{U}\left(b, \pm x\right) &=& \frac{\Gamma\left(\frac{1}{4} - \frac{1}{2}b\right)\sin\left(\frac{3}{4}\pi - \frac{1}{2}b\pi\right) e^{-x^2/4}}{2^{(2b+1)/4}\; \sqrt{\pi}} \;
{}_1 F_{1}\left(\frac{1}{2}b + \frac{1}{4}; \frac{1}{2}; \frac{1}{2}x^2\right) \nb\\
&&  \mp  \frac{\Gamma\left(\frac{3}{4} - \frac{1}{2}b\right)\sin\left(\frac{5}{4}\pi - \frac{1}{2}b\pi\right) x e^{-x^2/4}}{2^{(2b-1)/4}\; \sqrt{\pi}}\;
{}_1 F_{1}\left(\frac{1}{2}b + \frac{3}{4}; \frac{3}{2}; \frac{1}{2}x^2\right).
\eqn
From the above definitions, it can be shown that
\bqn
\lb{eq5.62}
{U}\left(b, -x\right) &=& \cos(b\pi)\; \bar{U}\left(b, x\right) - \sin(b\pi)\; {U}\left(b, x\right),\nb\\
\bar{U}\left(b, -x\right) &=& \sin(b\pi)\; \bar{U}\left(b, x\right) + \cos(b\pi)\; {U}\left(b, x\right).
\eqn
For $x \gg 1$ we have
\bqn
\lb{eq5.60}
U\left(b, x\right) &\simeq& x^{-b-1/2} \; e^{-x^2/4}, \quad U'\left(b, x\right) \simeq -\frac{1}{2}x^{-b+1/2}\; e^{-x^2/4}, \nb\\
\bar{U}\left(b, x\right) &\simeq& \sqrt{\frac{2}{\pi}}\; \Gamma\left(\frac{1}{2} - b\right) \; x^{b-1/2} \; e^{x^2/4}, \quad
\bar{U}'\left(b, x\right) \simeq \sqrt{\frac{1}{2\pi}}\; \Gamma\left(\frac{1}{2} - b\right) \; x^{b+1/2} \; e^{x^2/4},\nb\\
\lb{eq5.61}
U\left(b, -x\right) &\simeq& \frac{(2\pi)^{1/2}}{\Gamma\left(\frac{1}{2} + b\right)} \; x^{b-1/2} \; e^{x^2/4}, \quad 
U'\left(b, - x\right) \simeq - \frac{\pi^{1/2}}{\sqrt{2} \; \Gamma\left(\frac{1}{2} + b\right)} x^{b+1/2}\; e^{x^2/4},\nb\\
\bar{U}\left(b, -x\right) &\simeq& \sqrt{\frac{2}{\pi}}\; \Gamma\left(\frac{1}{2} - b\right)\sin(b\pi) \; x^{b-1/2} \; e^{x^2/4}
+ \cos(b\pi)x^{-b-1/2} \; e^{-x^2/4},\nb\\
\bar{U}'\left(b, -x\right) &\simeq& \sqrt{\frac{1}{2\pi}}\; \Gamma\left(\frac{1}{2} - b\right)\sin(b\pi) \; x^{b+1/2} \; e^{x^2/4}
-\frac{1}{2} \cos(b\pi)x^{-b+1/2} \; e^{-x^2/4}.
\eqn
With the above asymptotical behaviors of $U\left(b, \pm x\right)$ and $\bar{U}\left(b, \pm x\right)$ let us consider the junction conditions across each of the boundaries of the four regions, I - IV, separately.

\subsubsection{Junction Conditions Across Regions I - II}

To match the mode function $\nu_k^{\text{II}}$ given by Eq.(\ref{eq5.57}) to the one  $\nu_k^{\text{I}}$ given by  Eq.(\ref{eq4.11}), we first note the relation between $y$ and $\xi$ for $y \ll 1$ (or $\zeta \ll - \zeta_0$) given by Eq.(\ref{eq5.55}), from which we find that 
\bqn
\lb{eq5.66}
y^{-3/2} \simeq x^b e^{\frac{1}{4}x^2}, \quad
b \equiv -\frac{1}{2} \xi_0^2, \quad x = - \sqrt{2}\xi \gg 1.
\eqn
Then, from Eq.(\ref{eq5.61}) we find that  
\bqn
\lb{eq5.67}
\nu_{k}^{\text{II}}(y) \simeq \left(\frac29\right)^{1/4}\Bigg[\tilde{a}_k \frac{(2\pi)^{1/2}}{\Gamma\left(\frac{1}{2} + b\right)}
+ \tilde{b}_k\sqrt{\frac{2}{\pi}}\;\Gamma\left(\frac{1}{2} - b\right)\sin(\pi b)\Bigg] \frac{1}{y} + \left(\frac29\right)^{1/4}\tilde{b}_k\cos(\pi b) \; y^2,\nb\\
\eqn
Hence, we find that the matching conditions $\nu_{k}^{\text{II}}(y) = \nu_{k}^{\text{I}}(y)$ lead to
\bqn
\lb{eq5.68}
\tilde{a}_k &=& \frac{\Gamma\left(\frac{1}{2} + b\right)}{\sqrt{2\pi}}\Bigg[\left(\frac92\right)^{1/4}i\left(a_k - b_k\right) - \frac{\Gamma\left(\frac{1}{2} - b\right)\tan(\pi b)}{\sqrt{\pi}} \left(\frac29\right)^{1/4}\left(a_k + b_k\right)\Bigg], \nb\\
\tilde{b}_k &=& \left(\frac{1}{18}\right)^{1/4}\frac{1}{\cos(b\pi)}\left(a_k + b_k\right).  
\eqn

\subsubsection{Junction Conditions Across Regions II - III}

For $y > y_1$ from Eq.(\ref{eq5.56}) we find that 
\bqn
\lb{eq5.69}
{\cal{F}}_1(y) \equiv \int_{y_1}^{y}\sqrt{g(y)} dy \simeq \frac{1}{4}x^2 + b\ln\left(\sqrt{2}\;x\right),
\eqn
but now with $x \equiv \sqrt{2}\xi > 0$, while the parameter $b$ is still defined as that in Eq.(\ref{eq5.66}). Then, using the asymptotical behavior of 
$U(b, x)$ and $\bar{U}(b, x)$ for $x \gg 1$ givne by Eq.(\ref{eq5.60}), we find that
\bqn
\lb{eq5.70}
\nu_{k}^{\text{II}}(y) &\simeq& \frac{1}{(2g)^{1/4}}\Bigg[\tilde{a}_k e^{-{\cal{F}}_1(y)}  
+ \tilde{b}_k\sqrt{\frac{2}{\pi}}\;\Gamma\left(\frac{1}{2} - b\right) e^{{\cal{F}}_1(y)}\Bigg],\;
(y \gg 1).
\eqn
On the other hand, for $y_b < y \leq y_2$,   we have 
\bqn
\lb{eq5.71}
{\cal{F}}_2(y) \equiv \int_{y}^{y_2}\sqrt{g(y)} dy  = - \int^0_{\zeta}\sqrt{\zeta} d\zeta = \frac{2}{3}\zeta^{3/2}.
\eqn
 Considering the asymptotical behavior of $\rm{Ai}(\zeta)$ and $\rm{Bi}(\zeta)$ given by Eq.(\ref{eq4.9}) for $\zeta \gg 1$, we find that 
\bqn
\lb{eq5.72}
\nu_{k}^{\text{III}}(y) &\simeq& \frac{1}{2\pi^{1/2} g^{1/4}}\left[\tilde{\alpha}_k   e^{-{\cal{F}}_2(y)}  
+ 2 \tilde{\beta}_k  e^{{\cal{F}}_2(y)}\right],\; (y \gg 1). 
\eqn
Then, the continuous condition $\nu_{k}^{\text{II}}(y) = \nu_{k}^{\text{III}}(y)$ leads to
\bqn
\lb{eq5.73}
\tilde{\alpha}_k &=& 2^{5/4} \Gamma\left(\frac{1}{2} - b\right) e^{\Phi}\; \tilde{b}_k, \quad
\tilde{\beta}_k = \left(\frac{\pi^{2}}{2}\right)^{1/4} e^{-\Phi}\; \tilde{a}_k,
\eqn
where
\bqn
\lb{eq5.74}
\Phi \equiv  \int_{y1}^{y_2}\sqrt{g(y)} \; dy.  
\eqn

 It should be noted that when $|y_2 - y_1|$ is not very large, the asymptotic matching between $\nu^{\text{II}}_k(y)$ and $\nu^{\text{III}}_k(y)$ presented above
cannot be applied, and one can simply impose the continuous conditions across the point $y = y_b$ by
\bq
\lb{eq5.74aa}
\nu^{\text{II}}_k(y_b) = \nu^{\text{III}}_k(y_b), \quad \frac{d\nu^{\text{II}}_k(y_b)}{dy} = \frac{d\nu^{\text{III}}_k(y_b)}{dy},
\eq
where $y_1 < y_b < y_2$, as shown in Fig. \ref{fig17:g-k087}.

\subsubsection{Junction Conditions Across Regions III - IV}

Finally, let us consider the junction conditions across Regions III and IV. let us first notice that for $y \gg y_2$, we have
\bqn
\lb{eq5.75}
\int_{y_2}^y\sqrt{-g(y)} \; dy \simeq y - y_2 = \frac{2}{3}\left(-\zeta\right)^{3/2}.
\eqn
Then, from the asymptotical behavior of $\rm{Ai}(-x)$ and $\rm{Bi}(-x)$ for $x \gg 1$ given by Eq.(\ref{eq4.15}),  we find that
\bqn
\lb{eq5.76}
\nu_{k}^{\text{III}}(y) &\simeq& \frac{1}{\pi^{1/2}}\Bigg\{\left[\tilde{\alpha}_k \cos\hat{y}_2 +   \tilde{\beta}_k  \sin\hat{y}_2\right]\cos y
+ \left[\tilde{\alpha}_k \sin\hat{y}_2 -   \tilde{\beta}_k  \cos\hat{y}_2\right]\sin y\Bigg\}, 
\eqn
where $\hat{y}_2 \equiv y_2 + \pi/4$. Matching it with the one given by Eq.(\ref{sol3A}) we find that 
\bqn
\lb{eq5.77}
\alpha_k &=& \sqrt{\frac{k}{2\pi}} \left(\tilde{\alpha}_k e^{i\hat{y}_2} - \tilde{\beta}_k e^{-i\hat{y}_2}\right), \quad
\beta_k = \sqrt{\frac{k}{2\pi}} \left(\tilde{\alpha}_k e^{-i\hat{y}_2} + \tilde{\beta}_k e^{i\hat{y}_2}\right).
\eqn

Inserting Eqs.(\ref{eq5.68}) and (\ref{eq5.73}) into the above equation, we can finally write $\alpha_k$ and $\beta_k$ in terms of $a_k$ and $b_k$. 
Thus, once the initial condition ($a_k, b_k$) is given, we can find $\alpha_k$ and $\beta_k$, which uniquely determine the mode function $\nu_k$. 

In Fig. \ref{fig18:num-vs-uaa-k087}, we plot the UAA solution together with the numerical one for $k = 0.87 m_p$, for which we find that the three roots of $g(y) = 0$
are given respectively by $y_0 \simeq 1.42929$ ($t_0 \simeq -0.823261$), $y_1 \simeq 1.62883$ ($t_1 \simeq -0.518961$) and $y_2 \simeq 2.31094$ ($t_2 \simeq 0.31563$).
From the figure, we can see again that our UAA solution traces the numerical one well, and captures the main features of the numerical solutions.

\begin{figure}[htp]
{
\includegraphics[width=8cm]{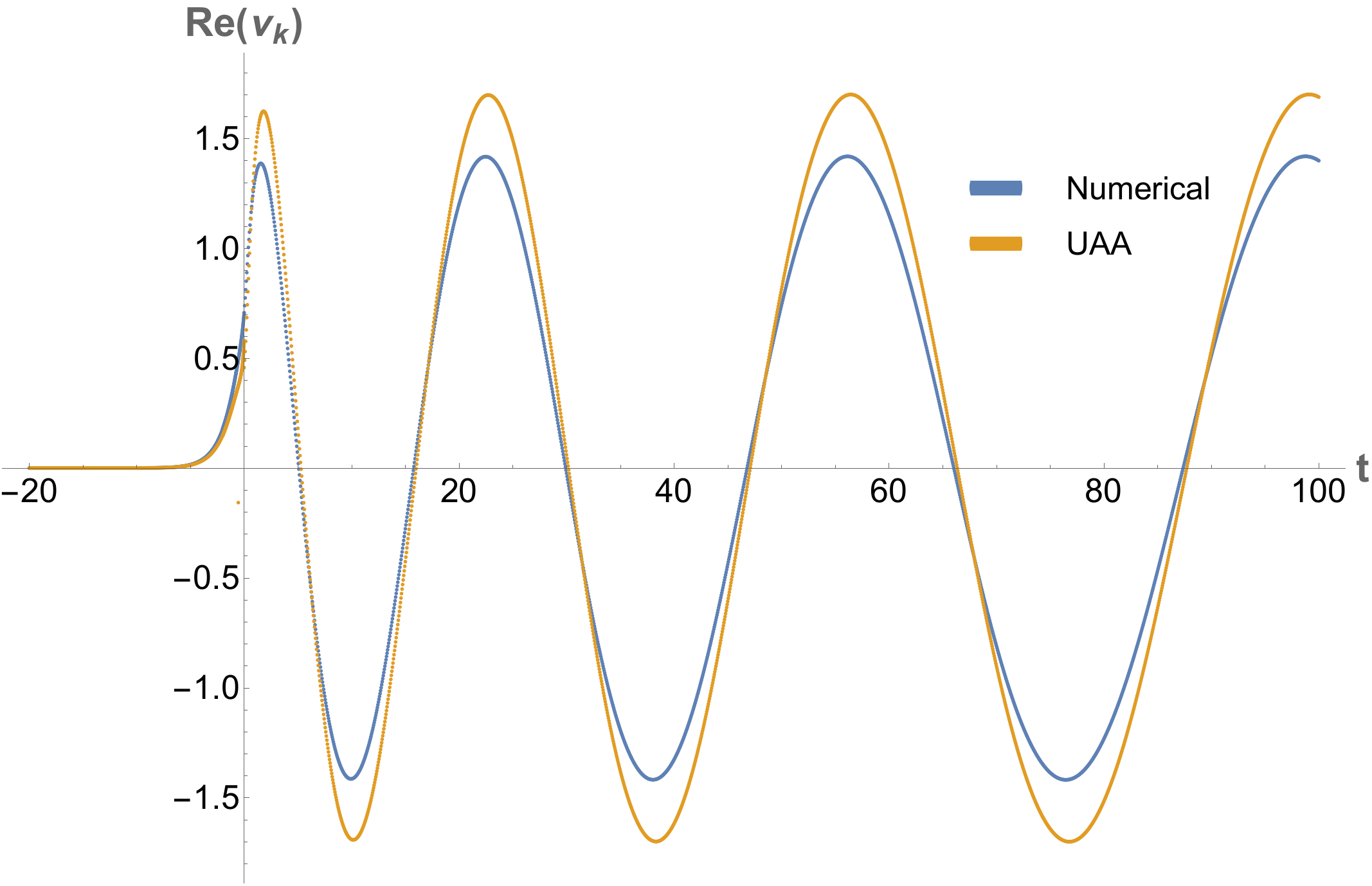}
\includegraphics[width=8cm]{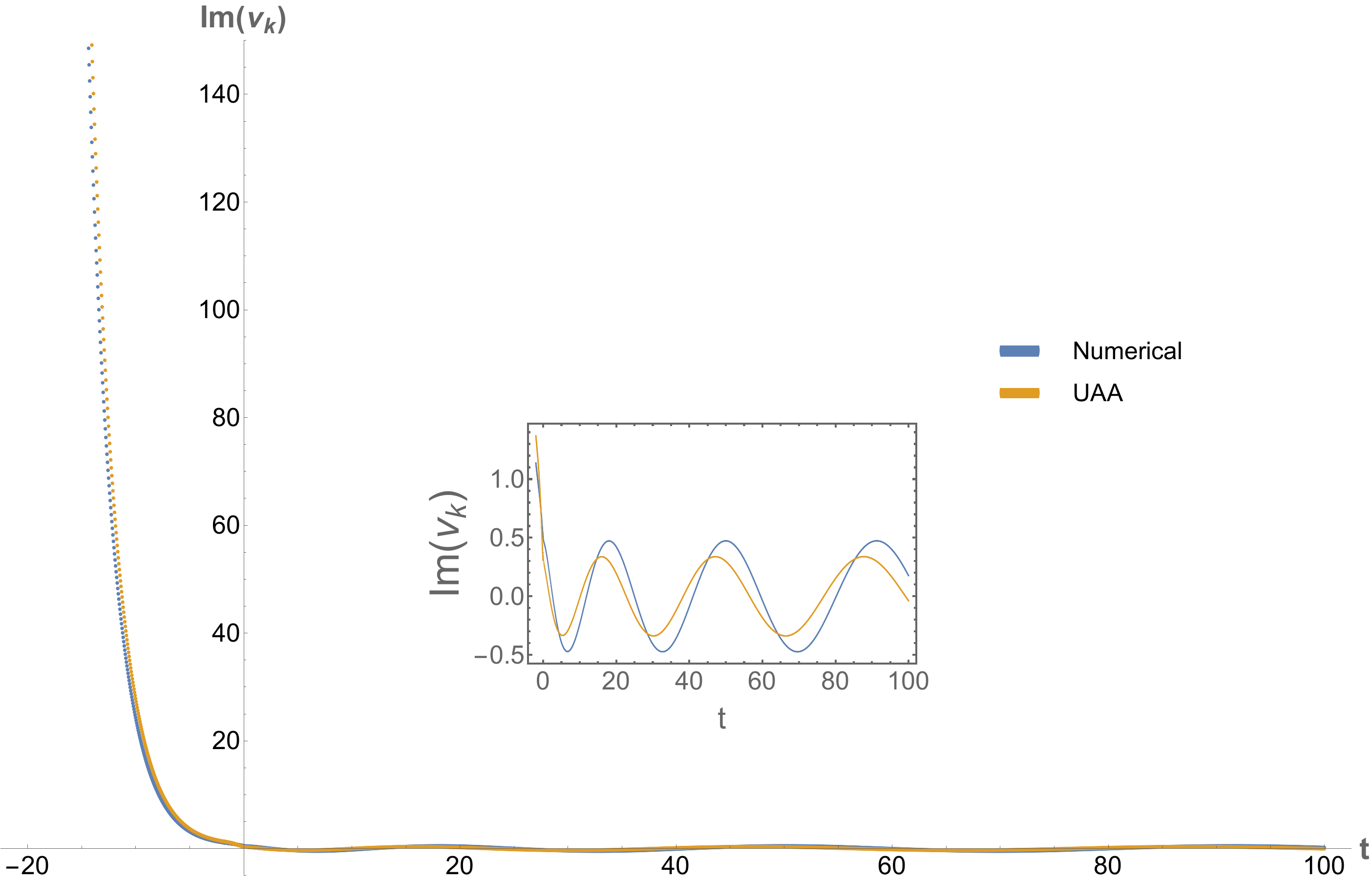}
}
\caption{The numerical and UAA solutions with $k = 0.87m_p$ and $b_k = 0$. The corresponding turning points are $y_0 \simeq 1.42929$ ($t_0 \simeq -0.823261$), $y_1 \simeq 1.62883$ ($t_1 \simeq -0.518961$) and $y_2 \simeq 2.31094$ ($t_2 \simeq 0.31563$). The numerical solutions are obtained by integrating the MS with the initial conditions at $t_i = -20 t_p$. The UAA solutions correspond to Eq.~\eqref{eq5.51} and Eq.~\eqref{eq5.57}.}
\lb{fig18:num-vs-uaa-k087}
\end{figure}

To extend our above studies to the case $k < k_1$, in which $g(y) = 0$ has only one real root, as shown in Fig. \ref{fig8:g5k}, we can consider the two additional roots complex roots of  $g(y) = 0$, as we did in Eqs.(\ref{eq4.23aa}) and (\ref{eq4.22aa}), bu now with 
\bq
\lb{eq5.77aa}
\xi\left(y_0\right) = - i\left|\xi_0\right|, \quad \xi\left(y_1\right) =  i\left|\xi_0\right|, 
\eq
where $\xi_0^2 < 0$, given by  
\bqn
\lb{eq5.77bb}
\xi_0^2 \equiv - \frac{2}{\pi} \left|\int_{y_0}^{y_1} \sqrt{-g(y)} \; dy\right| = -  \frac{2}{\pi} \left|\int_{-i|\xi_0|}^{i|\xi_0|} \sqrt{\xi^2 + |\xi_0^2|} \; d\xi\right|,
\eqn
where the integrations are performed along the imaginary axes \cite{Olver:1975a}. Hence, the parameter $b$ introduced in Eq.(\ref{eq5.66}) now becomes positive
\bq
\lb{eq5.77cc}
b = - \frac{1}{2}\xi_0^2 = \frac{1}{2}\left|\xi_0\right|^2.
\eq
Except for this difference, all the formulas developed above remain the same forms, including our UAA solutions.

\section{Initial Conditions and Power Spectrum of Cosmological Perturbations}
\lb{InitialConditions}

In LQC and modified loop quantum cosmologies (mLQCs), including mLQC-I, an important question is how to impose the initial conditions \cite{Kowalczyk:2024ech}. This problem includes two parts: (a) the time when the initial conditions are imposed, and (b) what are the initial conditions that should be imposed.
Certainly, these two parts are not independent and closely related to each other. In particular, the initial conditions depend crucially on the choice of the initial time. 
In the literature, three different times have been often considered: 1) time in the remote contracting phase $(t_i \ll t_B)$ \cite{Agullo:2017eyh,Zhu:2017jew,Li:2021mop,Li:2024xxz}; 2) the bounce time $(t_i =  t_B)$ \cite{Agullo:2013ai,Kowalczyk:2025qbi}; and 3) the silent point $t_i = t_{\text{silent}}$ \cite{Mielczarek:2014kea,Li:2018vzr}. In addition to these moments,  conditions imposed in the whole process of the evolution of the perturbations have been also considered, including the ones: i) generalized Penrose's vanishing Weyl curvature hypothesis \cite{Ashtekar:2016pqn,Ashtekar:2016wpi},   ii) non-oscillatory power spectrum \cite{deBlas:2016puz,CastelloGomar:2017kbo}; and more recently iii) low energy states \cite{Martin-Benito:2021szh,Navascues:2021qcp}.

In this paper, we choose to impose the initial conditions in the remote contracting phase, that is, $\eta_i \ll 1$. 
Then, the question reduces to what conditions we should impose at $\eta_i$. As mentioned above, in the remote contracting phase, some of the physically relevant modes are outside of the Hubble horizon, and the adiabatic vacuum cannot be applied, including  the BD vacuum \cite{Baumann:2009ds}. To see this more clearly, let us first recall how one impose the BD vacuum in GR.

\subsection{The BD Vacuum in Classical Slow-Roll Inflationary Models} 

To understand the problem properly, let us first briefly review the BD vacuum often imposed in classical slow-roll inflationary models, in which the 
backgroubnd is well-approximed by the expanding de Sitter spacetime
\bqn
\lb{eq6.1}
ds^2 = -d\hat{t}^2 + e^{2H_{\Lambda}\hat{t}}d^2\vec{x} = C(\hat\eta)\left(-d\hat\eta^2 + d^2\vec{x}\right),
\eqn
where $d^2\vec{x} \equiv dx^2 + dy^2 + dz^2$, $C(\hat\eta) \equiv a^2(\hat\eta) = 1/(H_{\Lambda}\hat\eta)^2$, and 
\bqn
\lb{eq6.2}
\hat\eta \equiv \int{\frac{dt}{a(\hat{t})}} = - \frac{1}{H_{\Lambda}a(\hat{t})} %e^{-H_{\Lambda}\hat{t}} 
= \begin{cases}
  0, & \hat{t} \rightarrow \infty, \cr
    -\infty, & \hat{t} \rightarrow -\infty. \cr
\end{cases}
\eqn
The initial conditions in classical relativity are imposed in the very early time of the universe, at which we have $a(\hat{t}) \simeq 0$, which corresponds to  $\hat{t} \simeq -\infty$. In the Penrose diagram of the de Sitter spacetime, given by Fig. \ref{fig1:Penrose}, it corresponds to the diagonal line BC, which is null. On the other hand, the horizontal line AB corresponds to $\hat{t} = \infty$, which is a spacelike hypersurface. 

The mode function in the de Sitter background satisfies the following MS equation \cite{Baumann:2009ds}
\bqn
\lb{eq6.3}
\frac{d^2\nu_k(\hat\eta)}{d^2\hat\eta} + \left(k^2 - \frac{2}{\hat\eta^2}\right)\nu_k(\hat\eta) = 0,   
\eqn
which has the general solutions,
\bqn
\lb{eq6.4}
 \nu_k = \hat\alpha_k e^{-ik\hat\eta} \left(1 - \frac{i}{k\hat\eta}\right) + \hat\beta_k e^{ik\hat\eta} \left(1 + \frac{i}{k\hat\eta}\right),   
\eqn
where $\hat\alpha_k$ and $\hat\beta_k$ are two integration constants and shall be determined by the initial conditions. Since
\bqn
\lb{eq6.5}
\frac{d^n}{d\hat\eta^n}\left(\frac{C_{,\hat\eta}}{C}\right)  =  \frac{2(n!)}{(-\hat\eta)^{n+1}} \rightarrow 0,
\eqn
as $\hat\eta \rightarrow -\infty$, the adiabatic condition is satisfied \cite{Birrell:1982ix}. As a result, one can use the adiabatic approximations
\bqn
\lb{eq6.6}
\nu_k(\hat\eta) &=& \frac{1}{\sqrt{2W_k(\hat\eta)}}\exp\left\{- \int{W_k(\hat\eta)d\hat\eta}\right\},\nb\\
W_k^2(\hat\eta) &=& \omega_k^2 - \frac{1}{2}\left(\frac{W_k''}{W_k} - \frac{3{W_k'}^2}{2 W_k^2}\right),  
\eqn
where $\omega_k^2 \equiv k^2 - 2/\hat\eta^2$. In particular, the BD vacuum (which is also called the Euclidean vacuum) 
\bqn
\lb{eq6.7}
 \nu_k^{\text{initial}}{}_{\text{GR}}(\hat\eta) = \frac{1}{\sqrt{2k}} e^{-ik\hat\eta},   
\eqn
is commonly chosen as the initial condition imposed at $\hat\eta \simeq -\infty$. Clearly, this corresponds to the choice  
\bqn
\lb{eq6.8}
\hat\alpha_k = \frac{1}{\sqrt{2k}}, \quad \hat\beta_k = 0,      
\eqn
of the general solution given by Eq.(\ref{eq6.4}).

\subsection{The Initial Conditions in mLQC-I} 

In mLQC-I, as shown above, the spacetime is also asymptotically de Sitter as $t \rightarrow -\infty$. However, since now the universe is in the contracting phase, we have 
\bqn
\lb{eq6.9}
ds^2 = -d{t}^2 + e^{-2H_{\Lambda}{t}}d^2\vec{x} = C(\eta)\left(-d\eta^2 + d^2\vec{x}\right),
\eqn
where $C(\eta) = a^2(\eta)$ and
\bqn
\lb{eq6.10}
a(t) = e^{-H_{\Lambda} t} = \frac{1}{H_{\Lambda}\eta}
= \begin{cases}
  \infty, & \eta \rightarrow 0, \cr
    0, & \eta \rightarrow \infty. \cr
\end{cases}
\eqn
Comparing the two metrics given by Eqs.(\ref{eq6.1}) and (\ref{eq6.9}) we can see that one can get one from the other by setting $\hat{t} = -t$. 
In particular, imposing the initial conditions at the remote contracting phase, $t = - \hat{t} = -\infty$, implies imposing the initial conditions at the spacelike horizontal line AB of Fig. \ref{fig1:Penrose}. This is completely different from the moment $\hat{t} = -\infty$, denoted by the diagonal line BC, at which the initial conditions in the classical inflationary models are usually imposed. In addition, now we have
\bqn
\lb{eq6.11}
{\frac{d^n}{d\eta^n}\left(\frac{C_{,\eta}}{C}\right)} =  \frac{2(n!)}{\eta^{n+1}} \rightarrow \infty,
\eqn
as $\eta \rightarrow 0$ (or $t \rightarrow -\infty$). Then, the adiabatic approximations (\ref{eq6.6}) cannot be applied to the current case. 

However,   following Birrell and Davies \cite{Birrell:1982ix} let us introduce a new variable $\eta_1 = \eta/T$, where $T$ is called the adiabatic parameter. Then, in terms of $\eta_1$ the MS equation (\ref{eq6.3})  takes the form
\bqn
\lb{eq6.12}
\nu_k''(\eta_1) + \Omega_k^2\nu_k(\eta_1) = 0,   
\eqn
where 
\bqn
\lb{eq6.13}
\Omega_k^2 \equiv T^2\omega_k^2 = T^2 k^2 - \frac{2}{\eta_1^2}.
\eqn
Eq.(\ref{eq6.11}) has the general solutions
\bqn
\lb{eq6.13aa}
\nu_k(\eta_1) = -a_ke^{-ikT\eta_1}\left(1 - \frac{i}{kT\eta_1}\right) - b_k e^{ikT\eta_1}\left(1 + \frac{i}{kT\eta_1}\right).
\eqn

Since $dC(\eta_1)/d\eta = T^{-1}dC(\eta_1)/d\eta_1$, we find that 
\bqn
\lb{eq6.14}
\frac{d^n}{d\eta^n}\left(\frac{C_{,\eta}(\eta_1)}{C(\eta_1)}\right)  =  \frac{2(-1)^n(n!)}{T^{n+1}\eta^{n+1}} \rightarrow 0,
\eqn
as $T \rightarrow \infty$ for any given finite and non-zero $\eta_1$. That is, $C(\eta_1)$ and all its derivatives with respect to $\eta$ vary infinitely slowly
for a very large $T$. In particular, $\Omega_k^2$ defined by Eq.(\ref{eq6.13}) is large with respect to the derivatives of $C_{,\eta}/C$ 
\bqn
\lb{eq6.17}
\Omega_k^2(\eta_1) \gg \frac{d^n}{d\eta^n}\left(\frac{C_{,\eta}(\eta_1)}{C(\eta_1)}\right),\; (n > 0),
\eqn
as $T \rightarrow \infty$ for fixed $\eta_1$. As a result, the adiabatic modes given by Eq.(\ref{eq6.6}) become good approximations.  In review of Eq.(\ref{eq6.13aa}) we can see that the adiabatic positive frequency modes correspond to $b_k = 0$. Then, the Wronskian (\ref{Wcd}) leads to 
\bq
\lb{eq6.15}
a_k = - \frac{1}{\sqrt{2k}}, \quad b_k = 0,
\eq 
so that
\bq
\lb{eq6.16} 
 \nu^{\text{initial}}_k(\eta) \simeq  \frac{1}{\sqrt{2k}}  e^{-ik\eta}\left(1 - \frac{i}{k\eta}\right), \; (\eta \ll 1).
\eq
Now several remarks are in order:
\begin{itemize}
    \item  First,  the above condition is different from that of Eq.(\ref{eq6.7}) commonly adopted in GR, 
as now we have $(k\eta)^{-1} \gg 1$ for sufficient early time ($\eta \ll 1$), so the term $i/(k\eta)$ in Eq.(\ref{eq6.16}) is usually not negligible. To distinguish these two vacuums, in \cite{Li:2021mop} the one given by Eq.(\ref{eq6.16}) is called {\em the de Sitter vacuum}.

\item Second, lately a new formalism to define vacuum states for a single field is introduced \cite{Penna-Lima:2022dmx}, in which distances between neighbor points are shown to be proportional to the Bogoliubov coefficients. The vacuum state for each mode is then defined as the unique trajectory from which all mapped phase space solutions move within thin annular regions around it. Such defined vacuum states are stable, in the sense that solutions evolved from a point in the trajectory stay close to it and the particle creation is minimized. Applying this new definition to the de Sitter spacetime, it 
is shown that the initial condition (\ref{eq6.15}) holds even for the non-adiabatic modes -  modes outside of the Hubble horizons in the contracting phase as shown in Fig. \ref{fig6:lamdaH}. 

\item Finally, an asymptotic Hamiltonian diagonalization (AHD) prescription was proposed in \cite{ElizagaNavascues:2019itm} in order to fix the initial condition of the perturbations, and it was found that it also leads to the initial condition (\ref{eq6.15}) in the de Sitter background. For further applications of AHD prescription, see, for example, Refs. \cite{Kowalczyk:2024ech,Kowalczyk:2025qbi} and references therein.

\end{itemize}

\section{Conclusions}
\lb{Sec:summary}

In this paper, we applied the UAA method \cite{Olver:1997} to find the first-order approximation solution of the mode function that satisfies the generalized MS equation (\ref{MS_equation}) in the dressed metric approach within the framework of mLQC-I \cite{Yang:2009fp,Li:2021mop}. Thanks to the universal and analytical background solutions given by 
Eqs.(\ref{eq2.21}), (\ref{eq2.23}) and (\ref{eq2.26}), which made such analytical analyses possible \footnote{It should be noted that the universal behavior of the background evolution holds only for initially kinetic-energy-dominated  conditions at the bounce, $\dot\phi_B^2 \gg 2V(\phi_B)$. For other initial conditions, such as initially potential-energy-dominated  ones, such universality gets lost and inflationary phase may not occur \cite{Li:2018fco,Li:2018opr,Li:2019ipm}. This is true not only in mLQC-I but also in LQC and other models \cite{Li:2021mop,Bonga:2015xna,Zhu:2017jew}. However, since inflation happens generically not only in LQC \cite{Ashtekar:2011rm} but also in mLQC-I and other models \cite{Li:2019ipm}, without loss of the generality, in this paper we have focused on the initial conditions that lead generically to inflation in the post-bounce region.}.

In particular, with the extension of the effective potential across the quantum bounce into the deep remote contracting phase considered in \cite{Li:2020mfi}, we analyzed the effective mass function $m_{\text{eff}}$ defined by Eqs.(\ref{eq3.9aaa}) and (\ref{eq3.9bbb}) in detail. As shown in Fig. \ref{fig5:meff}, the effective mass square  is always negative, so the effective Hubble horizon $\lambda_H^2 \equiv - m_{\text{eff}}^{-2}$ is positive and decreases exponentially as $t \rightarrow -\infty$ [cf. Eqs.(\ref{eq3.11}) and (\ref{eq3.12})], while in the post-bounce phases, $\lambda_H$ exhibits  complicated behaviors. Dividing the whole evolution into four different phases, pre-de Sitter, (extended) bouncing, transition and inflation, the mode function is given analytically by Eq.(\ref{sol1}) in the pre-de Sitter phase, and Eq.(\ref{sol3A}) in the transition and inflationary phases. Therefore, the mode function is not known only during the bouncing phase. 

To find analytical solution in this bouncing phase, we applied the UAA method \cite{Olver:1997} to the generalized MS equation (\ref{MS_equation}), and found that three cases need to be considered separately [cf. Fig. \ref{fig8:g5k}], depending on the number and nature of the turning points, the roots of the equation $g(y) = 0$ defined by Eq.(\ref{eq4.4}).   After expressing the mode function in terms of the well-known mathematical functions in each of the three cases, we matched the solutions together and expressed the two integration constants $\alpha_k$ and $\beta_k$ of the mode function given by Eq.(\ref{sol3A}) in the inflationary phase in terms of $a_k$ and $b_k$ appearing in the mode function in the pre-de Sitter phase. As a result, once the initial condition is chosen in the pre-de Sitter, the constants $\alpha_k$ and $\beta_k$ will be uniquely determined, so is the corresponding power spectrum given by Eq.(\ref{scalarpower}). Such determined $\alpha_k$ and $\beta_k$ will clearly depend on the comoving wavenumber $k$, that is, the resultant power spectrum can be expressed in the form
\bq
\lb{eq7.1}
P_\mathcal R=f(k) P^{GR}_\mathcal R,
\eq
where $P^{GR}_\mathcal R$ stands for the primordial power spectrum predicted in the classical general relativity. The above expression is exactly in the form found in \cite{Ashtekar:2020gec,Ashtekar:2021izi}, for which several anomalies found in the CMB observations can be alleviated.  

  {The function $f(k)$ certainly depends on the choice of the initial conditions. In Sec. VI,  using  a method developed by Birrell and Davies \cite{Birrell:1982ix}, we identify an initial state  in the remote contracting phase given by Eq.(\ref{eq6.16}), which turns out 
to be stable, minimize particle creations \cite{Penna-Lima:2022dmx}, and diagonalize the Hamiltonian \cite{ElizagaNavascues:2019itm}, despite the fact that at this time some modes may be still outside of the Hubble horizon and not in their adiabatic states.  It would be also very interesting to compare such obtained $f(k)$ with the ones obtained by other initial conditions \cite{Agullo:2013ai,Kowalczyk:2024ech,Kowalczyk:2025qbi,Ashtekar:2016pqn,Ashtekar:2016wpi,deBlas:2016puz,CastelloGomar:2017kbo,Martin-Benito:2021szh,Navascues:2021qcp,MenaMarugan:2024zcv,Alonso-Serrano:2023xwr}.}

In the forthcoming work, we shall calculate $f(k)$ explicitly, after    generalizing the current firs-order approximation to high-order ones, so that the approximate solutions  will have enough accuracy \cite{Zhu:2014aea,Zhu:2016srz},  in order to compare our theoretical predictions with current and forthcoming observations  \cite{Abazajian:2013vfg}.

To have the mLQC-I model physically viable, several other important issues also need to be addressed, including particle creation, anisotropy and non-Gausianity \cite{Agullo:2023rqq}. We wish to come back to these issues soon. 

 Finally, we note that in the studies of cosmological perturbations in LQC/mLQCs, there exist two main approaches in the literature, {\em the dressed metric} \cite{Agullo:2012sh,Agullo:2012fc,Agullo:2013ai}, and {\em the hybrid} \cite{ElizagaNavascues:2020uyf}. In the current paper,  we have mainly focused ourselves on the dressed metric approach. It would be very interesting to extend such studies to the  hybrid approach, of which we just studied numerically in \cite{Li:2020mfi} and observed some differences. It is important to understand such differences analytically, and then compare them with observations.

\section*{Acknowledgments}

 We thank Prof. Robert Brandenberger for reading our manuscript carefully, and valuable discussions, suggestions and comments. We also thank Dr. Bao-Fei Li for the early stage collaboration of the project and Drs. We also thank Qiang Wu and Tao Zhu for their kind help, suggestions, and comments. R.P. and A.W. are partially supported by the US NSF grant, PHY-2308845, while J.S. is supported by Baylor through the  Baylor Physics graduate program.

\bibliography{BibNVar}
\bibliographystyle{h-physrev5}

\end{document}